\documentclass[a4paper,fleqn,usenatbib]{mnras}

\usepackage{newtxtext,newtxmath}
\usepackage[T1]{fontenc}
\usepackage{ae,aecompl}

\usepackage{graphicx}	% Including figure files
\usepackage{amsmath}	% Advanced maths commands
\usepackage{amssymb}	% Extra maths symbols
\usepackage{bm}
\usepackage{graphics}
\usepackage{enumitem}
\usepackage{color}
\usepackage{xspace}
\usepackage{float}
\usepackage{listings}
\newcommand{\be}{\begin{equation}}
\newcommand{\ee}{\end{equation}}
\newcommand{\bea}{\begin{eqnarray}}
\newcommand{\eea}{\end{eqnarray}}
\newcommand{\bi}{\begin{itemize}[leftmargin=3.5mm]\setlength{\itemindent}{-1.6mm}}
\newcommand{\ei}{\end{itemize}}
\renewcommand{\r}{\mathbf{r}}

\newcommand{\msun}{{\rm M}_{\odot}}
\newcommand{\Dx}{\Delta x}
\newcommand{\dx}{dx}
\newcommand{\dftools}{\texttt{dftools}\xspace}
\newcommand{\hyperfit}{\texttt{hyper.fit}\xspace}
\newcommand{\veff}{V}%{V_{\rm eff}}
\newcommand{\vefflss}{V_{\rm LSS}}%{V_{\rm eff}^{\rm LSS}}
\newcommand{\vmax}{V_{\rm max}}

\renewcommand{\L}{\mathcal{L}}
\newcommand{\x}{\mathbf{x}}
\newcommand{\unitj}{{\rm kpc\,km\,s^{-1}}}
\newcommand{\fig}[1]{Fig.~\ref{fig:#1}}
\newcommand{\eq}[1]{equation~(\ref{eq:#1})}
\newcommand{\Eq}[1]{Equation~(\ref{eq:#1})}
\newcommand{\s}[1]{Section~\ref{s:#1}}
\renewcommand{\ss}[1]{Section~\ref{ss:#1}}
\renewcommand{\a}[1]{Appendix~\ref{a:#1}}
\newcommand{\ie}{i.e.\xspace}
\newcommand{\eg}{e.g.\xspace}
\newcommand{\cf}{c.f.\xspace}
\newcommand{\cov}{{\rm cov}}
\newcommand{\para}{{\bm{\theta}}}
\newcommand{\ha}{H{\sc\,i}\xspace}
\definecolor{darkgreen}{rgb}{0,0.7,0}

\title[Eddington's Demon]{Eddington's Demon: Inferring Galaxy Mass Functions and other Distributions from Uncertain Data}

\author[D. Obreschkow et al.]{
D. Obreschkow,$^{1,2}$\thanks{E-mail: danail.obreschkow@icrar.org}
S. G. Murray,$^{2,3}$
A. S. G. Robotham,$^1$
T. Westmeier,$^1$
\\
\\
$^{1}$International Centre for Radio Astronomy Research (ICRAR), University of Western Australia, 35 Stirling Highway,\\
~Crawley WA 6009, Australia\\
$^{2}$ARC Centre of Excellence for All-Sky Astrophysics (CAASTRO), Australia\\
$^{3}$International Centre for Radio Astronomy Research (ICRAR), Curtin University, 1 Turner Ave, Bentley WA 6102, Australia
}

\date{Accepted 2017 December 2. Received 2017 December 2; in original form 2017 September 25}

\pubyear{2015}

\begin{document}
\label{firstpage}
\pagerange{\pageref{firstpage}--\pageref{lastpage}}
\maketitle

\begin{abstract}
We present a general modified maximum likelihood (MML) method for inferring generative distribution functions from uncertain and biased data. The MML estimator is identical to, but easier and many orders of magnitude faster to compute than the solution of the exact Bayesian hierarchical modelling of all measurement errors. As a key application, this method can accurately recover the mass function (MF) of galaxies, while simultaneously dealing with observational uncertainties (Eddington bias), complex selection functions and unknown cosmic large-scale structure. The MML method is free of binning and natively accounts for small number statistics and non-detections. Its fast implementation in the $R$-package \dftools is equally applicable to other objects, such as haloes, groups and clusters, as well as observables other than mass. The formalism readily extends to multi-dimensional distribution functions, \eg a Choloniewski function for the galaxy mass--angular momentum distribution, also handled by \dftools. The code provides uncertainties and covariances for the fitted model parameters and approximate Bayesian evidences. We use numerous mock surveys to illustrate and test the MML method, as well as to emphasize the necessity of accounting for observational uncertainties in MFs of modern galaxy surveys.
\end{abstract}


%%%%%%%%%%%%%%%%%%%%%%%%%%%%%%%%%%%%%%%%%%%%%%%%%%

%%%%%%%%%%%%%%%%% TALK %%%%%%%%%%%%%%%%%%

%%%%%%%%%%%%%%%%% BODY OF PAPER %%%%%%%%%%%%%%%%%%

\section{Introduction}\label{s:introduction}

Few challenges have incited more publications by astrophysicists of several generations than the task of fitting a luminosity function (LF) or mass function (MF) to samples of stars, galaxies, groups and clusters \citep[\eg][]{Schmidt1968,Lynden-Bell1971,Turner1979,Sandage1979,Kirshner1979,Davis1982,Efstathiou1988,Zwaan2003,Teerikorpi2004,Cole2011,Loveday2015,Weigel2016}. LFs and MFs quantify the space density of these objects in the universe as a function of their luminosity and mass, respectively. They are fundamental observables that summarize the statistical outcome and scale-dependence of complex formation processes \citep[\eg][]{Croton2006,Murray2013}. LFs and MFs thus constitute a crucial bridge between astrophysical observations and theory, having shaped their synergetic progress for more than half a century.

For simplicity, this paper will talk about the `galaxy MF', however the concepts, formalisms and implementations are readily transferable to other objects, e.g.~stars, star clusters, galaxy clusters and dark haloes. Likewise, the galaxy mass $M$ is not further specified to emphasize its applicability to any choice, \eg stellar mass, gas mass, or dynamical mass. In fact, $M$ can be substituted for any other observable, including luminosity, absolute magnitude and even multi-dimensional observables. This article hence treats galaxy MFs as an example of the general problem of inferring a distribution function (DF) generating empirical data subject to measurement uncertainties and sample biases.

Faint galaxies are far more common than luminous, massive ones. In a fixed cosmic volume, the number of galaxies per unit mass approximately declines as a power law up to a cut-off scale, beyond which the number density declines almost exponentially. In appreciation of the power law behaviour, it is convenient to define the MF $\phi(x)$ as the the density of galaxies per unit volume and unit of logarithmic mass $x=\log_{10}(M/\msun)$. Formally, in a cosmic volume $V$, the expected number of galaxies in an interval $[x,x+dx]$ is
\be\label{eq:mf}
	dN = \phi(x)\,V\,dx.
\ee
The choice of $\log_{10}$-units is a subjective preference and easily converts to the natural-log MF, $\phi'(x)=\phi(x)/\ln10$, or linear MF, $\Phi(M)=\phi(x)/(M\ln10)$.

Many analytical parametric functions have been proposed to fit observed MFs. They can all be writen as $\phi(x|\para)$, where $\para$ is a vector of $P$ scalar model parameters. A common example is the Schechter function \citep{Schechter1976},
\be\label{eq:Schechter}
	\phi(x|\para)_{\rm Schechter} = \ln(10)\phi_\ast\mu^{\alpha+1}e^{-\mu},
\ee
where $\mu= M/M_\ast=10^x\msun/M_\ast$. This model depends on three parameters: the amplitude $\phi_\ast$, the break mass $M_\ast$ and the power law slope $\alpha$. The Schechter function is the best-known MF model that captures the truncated power law behaviour. We will use the Schechter function for most illustrations, but the formalism and implementations remain applicable to any DF model, including quasi non-parametric (`stepwise') models, discretized into a custom number of bins or vertices.

The measurement of MFs requires a redshift survey, providing a sample of galaxies with approximate distances and hence a means of converting apparent to absolute magnitudes and intrinsic mass. Fitting a MF model $\phi(x|\para)$ to these data is not trivial. The most basic and intuitive approach \citep{Schmidt1968} is to bin the data in mass and estimate an observed space density $\phi_j$ in each bin $j$, by dividing the number of detections by the bin width and the maximal volume $\vmax$, in which galaxies of that bin could have been detected ($1/\vmax$-method). A model function $\phi(x|\para)$ is then fitted to the bin values $\phi_j$. This method suffers from a list of limitations:
\bi
	\item The fit depends on the choice of binning (bin centres and spacing).
	\item It is not clear how $\phi(x|\para)$ is best fitted to the binned data. Given the varying number of galaxies per bin and Poisson statistics, least-square minimization is inaccurate.
	\item The inclusion of non-detections, \ie mass bins which happen to contain no galaxy, is often cumbersome because of the impossibility to assign Poisson errors to such bins.
	\item Dealing with observational uncertainties in the galaxy masses is difficult.
	\item Complex detection limits with source-dependent completeness and reliability (defined in \ss{gradual}) make the choice of $\vmax$ ambiguous.
	\item Cosmic large-scale structure (LSS) can introduce systematic errors (see \ss{lsstheory}).
\ei

Most of these challenges cannot be overcome by brute force alone, \eg by simply observing more galaxies, and hence remain an issue for modern spectroscopic redshift surveys detecting hundreds of thousands  \citep{Colless2001,York2000,Liske2015,Drinkwater2010,Grazian2015,Davidzon2017} to millions \citep{Dawson2013,Amiaux2012} of galaxies. Some challenges, such as small number statistics and LSS are particularly pronounced in small samples ($\lesssim10^3$ galaxies) and samples with a strong mass bias. Such samples often arise, by construction, when subsamples are drawn from larger sets to address cutting-edge topics (\eg highest redshifts, satellite population, rare environments).

All the caveats above have been addressed from different angles in the literature (\eg references in the first paragraph). The persisting problem is to overcome all of them at once (see \citealp{Pihajoki2017} for the related case of non-linear model fitting). In particular, approaches dealing with measurement uncertainties \cite[\eg][]{Teerikorpi2004}, especially prominent in cluster studies \citep{Mortonson2011,Evrard2014}, are hard to reconcile with approaches addressing all the other issues. Perhaps for this reason, a surprising number of MF and LF studies in modern galaxy surveys neglect measurement uncertainties or do not deal with them in a statistically accurate way. This bias is nonetheless significant in spite (or rather because) of the large number of detections in modern surveys. Another problem is that MF fitting methods are quite tricky or at least time-consuming to implement due to their inevitable mathematical complexity. Thus, most authors of MF papers of redshift surveys have spent considerable time developing/publishing their own techniques prior to processing the actual data.

The objective of this paper is to derive, demonstrate and implement a method that simultaneously overcomes all the caveats above, building on the extensive literature. This method, derived in \s{method}, is based on a variation of the maximum likelihood (ML) method -- with the statistical axioms that this approach entails (\a{exactsolution}) -- and makes only a few assumptions on the nature of the data and model to be fitted. It can be used to infer the most likely model parameters $\para$ of any DF model $\phi(x|\para)$, while accounting for generic measurement uncertainties and complex selection functions. \s{implementation} presents a fast algorithm for this method and describes its numerical implementation in the $R$ statistical language. Sections~\ref{s:examples} and~\ref{s:advanced} test and illustrate the method using controlled mock data. For clarity and simplicity, we avoid the use of real observations in this paper. \s{conclusion} concludes with a critical summary and outlook.

%%%%%%%%%%%%%%%%%%%%%%%%%%%%%%%%%%%%%%%%%%%%%%%%%%%%%%%%%%%%%%%%%%%%%%%%%%%%%%%%%%%%%%

\section{Mathematical method}\label{s:method}

\subsection{Generative model search}\label{ss:likelihood}

Let us consider a galaxy survey detecting $N$ galaxies $i=1,...,N$ with masses $M_i=10^{x_i}\msun$. We temporarily assume that these mass measurements are exact and bin them into a finite number of bins $j$, equally spaced in log-mass $x$, with bin widths $\Dx$ and bin centres $x_j$. In this case, the measurement can be summarized via the discrete \textit{source counts} $n_j\in\mathbb{N}_0$, where $n_j$ is the number of galaxies with masses in the interval $x\in\{x_j\pm\Dx/2\}$.

We then assume that the space density of the whole galaxy population is described by a MF $\phi(x|\para)$, \eg a Schechter function (\eq{Schechter}). The objective is to find the most likely model parameters $\para$ generating the observed source counts $n_j$. To do so, we note that the predicted number of galaxies detected in the mass bin $j$ is
\be\label{eq:expectednb}
	\lambda_j(\para) = \int_{x_j-\Dx/2}^{x_j+\Dx/2}\phi(x|\para)\veff(x)\dx.
\ee
For the moment, the effective volume $\veff(x)$ can be thought of as the volume probed by galaxies of log-mass $x$ in terms of a sharp detection limit: every galaxy of log-mass $x$ is detected if and only if it lies inside this volume and there are no false detections. Explicit expressions for $\veff(x)$ in the presence of general selection functions will be derived in \ss{gradual}.

Given the expected source counts $\lambda_j(\para)$, the likelihood $L_j(\para)$ of detecting $n_j$ galaxies in bin $j$ is assumed to be given by the Poisson distribution function
\be
	L_j(\para) = \frac{e^{-\lambda_j(\para)}\lambda_j(\para)^{n_j}}{\Gamma(n_j+1)}.
\ee
The total likelihood function $L=\prod_j L_j$ is conveniently written as logarithm (akin to the photon statistics of \citealp{Cash1979}),
\be\label{eq:lraw}
	\ln L(\para) = \sum_j\big(n_j\ln\lambda_j(\para)-\lambda_j(\para)-\ln\Gamma(n_j+1)\big)
\ee
with the convention $0\ln0=0$ to account for bins expected to be empty. This likelihood (without the parameter-independent last term) constitutes the core of numerous MF papers since its introduction for parametric \citep{Sandage1979} and non-parametric \citep{Efstathiou1988} MFs.

We now generalize \eq{lraw} to account for statistical measurement errors. To this end, each datum $x_i$ is replaced by a probability distribution function (PDF) $\rho_i(x)$ (with normalization $\int\rho_i(x)\dx=1$). This PDF represents the probability that the galaxy $i$ has a true log-mass $x$, based solely on the measurement, that is \emph{assuming a flat prior} without using any knowledge on the underlying MF $\phi(x|\para)$ and selection $\veff(x)$. A typical example is the case where each object $i$ has a measured value $x_i$ with a normally distributed uncertainty. In this case, $\rho_i(x)$ is a Gaussian centred at $x_i$. The evaluation of $\rho_i(x)$ can be a subtle task, which depends on whether the underlying error model is conditional on the true or the observed mass. A non-trivial example will be provided in \ss{varsigma}.

Interestingly, in an uncertain measurement, the mode of $\rho_i(x)$ is not a good proxy for the true log-mass if the MF is steep (\ie varies considerably across the width of $\rho_i(x)$). This feature, known as \emph{Eddington bias}, can be accounted for using Bayes theorem to write the bias-corrected PDFs as
\be\label{eq:debias}
	\tilde{\rho}_i(x|\hat\para) = \frac{\rho_i(x)\phi(x|\hat\para)\veff(x)}{\int\rho_i(x)\phi(x|\hat\para)\veff(x)\dx},
\ee
where $\hat\para$ is the vector of the most likely model parameters. The problem with \eq{debias}, of course, is that the most likely model parameters $\hat\para$ are a priori unknown, which will lead to an interesting optimization problem.

Note that \eq{debias} is an approximation, because the exact posteriors $\tilde{\rho}_i(x)$ would require integrating over the full posterior PDF of $\para$ instead of using the mode $\hat\para$. However, we will show later that this approximation does not in fact change the solution. Another subtle point is that \eq{debias} assumes that the observational uncertainty is introduced to the data after drawing it from the population source counts $\phi(x|\hat\para)\veff(x)$, that is \emph{after} applying the selection encoded in the effective volume $\veff(x)$. In astrophysical observations it is not uncommon that some scatter is introduced to the data already before applying the selection function, or that there are multiple layers of scattering events and selection processes. For the moment we assume that these cases can be recast into a single selection function $\veff(x)$, followed by an uncertain measurement. Further discussion and justification, along with an example of fitting scatter followed by selection are provided in \a{ordering}.

We then split each bias-corrected mass measurement into the mass bins $j$ via
\be\label{eq:continuousn}
	n_j(\hat\para)=\sum_i\int_{x_j-\Dx/2}^{x_j+\Dx/2}\tilde{\rho}_i(x|\hat\para)\dx.
\ee
In this way, the source counts $n_j\in\mathbb{R}$ become non-integers, but the normalization $\sum_j n_j=N$ persists.

Finally, we let $\Dx$ become infinitesimal and rewrite the \textit{predicted} (\eq{expectednb}) and \textit{observed posterior} (\eq{continuousn}) source count densities $\lambda_j/\Dx$ and $n_j/\Dx$ as
\begin{eqnarray}
	\lambda(x|\para) & = & \phi(x|\para)\veff(x), \label{eq:deflambda} \\
	n(x|\hat\para) & = & \sum_i\tilde{\rho}_i(x|\hat\para). \label{eq:defn}
\end{eqnarray}
The likelihood function (\eq{lraw}) then becomes
\be\label{eq:likelihood}
	\ln \L(\para,\hat\para) = \int\Big(n(x|\hat\para)\ln\lambda(x|\para)-\lambda(x|\para)\Big)\dx,
\ee
where we have dropped a constant\footnote{\Eq{lraw} diverges as $\Dx\rightarrow0$. The trick to avoid this divergence is to subtract the term $\sum_j(n_j\ln \Dx-\ln\Gamma(n_j+1))$ before letting $\Dx\rightarrow0$, which is possible because this term does not depend on the parameter-vector $\para$ to be fitted.} that does not depend on $\para$. We refer to $\L(\para,\hat\para)$ as the \emph{modified likelihood} function to emphasize its subtle difference to the true likelihood function $L(\para,\hat{x}_i)$, \ie the likelihood of the full two-stage Bayesian hierarchical model \citep{Allenby2005} that treats the true values $\hat{x}_i$ leading to the uncertain measurements $x_i$ as $N$ additional model parameters. In a self-consistent ML solution, the parameter-vector $\para$ maximising \eq{likelihood} (while keeping $\hat\para$ fixed) must be equal to $\hat\para$ used for debiasing the observations. In other words, we are looking for the parameter-vector $\hat\para$ satisfying
\be\label{eq:masterequation}
	\hat\para = \mathcal{F}(\hat\para),
\ee
where
\be
	\mathcal{F}(\hat\para)=\underset{\para}{\rm argmax}~\L(\para,\hat\para).
\ee
We call this approach the modified maximum likelihood (MML) method and refer to the solution $\hat\para$ of \eq{masterequation} as the MML estimator (MMLE).

Importantly, it can be shown analytically (see \a{exactsolution}) that if a unique solution exists for the standard ML estimator (MLE) of the full Bayesian hierarchical model, then this same solution exists uniquely also for \eq{masterequation}. This identity of the MMLE and MLE is the central theorem of this work that ultimately justifies the MML method and puts it on a robust mathematical basis. By virtue of this theorem the MMLE inherits all of the interesting features of the MLE. In particular, it is an asymptotically unbiased, minimum-variance and normally distributed estimator \citep{Kendall1979}. These are precisely the properties that one would naturally request from an optimal estimator. Incidentally, if the ML solution is \emph{not} unique, \ie if the true likelihood has multiple maxima, the MMLE is also not unique. In practice, different solutions $\hat\para$ can then be found by sampling the initial parameters $\hat\para_0$ (see \ss{alg}), for instance as part of an MCMC algorithm, and the solution that maximizes $ \L(\hat\para,\hat\para)$ is the most likely model. However, in the wide range of mock examples considered in this work, such an approach has never been necessary.

Solving \eq{masterequation} is much easier than maximising the full likelihood in the presence of general observational uncertainties, but remains nonetheless a challenging optimization problem. A fast and stable algorithm is presented in \ss{alg} and implemented in \ss{package}.

\subsection{Selection function}\label{ss:gradual}

This section elaborates on how to evaluate the effective volume $\veff(x)$, so far assumed to be given.

Most galaxy surveys do not have sharp sensitivity limits. There is no well-defined maximum volume  inside which all galaxies of a fixed mass are detected, while outside none are detected. Instead, the fraction of detected sources (true positives) decreases gradually when reaching the detection limit, while the fraction of wrong detections (false positives) increases. These fractions are typically quantified via the \textit{completeness} $C\in[0,1]$, defined as the probability of a real source to be detected, and the \textit{reliability} $R\in[0,1]$, defined as the probability of a detection corresponding to a real source. Both fractions generally depend on the mass and the position $\r\in\mathbb{R}^3$ in the survey volume, here defined as the comoving position relative to the observer. Sometimes $C$ and $R$ also depend on known or unknown extra properties (here labelled $...$), such as the galaxy inclination. For what follows, it is convenient to introduce the \textit{selection function}
\be\label{eq:selfct}
	f(x,\r)=\left\langle\frac{C(x,\r,...)}{R(x,\r,...)}\right\rangle,
	% C = (true positives)/trues
	% R = (true positives)/positives
\ee
where the expectation $\langle\rangle$ averages over the extra variables. In this way, $f(x,\r)$ always equals the expected ratio between the number of detections (including false positives) and the number of true sources (whether detected or not). Hence, the expected number density of detections per unit log-mass $x$ and comoving volume is $\phi(x|\para)f(x,\r)$ and the expected number density over the whole survey volume is
\be\label{eq:lambdagradual}
	\lambda(x|\para) = \phi(x|\para)\int f(x,\r) d^3r.
\ee
Matching this equation to \eq{deflambda} implies that the MML formalism (equations~\ref{eq:deflambda}--\ref{eq:masterequation}) applies to arbitrary selection functions, upon defining the effective volume as
\be\label{eq:veffgradual}
\begin{split}
	\veff(x) &= \int f(x,\r)d^3r\\
	& = \int_0^{2\pi}\!\!\!\!\!\!d\alpha\!\int_{\!-\!\pi/2}^{\pi/2}\!\!\!\!\!\!\!\!d\delta\int_0^{\infty}\!\!\!\!\!\!dr\,r^2\!\cos(\delta) f(x,\alpha,\delta,r),
\end{split}
\ee
where the second line is the explicit form in spherical coordinates: right ascension $\alpha$, declination $\delta$ and distance $r=|\r|$.

Two common cases, worth expanding, are those where the selection function only depends on $x$ and $r$ (Section~\ref{sss:fz}) and where the effective volume is only given on a galaxy-by-galaxy basis (Section~\ref{sss:onlyvi}).

\subsubsection{Isotropic selection function}\label{sss:fz}

Often the selection function is independent of the direction ($\alpha$ and $\delta$) and only varies with the comoving distance $r$. Such \textit{isotropic} selection functions $f(x,r)$ reduce \eq{veffgradual} to
\be\label{eq:veffgradual2}
	\veff(x) = \int\!V^{\prime}(r) f(x,r)dr,
\ee
where $V^{\prime}(r)$ is the derivative of the total observed comoving volume $V(r)$. For instance, a survey of redshift-independent solid angle $\Omega$ encompasses a volume $V(r)=\Omega r^3/3$, \ie $V'(r)=\Omega r^2$ and hence
\be\label{eq:veffgradual3}
	\veff(x) = \Omega\int r^2 f(x,r) dr.
\ee

Isotropic selection functions are sometimes expressed as $f(x,z)$, where $z$ is the cosmological redshift corresponding to the distance $r$. To write equations~(\ref{eq:veffgradual2}) and (\ref{eq:veffgradual3}) in terms of $z$, it suffices to substitute $r$ for its expression as a function of $z$. For instance, in the local universe, $r=cz/H_0$ with Hubble constant $H_0$ and speed of light $c$, and hence $V(z)^\prime=\Omega c^3 H_0^{-3}z^2$.

\subsubsection{Volume given for each galaxy}\label{sss:onlyvi}

It is quite common for galaxy surveys to store a single effective volume $V_i=\veff(x_i)$ for each galaxy $i$, without specifying the continuous functions $\veff(x)$ or $f(x,\r)$ needed in the MML formalism. In practice, the $V_i$ are often computed from the maximal volume $\vmax(x_i)$ in which a galaxy of log-mass $x_i$ could have been detected, corrected for the estimated completeness and reliability of the source, $V_i = C_i R_i^{-1} \vmax(x_i)$. The important point is that the galaxy $i$ contributes a density $V_i^{-1}$ to the galaxy MF at its particular mass.

The challenge consists in reconstructing the continuous function $\veff(x)$ from the finite set $\{V_i\}$. This challenge also has to face the fact that two (or more) galaxies of identical mass $x$ can have different values $V_i$. For example, some galaxies might be seen edge-on with strong dust extinction, while others are seen face-on with little extinction. Thus, the first class is harder to detect and hence admits a smaller effective volume than the second one. So what value should be chosen for $\veff(x)$? The answer comes from the requirement that space densities must add up (following from conservation of mass), which implies that $\veff(x)$ is the harmonic mean of the individual effective volumes of all detections. Formally,
\be\label{eq:veffmean}
	\veff(x_i)=\big\langle V_i^{-1}\big\rangle^{-1},
\ee
where the expectation goes over all sources of the same log-mass $x_i$ and hence marginalizes over all other variables. A simple example demonstrating the applicability of this harmonic mean is given in \a{harmonic}.

The reasoning behind \eq{veffmean} implies that $\veff^{-1}(x)$ can be interpolated linearly from the given values $V^{-1}(x_i)$.
%Explicitly, if we order the galaxies by mass, such that $x_1\leq x_2\leq...\leq x_N$, then for any $x$ satisfying $x_i<x<x_{i+1}$, $\veff^{-1}(x)$ is the linear interpolation in $x$ between $V_i^{-1}$ and $V_{i+1}^{-1}$. In the case of $q$ identical masses, $x=x_i=x_{i+1}=...=x_{i+q-1}$, \eq{veffmean} gives $\veff^{-1}(x)=(V_i^{-1}+V_{i+1}^{-1}+...+V_{i+q-1}^{-1})/q$.
For values $x$ outside the range of $\{x_i\}$, the function $\veff(x)$ can normally be extrapolated based on the survey specifications. For example, in a volume-limited survey where the most massive galaxy $i=N$ could have been detected anywhere in the survey volume, $\veff(x)=V(x_N)$ for all $x>x_N$.

Numerical tests using the implementation of \s{implementation} show that this way of interpolating $\veff(x)$ from the set $\{V_i=\veff(x_i)\}$ generally leads to MML solutions that are statistically consistent with those obtained using the exact $\veff(x)$.

\subsection{Cosmic large-scale structure}\label{ss:lsstheory}

\begin{figure}
\begin{center}
\includegraphics[width=\columnwidth]{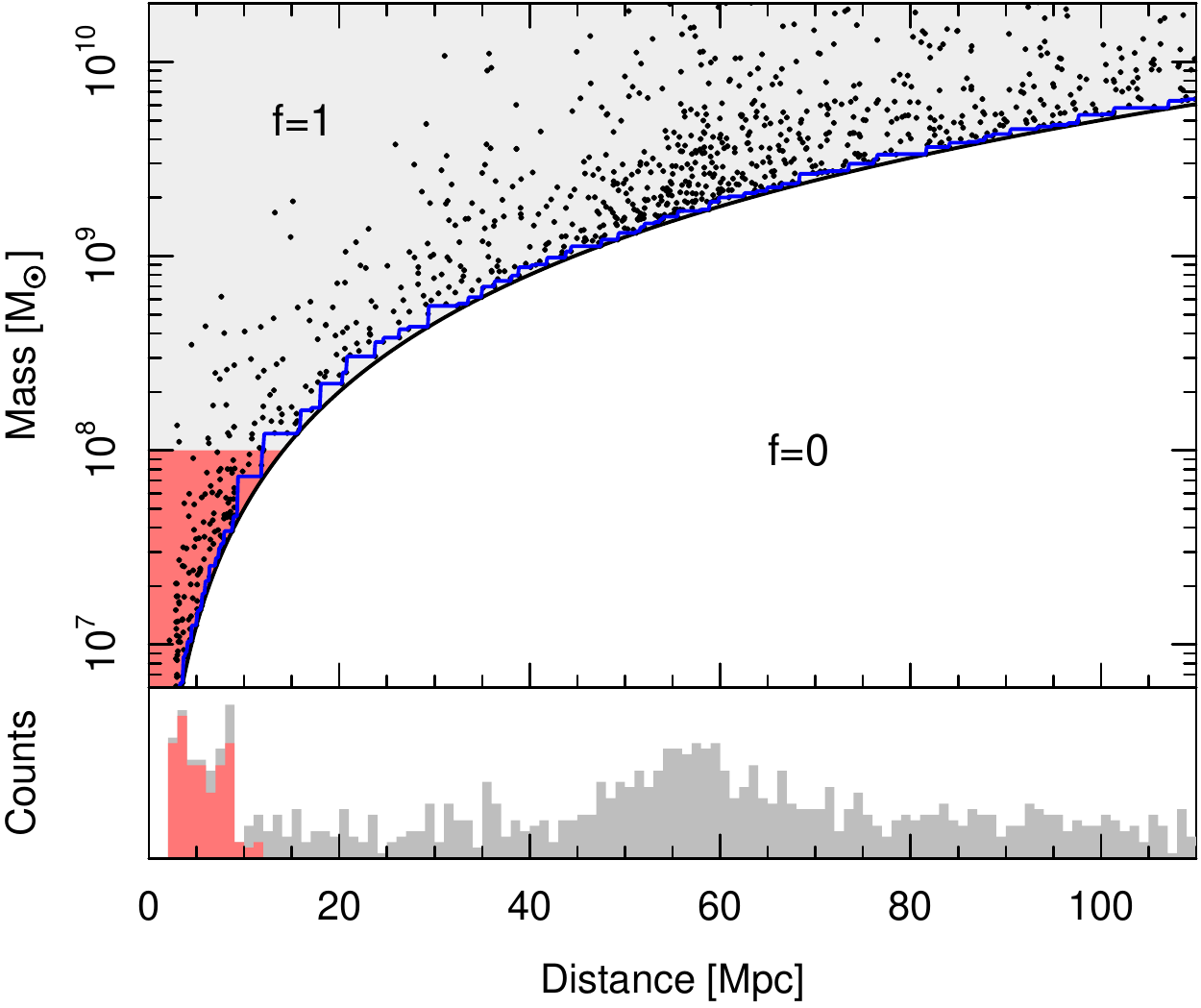}\vspace{-4mm}
\caption{Example of a sensitivity-limited mock galaxy survey, subject to LSS with a notable overdensity at distances smaller than 10~Mpc (red). This overdensity causes an over-average detection rate of galaxies with masses below $10^8~\msun$. See \ss{lsstheory} for details.}\label{fig:lss_illustration}
\end{center}
\end{figure}

Cosmic large-scale structure (LSS) can introduce systematic errors in the estimation of the MF. How these errors arise is illustrated in \fig{lss_illustration}. In this example, the average density within the survey volume is the mean density of the universe, hence the counts of massive galaxies that can be detected to the edge of this volume are insensitive to galaxy clustering. However, due to a nearby overdensity (red), the low-mass galaxies ($M<10^8\msun$), which can only be detected to a smaller distance ($\sim10\,\rm Mpc$) are over-represented. If not corrected, this selection bias will result in overestimating the steepness of the MF in the low-mass end.

LSS is suitably quantified by the relative density $g(r)$, defined as the mean density in the survey volume at comoving distance $r=|\r|$, relative to the mean density of the universe. Assuming that the number density of galaxies of any mass scales with $g(r)$, the expected number density of detections per unit log-mass $x$ and comoving volume is $\phi(x|\para)f(x,\r)g(r)$ and \eq{veffgradual} becomes
\be\label{eq:veffgraduallss}
	\vefflss(x) = \int f(x,\r)g(r)d^3r.
\ee
In the absence of LSS, we expect $g(r)\equiv1$ and \eq{veffgraduallss} reduces to \eq{veffgradual}. If a model of $g(r)$ is known, for instance from a pre-existing survey, LSS is accounted for using \eq{veffgraduallss} in the MML method.

Interestingly, $g(r)$ can be estimated, up to an overall normalization factor, directly from the distance-distribution of the galaxies in the sample. Here, we restrict the discussion of this procedure to the case of an isotropic selection function $f(x,r)$ (see Section~\ref{sss:fz}). In \a{lss} we show that if $g(r)$ is derived from the data, the LSS bias-corrected effective volume of \eq{veffgraduallss} becomes
\be\label{eq:veffgradualautolss}
	\vefflss(x|\hat\para) = \sum_i\frac{f(x,r_i)}{\int\phi(\tilde x|\hat\para)f(\tilde x,r_i)d\tilde x}.
\ee
As with the posterior PDFs of the data, $\tilde\rho_i(x|\hat\para)$, the effective volume with LSS is determined at the MML solution $\hat\para$. Since this solution is a priori unknown, $\vefflss(x|\hat\para)$ is also evaluated iteratively as part of the algorithm introduced in \ss{alg}.

\Eq{veffgradualautolss} requires a selection function, $f(x,r)$. A special situation is the case of a sharp survey limit, $f(x,r)\in\{0,1\}$, such that $f(x,r)=0$ if and only if $x$ is smaller than a distance dependent threshold $x_{\rm min}(r)$, shown as the black line in \fig{lss_illustration}. In this case, \eq{veffgradualautolss} reduces to
\be\label{eq:veffgradualautolss2}
	\vefflss(x|\hat\para) = \!\!\!\!\!\!\!\sum_{i\in\{x\geq x_{\rm min}(r_i)\}}\!\left(\int_{x_{\rm min}(r_i)}^\infty\!\!\!\!\phi(\tilde x|\hat\para)d\tilde x\right)^{-1}\!\!.
\ee
If $x_{\rm min}(r)$ increases monotonically with $r$, one can use the approximation
\be\label{eq:xlimapprox}
	x_{\rm min}(r)=\min_{r_i\geq r}(x_i),
\ee
as illustrated in \fig{lss_illustration} (solid blue line). There is no need to use the approximations of equations~(\ref{eq:veffgradualautolss2}) and (\ref{eq:xlimapprox}) or variations thereof, unless the selection function $f(x,r)$ is unavailable.

It is important to stress that the overall normalization of $g(r)$ and therefore $\vefflss(x|\hat\para)$ cannot be derived from the data. It is simply impossible to know purely from a list of galaxies whether the survey volume represents an under-density or an over-density, relative to the rest of the universe. We must therefore make a choice of how to normalize $\vefflss(x|\hat\para)$, for instance by demanding that
\be\label{eq:lssnormalization}
	\int\phi(x|\hat\para)\vefflss(x|\hat\para)w(x)dx = \int\phi(x|\hat\para)\veff(x|\hat\para)w(x)dx,
\ee
where $w(x)$ is a weighing function. Choosing $w(x)\equiv1$ preserves the total \emph{number} of galaxies. If $w(x)=10^x$, the total \emph{mass} of the survey is conserved.

\Eq{veffgradualautolss2} sums over the cumulative MF, representing the total density of galaxies with log-masses above $x_{\rm min}(r)$. This appearance of the cumulative MF (or LF) is almost universal to all published approaches accounting for LSS. In our case, the cumulative MF naturally appears through the short analytical derivation in \a{lss} and the assumption of a monotonic selection function, but it is instructive to follow the alternative derivations presented in the original works of \cite{Lynden-Bell1971}, introducing the so-called $C^{-}$-method, \cite{Turner1979}, introducing the $\varphi/\phi$-method, \cite{Sandage1979} and \cite{Kirshner1979}. All these and derived `density-corrected' methods \citep[e.g.]{Baldry2012,Wright2017}, explicitly or implicitly make the same basic assumptions that led to equations~(\ref{eq:veffgradualautolss2}) and (\ref{eq:xlimapprox}) and account for LSS using the same idea of modelling $g(r)$ directly from the distance- or redshift-distribution of the data. The advantage of the current formalism consists in the MML framework, which simultaneously handles mass uncertainties.

\subsection{\!\!Multi-dimensional distributions}\label{ss:multi}

The entire formalism presented so far is straightforward to generalize to $D$-dimensional galaxy properties $\x\equiv(x_1,...,x_D)$ of any integer $D\geq1$. For instance, if we construct the distribution of galaxies in the mass-size plane ($D=2$), $x_1$ could represent the mass and $x_2$ the half-mass radius. Let us write the $D$-dimensional DF by generalising \eq{mf} to
\be\label{eq:mfp}
	dN = \phi(\x) V d^Dx,
\ee
where $dN$ denotes the expected number of galaxies in the infinitesimal volume $d^Dx$ around $\x$. Given a model $\phi(\x|\para)$ with model parameters $\para$, the expected number density of detections per unit of $x_1$, $x_2$, ... and $x_D$ then reads
\be
	\lambda(\x|\para) = \phi(\x|\para)\veff(\x), \label{eq:deflambdap}
\ee
where $\veff(\x)$ is calculated in analogy to the one-dimensional case. For instance, if we know the multi-dimensional selection function $f(\x,\r)$, then $\veff(\x)=\int f(\x,\r)d^3r$ (see \eq{veffgradual}; or, if we only know the effective volume $V_i$ of each galaxy, then $\veff(\x)^{-1}$ can be linearly interpolated between $V_i^{-1}$ (see Section~\ref{sss:onlyvi}). The bias-corrected observed source counts are (generalization of equations~\ref{eq:debias} and \ref{eq:defn}),
\be\label{eq:defnp}
	n(\x|\hat\para) = \sum_i\frac{\rho_i(\x)\lambda(\x|\hat\para)}{\int\rho_i(\x)\lambda(\x|\hat\para)d^Dx}. 
\ee
The modified likelihood function (\eq{likelihood}) then generalizes to
\be\label{eq:likelihoodp}
	\ln \L(\para,\hat\para) = \int\Big(n(\x|\hat\para)\ln\phi(\x|\para)-\lambda(\x|\para)\Big)d^Dx.
\ee
All comments on $\ln\L(\para,\hat\para)$ made following \eq{likelihood} also apply to \eq{likelihoodp}. A numerical example of the MMLE for $D=2$ is presented in \ss{2dexample}.

\subsection{Parameter uncertainties}\label{ss:uncertainties}

At second order, the covariance matrix $\cov(\hat\para)$ of the best fitting parameters $\hat\para$ can be approximated as (Laplace approximation)
\be\label{eq:cov}
	\cov(\hat\para) = -\Big(H(\hat\para)\Big)^{-1}
\ee
where $H$ is the Hessian matrix (second derivatives) of the log-likelihood. However, the Hessian of our modified log-likelihood $\ln\L(\para,\hat\para)$ (\eq{likelihood}), defined as
\be\label{eq:hessian}
	H(\hat\para)_{ij} = \left.\frac{\partial^2\ln\L(\para,\hat\para)}{\partial\para_i\partial\para_j}\right|_{\para=\hat\para},
\ee
is not identical to the Hessian of the standard log-likelihood (despite the identity of the MLE and MMLE), as proven in \a{exactsolution}. Hence \eq{cov} is not necessarily a good approximation. This inaccuracy is negligible if the measurement uncertainties of the data $\{x_i\}$ are small ($<50\%$) compared to the range (standard deviation) of all data $\{x_i\}$. It is possible to express the correct Hessian of the standard log-likelihood at the MMLE solution (see \eq{htrue}), but its numerical evaluation is rather laborious, requiring $3P^2ND$ integrals. Moreover, neither of the Hessian approaches normally accounts for parameter uncertainties due to the removal of the LSS bias (\ss{lsstheory}), which is itself uncertain. Also, if the model parameters are fully degenerate or if the likelihood is non-Gaussian (\eg non-linear parameter correlations), the Laplace approximation breaks down.

These limitations of the Hessian approach in MML can all be addressed by estimating $\cov(\hat\para)$ via a non-parametric bootstrapping \citep{Efron1993} approach that resamples the $N$ data points, treating them as the whole population. Explicitly, one performs $Q>1$ bootstrap iterations, labelled $q=1,...,Q$. Each iteration includes three steps: (1) choose a random number $N_q$ from a Poisson distribution with expectation $N$; (2) draw a new sample of $N_q$ data points (e.g. galaxy masses) from the original sample of $N$ points \emph{with replacement} (\ie allowing for repetitions); and (3) determine the MMLE $\hat\para_q$ of this new sample. The covariance matrix of the original MMLE $\hat\para$ is then approximated as the covariances of the $\{\hat\para_q\}$. Following \cite{Babu1983}, $Q\approx N(ln N)^2$ interations typically suffice for a good estimate. An explicit example of this method is provided in \ss{resampling}.

If the MML method is performed while correcting for LSS bias (\ss{lsstheory}), we can either refit $\vefflss(x|\hat\para_q)$ at each resampling iteration $q$ or fix this function to $\vefflss(x|\hat\para)$ across all iterations. Depending on this choice, the bootstrap parameter covariances respectively include or exclude the uncertainty of the cosmic LSS itself. Accounting for the limited knowledge of LSS generally increases the uncertainties of the model parameters, sometimes by a significant amount as illustrated in the example of \ss{lssexample}. When fitting real galaxy surveys it is hence advisable to quote both the uncertainties with and without LSS uncertainties.

\subsection{Estimator bias correction}\label{ss:bias}

The MMLE (or MLE) can be biased, meaning that its expectation $E[\hat{\para}]$, equal to the average $\hat{\para}$ for an infinite number of random samples from the same population, differs from the true population model $\para_{\rm true}$. This is a general property of the MLE: only as the sample size tends to infinity, is the MLE guaranteed to be unbiased (\citealp{Kendall1979} for theory; Appendix~A of \cite{Robotham2015} for an example).

It is possible to construct a bias-corrected MLE analytically using the higher order derivatives of the likelihood function \citep{Cordeiro1994}. This approach demonstrates that the leading bias term varies as $N^{-1}$, but the correction terms are rather cumbersome and depend on the choice of the MF model $\phi(x|\para)$. Here, we resort to a more generic jackknifing approach to approximately correct the bias to order $N^{-1}$ of any ML estimator. Following \cite{Efron1981}, the bias-corrected estimator reads
\be\label{eq:biascorrection}
	\tilde\para = N\hat\para-\frac{N-1}{N}\sum_{i=1}^N\hat\para_{(i)},
\ee
where $\hat\para_{(i)}$ is the ML estimator of the $N-1$ galaxies, in which object $i$ has been removed from the original sample of $N$ galaxies. Importantly, since the number of galaxies itself affects the overall normalization of the MF, the parameters $\hat\para_{(i)}$ must be estimated using the renormalized volumes
\be
	V_{(i)}(x) = \frac{N-1}{N}\veff(x).
\ee

As illustrated in \ss{small}, this bias correction performs remarkably well. That said, it is often debatable whether the true MMLE $\hat\para$ or the bias-corrected estimator $\tilde\para$ is the `better' solution. In many ways unbiased estimators exhibit less favourable properties (see \citealp{Hardy2002} for an example). The choice ultimately depends on the application. In any case, we will show (\eg \fig{small}) that the difference between $\hat\para$ and $\tilde\para$ only becomes appreciable for very small samples ($\lesssim10$ galaxies) and even then their difference is small compared to the overall parameter uncertainties.

%%%%%%%%%%%%%%%%%%%%%%%%%%%%%%%%%%%%%%%%%%%%%%%%%%%%%%%%%%%%%%%%%%%%%%%%%%%%%%%%%%%%%%

\section{Numerical implementation}\label{s:implementation}

\subsection{Optimization algorithm}\label{ss:alg}

Solving the implicit \eq{masterequation} is a tricky optimization problem. This is because of the subtlety that its solution $\hat\para$ is \emph{not} obtained by maximising $\L(\hat\para,\hat\para)$. In other words, the solution $\hat\para$ of \eq{masterequation} does not generally satisfy $(\partial_\para\L)(\hat\para,\hat\para)=0$ (\a{exactsolution}). To solve \eq{masterequation}, we developed a customized algorithm, referred to as the `fit-and-debias' algorithm: First, evaluate the observed source count function $n_0(x)=\sum_i\rho_i(x)$. Then repeat the following iteration for $k=1,2,3,...$:

\begin{enumerate}[leftmargin=7mm]
\setlength{\itemindent}{-1.6mm}
\item Find the parameter-vector $\hat\para_k$ that maximizes
\be\label{eq:fit_tk}
	\ln \L(\hat\para_k) = \int\Big(n_{k-1}(x)\ln\phi(x|\hat\para_k)-\phi(x|\hat\para_k)\veff(x)\Big)\dx.
\ee
\item Use $\hat\para_k$ as new estimator to de-bias the source counts,
\be\label{eq:debias_n}
	n_{k}(x) = \sum_i \frac{\rho_i(x)\veff(x)\phi(x|\hat\para_k)\dx}{\int\rho_i(x)\veff(x)\phi(x|\hat\para_k)\dx}.
\ee
\end{enumerate}

The algorithm can be stopped as soon as a certain convergence criterion is reached, for instance if
\be
	\left\|\hat\para_{k}-\hat\para_{k-1}\right\|\leq\Delta,
\ee
where $\Delta\in\mathbb{R}_+$ is a predefined tolerance. In this work, we set $\Delta$ equal to the relative precision error ($\sim10^{-15}$) of the double-precision floating-point representation (IEEE 754). If a guess of initial parameters $\hat\para_0$ is available, the fit-and-debias algorithm can be accelerated by evaluating $n_0$ via \eq{debias_n} instead of using $n_0=\sum_i\rho_i(x)$.

The algorithm can readily account for (unknown) cosmic LSS (theory in \ss{lsstheory}). It suffices to substitute $V(x)$ in equations (\ref{eq:fit_tk}) and (\ref{eq:debias_n}) for $\vefflss^{k-1}(x)$ and add a third step:
\begin{enumerate}[leftmargin=7mm]
\setlength{\itemindent}{-1.6mm}
\setcounter{enumi}{2}
\item Use $\hat\para_k$ to update the effective volume,
\be
	\vefflss^{k}(x) = A\sum_i\frac{f(x,r_i)}{\int\phi(\tilde x|\hat\para)f(\tilde x,r_i)d\tilde x},
\ee
\end{enumerate}
with a normalization factor $A$ computed via \eq{lssnormalization} at $\hat\para=\hat\para_k$. Computing  $\vefflss^{0}(x)$ requires an initial guess $\hat\para_0$.

\begin{figure}
\begin{center}
\includegraphics[width=1.02\columnwidth]{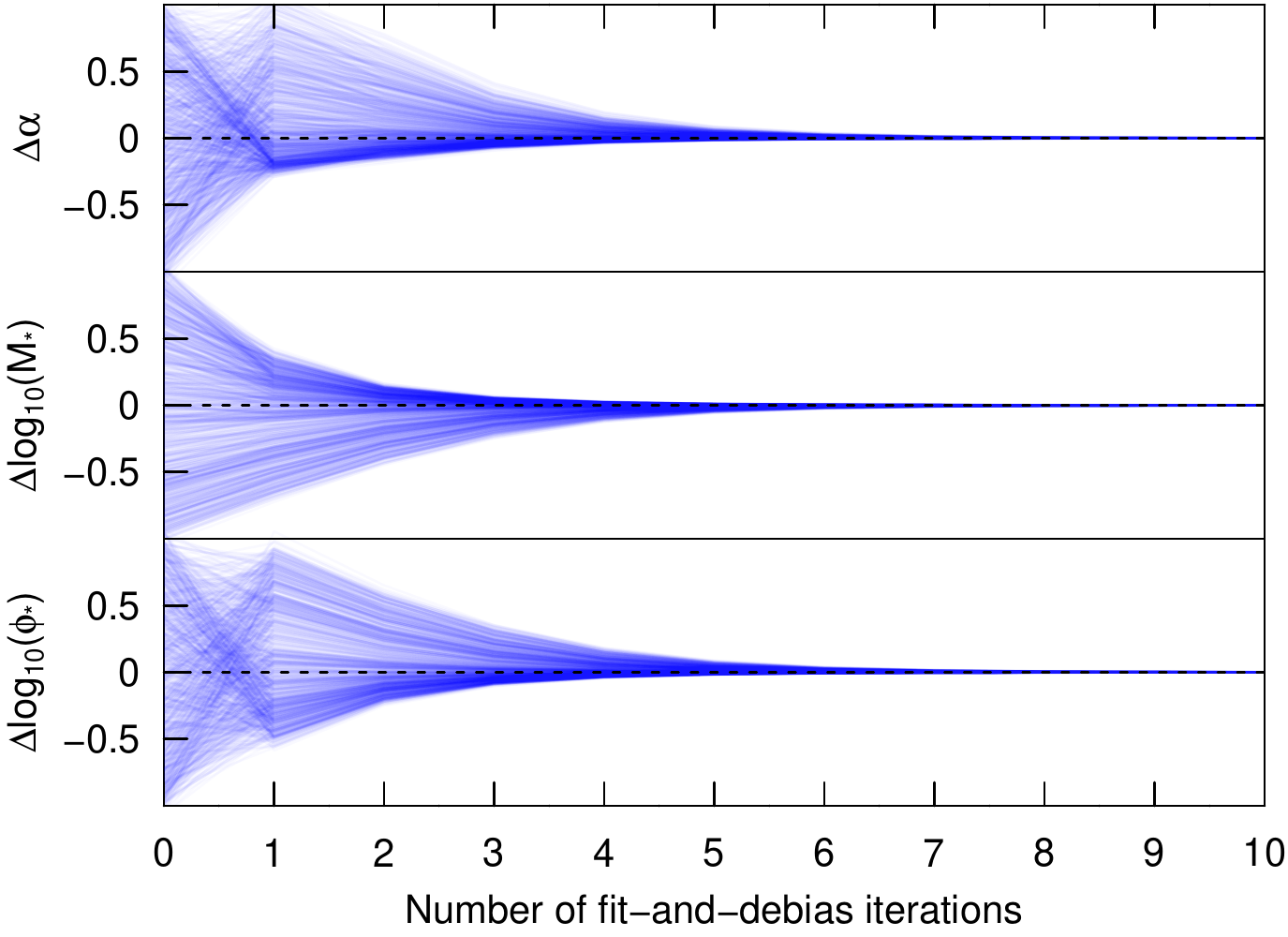}\vspace{0.4cm}\\
\includegraphics[width=1.02\columnwidth]{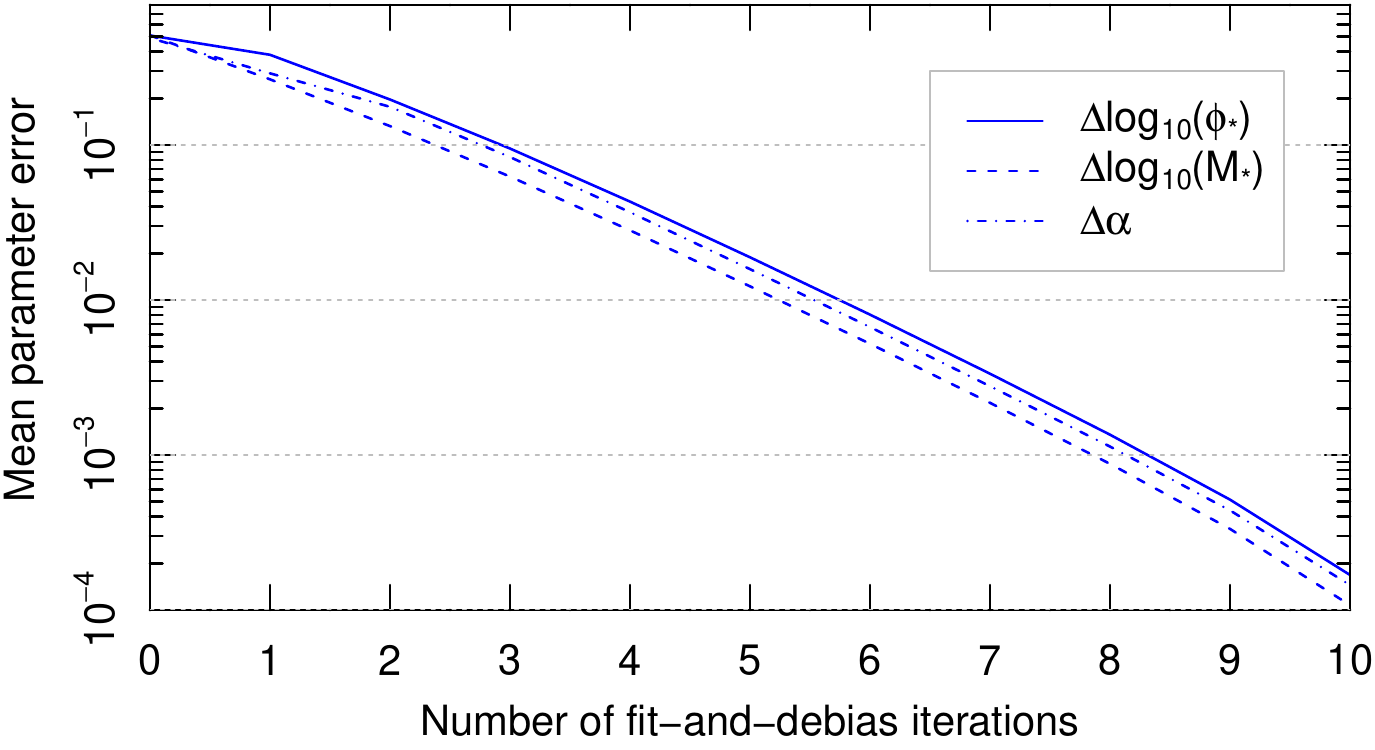}\vspace{-0.4cm}
\caption{Test of the `fit-and-debias' algorithm. Good (excellent) convergence is reached after five (ten) iterations through equations~(\ref{eq:fit_tk}) and (\ref{eq:debias_n}). See \ss{alg} for details.}\label{fig:convergence}
\end{center}
\end{figure}

\a{exactsolution} proves analytically that the fit-and-debias algorithm always converges towards the solution of \eq{masterequation}, which is itself unique and identical to the MLE of the true likelihood. To illustrate the typical convergence, we draw a random mock sample of $10^3$ galaxies from a fixed Schechter function (with parameters $\para_{\rm true}$ in \eq{true}) using a sensitivity-limited effective volume -- example discussed later in \ss{medium}. The Schechter function parameters are then inferred from the mock data using the fit-and-debias algorithm, starting with initial parameters $\hat\para_0$ that have been perturbed from $\para_{\rm true}$ by uniform random numbers in the interval $[-1,+1]$. This numerical experiment is repeated $10^3$ times. As shown in \fig{convergence} the algorithm quickly converges towards a stable solution in every single run. The top panel shows the evolution of the parameter errors $\Delta\para$, defined as the absolute difference between the parameters at a given iteration and their final value after 20 iterations. Only the first 10 iterations are shown, since, for all practical purposes, the parameters are sufficiently converged after 10 iterations. The bottom panel shows the evolution of the average parameter errors $\Delta\para$, revealing a monotonic decrease by a factor of $\sim2.5$ per iteration. In \ss{medium} we will demonstrate that the solutions of the $10^3$ random experiments are indeed consistent with the true parameters $\para_{\rm true}$.

\subsection{The $R$-package dftools}\label{ss:package}

The fit-and-debias algorithm for the MML method has been implemented in the package \dftools for the $R$-language, freely available for most operating systems (including Windows, MacOS, Linux). We refer the reader to the detailed documentation that comes with this package. In this documentation, all routines are explained alongside many examples. Here, we summarize the core functionality with some selected examples.

\begin{figure}
\begin{center}
\includegraphics[width=\columnwidth]{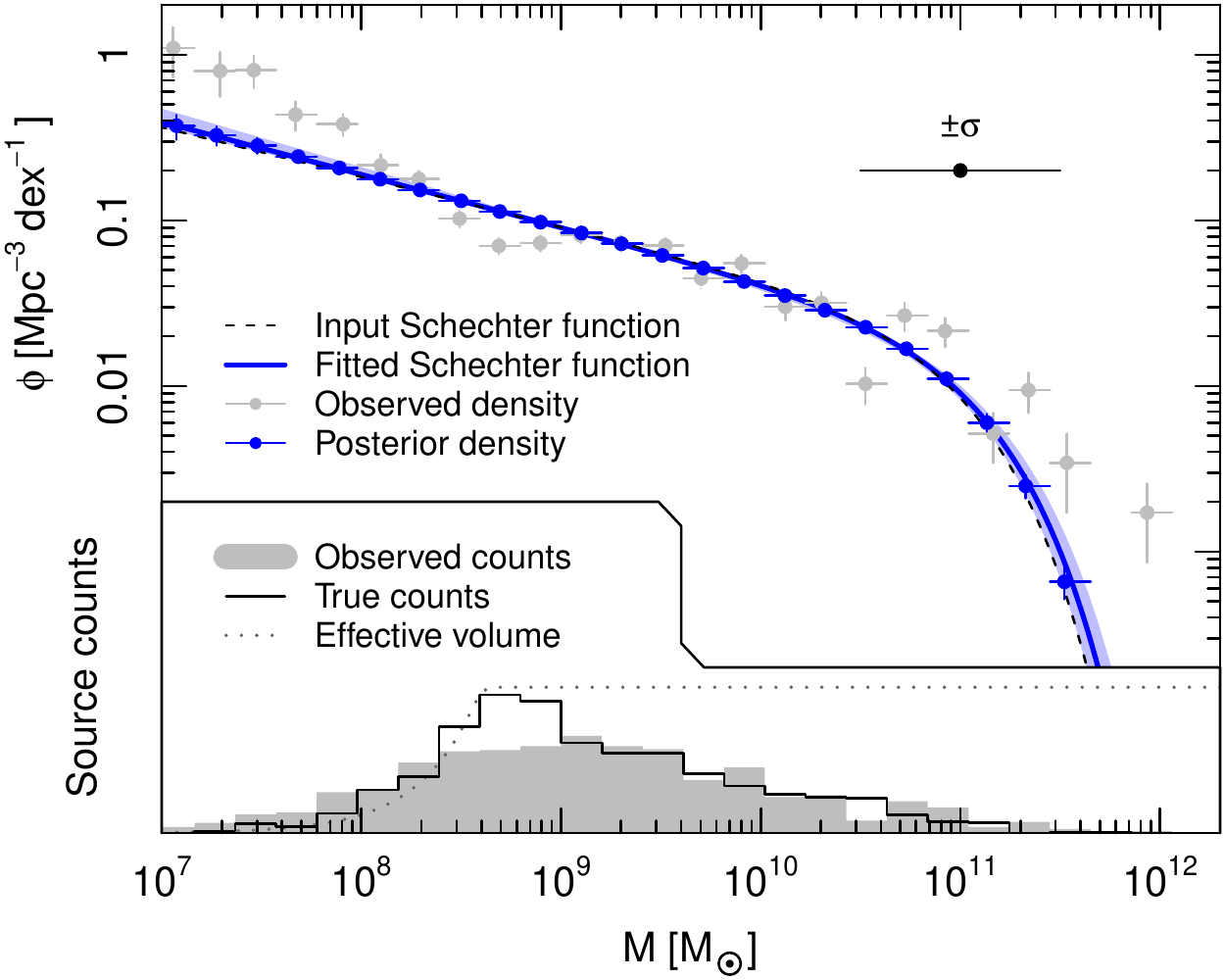}\vspace{-3mm}
\caption{Basic example of \dftools. $10^3$ mock galaxies are drawn from the input Schechter function, assuming the effective volume shown at the bottom and perturbing the masses by random log-normal errors of $\sigma=0.5$ dex standard deviation. The resulting source counts are shown as grey histogram and the corresponding raw MF (using the $1/\vmax$-method) is shown in bins as purple dots. Horizontal bars are the bin widths and vertical bars are Poisson errors. A Schechter function is then fitted to these mock data using the fit-and-debias algorithm (equations~\ref{eq:fit_tk} and \ref{eq:debias_n}), while accounting for the observational uncertainties. The best fitting solution (blue solid line) and its 68\% confidence regions (blue shading) are consistent with the input MF; and posterior masses (blue dots), computed from the grey data via \eq{debias}, align with the input model.}\label{fig:example}
\end{center}
\end{figure}

\begin{figure}
\begin{center}
\includegraphics[width=\columnwidth]{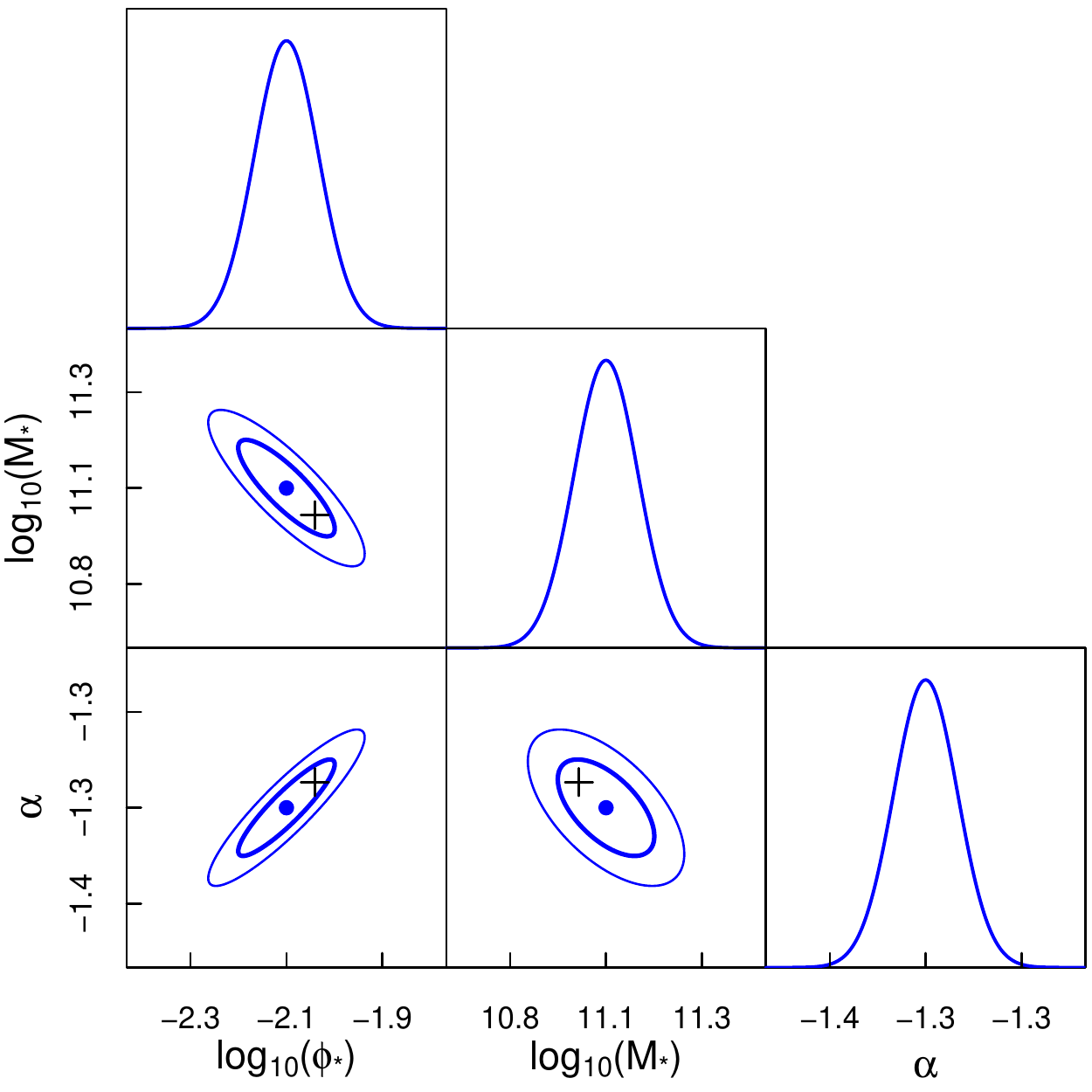}\vspace{-3mm}
\caption{Uncertainties and covariances of the best-fitting parameters for the example in \fig{example}. The best fitting values are plotted as blue dots with 68\% and 95\% confidence regions, in the Gaussian approximation, drawn as thick and thin blue lines. The true input parameters are given by the black crosses.}\label{fig:example_covariance}
\end{center}
\end{figure}

The \dftools package is distributed via GitHub. To install the package in $R$ use
{\color{blue}\begin{lstlisting}
install.packages("devtools")
library(devtools)
install_github("obreschkow/dftools")
\end{lstlisting}}
\noindent The package is then activated by calling
{\color{blue}\begin{lstlisting}
library(dftools)
\end{lstlisting}}
\noindent To view to inbuilt documentation, type
{\color{blue}\begin{lstlisting}
?dftools
\end{lstlisting}}
The package includes a routine to generate galaxy data given an arbitrary MF and selection function. For instance, to draw a sample of $10^3$ galaxies with Gaussian measurement errors of $\sigma=0.5$ (in $\log_{10}x$) from a Schechter function (\eq{Schechter}) with the default parameters $\para_{\rm true}=(\log_{10}\phi_\ast,\log_{10}M_\ast,\alpha)=(-2,11,-1.3)$ and a built-in sensitivity-limited selection function, use
{\color{blue}\begin{lstlisting}
dat = dfmockdata(n=1e3, sigma=0.5)
\end{lstlisting}}
All mock galaxies and survey specifications are stored in the list \texttt{dat}. For instance, the observed log-masses and their Gaussian uncertainties are stored in the vectors \texttt{dat\$x} and \texttt{dat\$x.err}, respectively, while the effective survey volume function $\veff(x)$ is stored in \texttt{dat\$veff}. Given this mock survey, the most likely generative MF can be fitted via
{\color{blue}\begin{lstlisting}
survey = dffit(dat$x, dat$veff, dat$x.err)
\end{lstlisting}}
\noindent This function executes the fit-and-debias algorithm (\ss{alg}), using the in-built \texttt{optim} function with the default algorithm \citep{Nelder1965} to maximize \eq{fit_tk}. The output argument \texttt{survey} is a list of several sub-lists, such as \texttt{survey\$data}, keeping track of the fitted galaxy data, and \texttt{survey\$fit}, containing the fitted parameters and their covariances. To visualize the fit and mock data, type
{\color{blue}\begin{lstlisting}
ptrue = dfmodel(output="initial")
mfplot(survey, xlim=c(1e7,2e12), p=ptrue)
\end{lstlisting}}
\noindent This command produces a plot similar to \fig{example} (without legend and true counts). The fit and its 68\%-uncertainty (light blue) is consistent with the input model. This can also be seen in the parameter covariance plot, obtained via
{\color{blue}\begin{lstlisting}
dfplotcov(list(survey,ptrue))
\end{lstlisting}}
\noindent Unless other graphical parameters are specified, this plot (\fig{example_covariance}) shows the best fitting parameters (blue dots) with their 68\% and 95\% confidence regions (ellipses) in the Gaussian approximation, as well as the input parameters (black crosses). The numerical values of the fitted parameters and their uncertainties can be displayed by calling
{\color{blue}\begin{lstlisting}
dfwrite(survey)
\end{lstlisting}}
\noindent For an extended discussion of the physics and mathematics conveyed by Figs.~\ref{fig:example} and~\ref{fig:example_covariance} we refer to the detailed examples in \s{examples}.

The \dftools package includes various example routines that can be executed via \texttt{dfexample(case)}. ss:2dexamplearying the integer argument \texttt{case} from 1 to 4 produces examples similar to those shown in Sections~\ref{ss:package}, \ref{ss:lssexample}, \ref{ss:nonpara} and \ref{ss:2dexample}. The open source code of \texttt{dfexamples} discloses the implementation of these examples. More pedagogical tutorials, called `vignettes' in $R$, will be included in the code and always updated along with the package.

%%%%%%%%%%%%%%%%%%%%%%%%%%%%%%%%%%%%%%%%%%%%%%%%%%%%%%%%%%%%%%%%%%%%%%%%%%%%%%%%%%%%%%

\section{Basic examples and benchmarks}\label{s:examples}

To benchmark the MML formalism and its implementation in the package \dftools, we generate and then fit mock surveys with precisely controlled input parameters. All surveys in this section assume that the true galaxy MF is a Schechter function (\eq{Schechter}) with parameters
\be\label{eq:true}
	\para_{\rm true}=(\log_{10}\phi_\ast,\log_{10}M_\ast,\alpha)=(-2,11,-1.3).
\ee
We will often drop the units, as they don't matter for the examples. However, upon adopting $\phi_\ast$ in units of $\rm Mpc^{-3}dex^{-1}$ and $M_\ast$ in units of $\msun$, the vector $\para_{\rm true}$ is consistent with the rounded parameters of the observed galaxy MF \citep{Bell2003b,Papastergis2012} for baryonic matter (stars and cold gas). Other MF models are considered in \s{advanced}.

Using this fixed input Schechter function, the following subsections consider widely different sample sizes to illustrate different effects. The largest sample ($N=10^5$ galaxies, \ss{large}) serves to isolate the effect of Eddington bias from other effects and demonstrate its removal. The mid-sized samples ($N\approx10^3$ galaxies, \ss{medium}) serve to test the uncertainties of the MMLE and illustrate their dependence on the selection function. Finally, an array of small galaxy samples ($N\approx10$ galaxies, \ss{small}) is picked to show the effect and removal of intrinsic estimator bias, only noticeable in such small samples.

\subsection{Large galaxy samples}\label{ss:large}

\begin{figure}
\begin{center}
\includegraphics[width=\columnwidth]{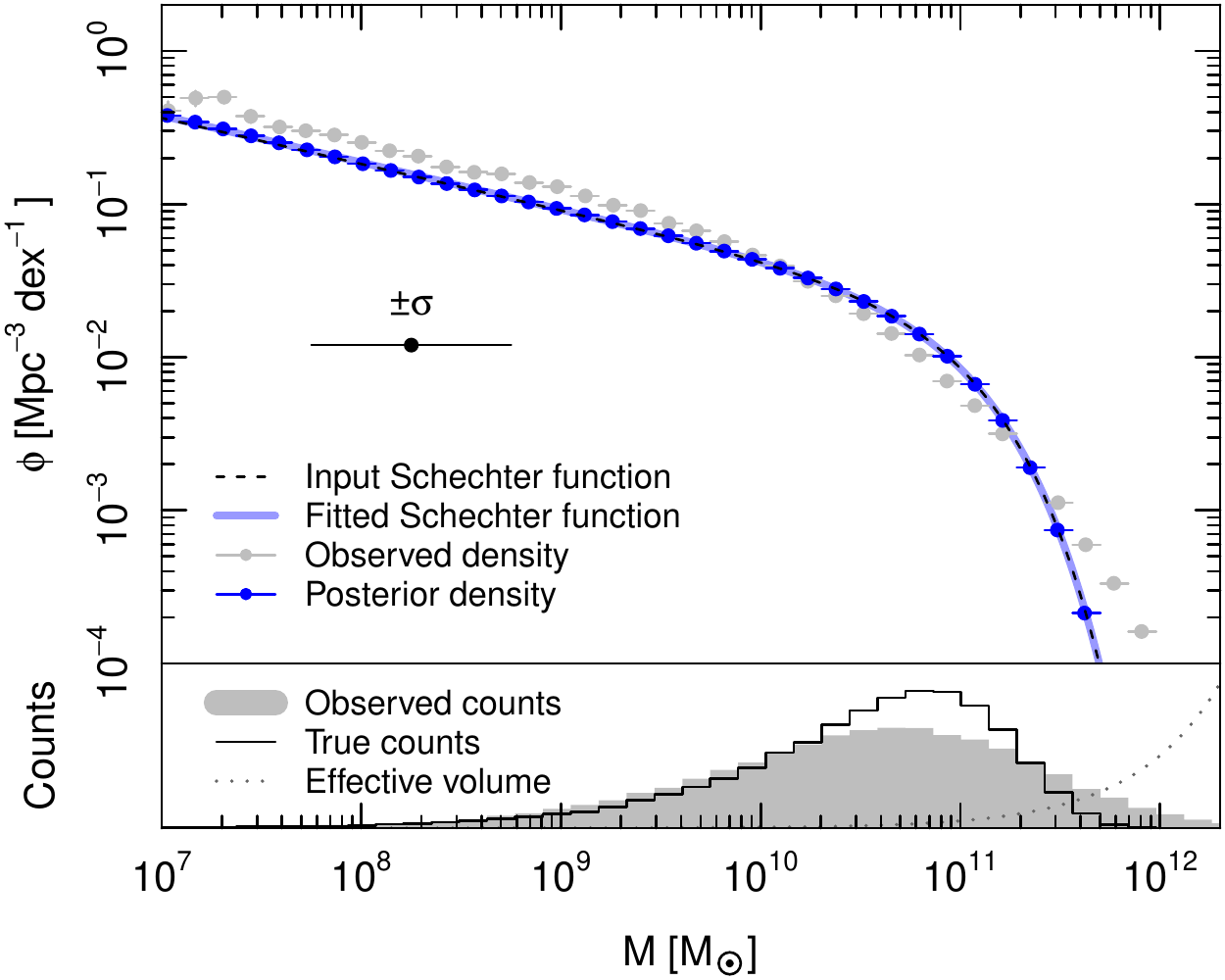}\vspace{-3mm}
\caption{Example of recovering a Schechter function from a mock data set with $10^5$ galaxies. Their large observing errors of standard deviation $\sigma=0.5~\rm dex$ give rise to significant Eddington bias, which is fully removed by the fit. Details are given in \ss{large}.}\label{fig:large}
\end{center}
\end{figure}

Let us model a galaxy survey that is purely sensitivity limited. For a constant mass-to-light ratio, this implies that a galaxy of mass $M$ is detectable to a maximum distance proportional to $M^{1/2}$. Hence the effective volume scales as
\be\label{eq:sensitivitylimit}
	\veff(x) \propto M^{3/2} \propto 10^{1.5x},
\ee
shown as the dotted line in the bottom panel of \fig{large}. We randomly pick $N=10^5$ galaxies (\ie log-masses $x_i$) from the expected source count function, \ie from the PDF $\phi(x|\para_{\rm true})\veff(x)$. Each log-mass $x_i$ is then perturbed by adding a random observing error $\Delta x_i$, drawn from a normal distribution with standard deviation $\sigma=0.5$ (large horizontal error bar in \fig{large}). The increase in $\veff(x)$ with mass makes the source count distribution of the sample (histogram in \fig{large}) biased towards high masses compared to the underlying MF (short-dashed line in upper panel) -- a typical feature known as Malmquist bias.

For purely illustrative purposes we bin the masses of this survey and compute the space density using the $1/\vmax$-method, \ie by dividing the number of galaxies in each bin by the mean volume $\veff(x)$ associated with this bin. The resulting MF (purple points in \fig{large}) differs significantly and systematically from the input MF: it clearly overestimates the number of low-mass ($M<10^9\msun$) and high-mass ($M>2\cdot10^{10}\msun$) galaxies, while underestimating the intermediate mass range. This important offset is due to Eddington bias: the declining MF makes it more likely that a galaxy with an observed log-mass $x$ is truely a lower-mass galaxy scattered upwards than a higher-mass galaxy scattered downwards. Overall this tends to smooth out the true MF, here by a Gaussian filter in log-mass $x$.
The challenge consists in recovering the true MF, given the Eddington biased mock data -- a key purpose of the MML method. The fit-and-debias algorithm converges in seven iterations in this example, taking only about 20s on a 3~GHz Intel Core i7 CPU, which is a very reasonable computation time for fitting $10^5$ uncertain measurements. The fit $\hat\para=(-2.000\pm0.004,10.998\pm0.002,-1.303\pm 0.003)$ is statistically consistent with and closely matched by the input parameters $\para_{\rm true}$. In fact, graphically the fitted MF in \fig{large} is indistinguishable from the input MF, illustrating the accurate removal of all Eddington bias.

We stress that the uncorrected Eddington bias (as seen in the poor fit of the purple data) is orders of magnitude larger than the uncertainties of the fit. Even if the observational uncertainties were as small as typical high-precision multi-wavelength stellar mass errors of $\sigma=0.1$ (in $\log_{10}M$, \citealp{Wright2017}), Eddington bias would still dominate over shot noise in a survey with $10^5$ (or more) galaxies. We therefore expect the MFs of modern galaxy surveys (references in \s{introduction}) to depend significantly on Eddington bias removal.

Finally, we can visualize the posterior data, computed as part of the MML method: \eq{debias} yields the posterior PDFs $\tilde\rho_i(x|\hat\para)$ for the log-mass of each galaxy $i$ individually, which can be summed up via \eq{defn} to obtain to posterior for the observed source counts, $n(x|\hat\para)$. The posterior observed MF is then computed as
\be
	\tilde\phi(x|\hat\para) = \frac{n(x|\hat\para)}{\veff(x)}.
\ee
For illustrative purposes, we chose to bin this function into the same mass-bins as the observed data, while defining the bin-values as the mean of $\tilde\phi(x|\hat\para)$ and the bin-centre as the $\tilde\phi$-weighted mean of $x$  in each bin. These binned posterior data (black points in \fig{large}) admit a similarly excellent agreement with the input MF as the best-fitting model itself.

\begin{figure*}
\begin{center}
\includegraphics[width=1\columnwidth]{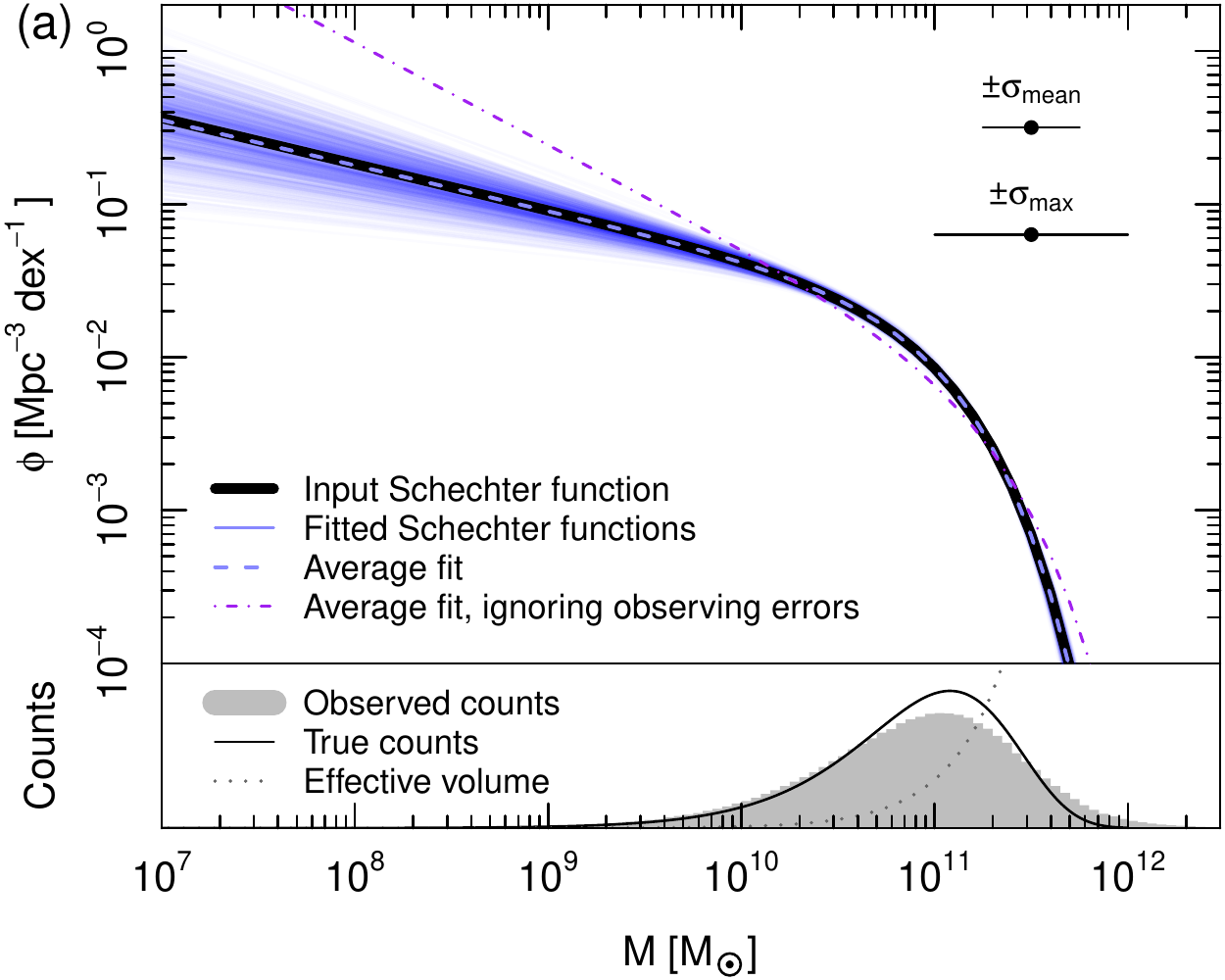}\hspace{5mm}
\includegraphics[width=1\columnwidth]{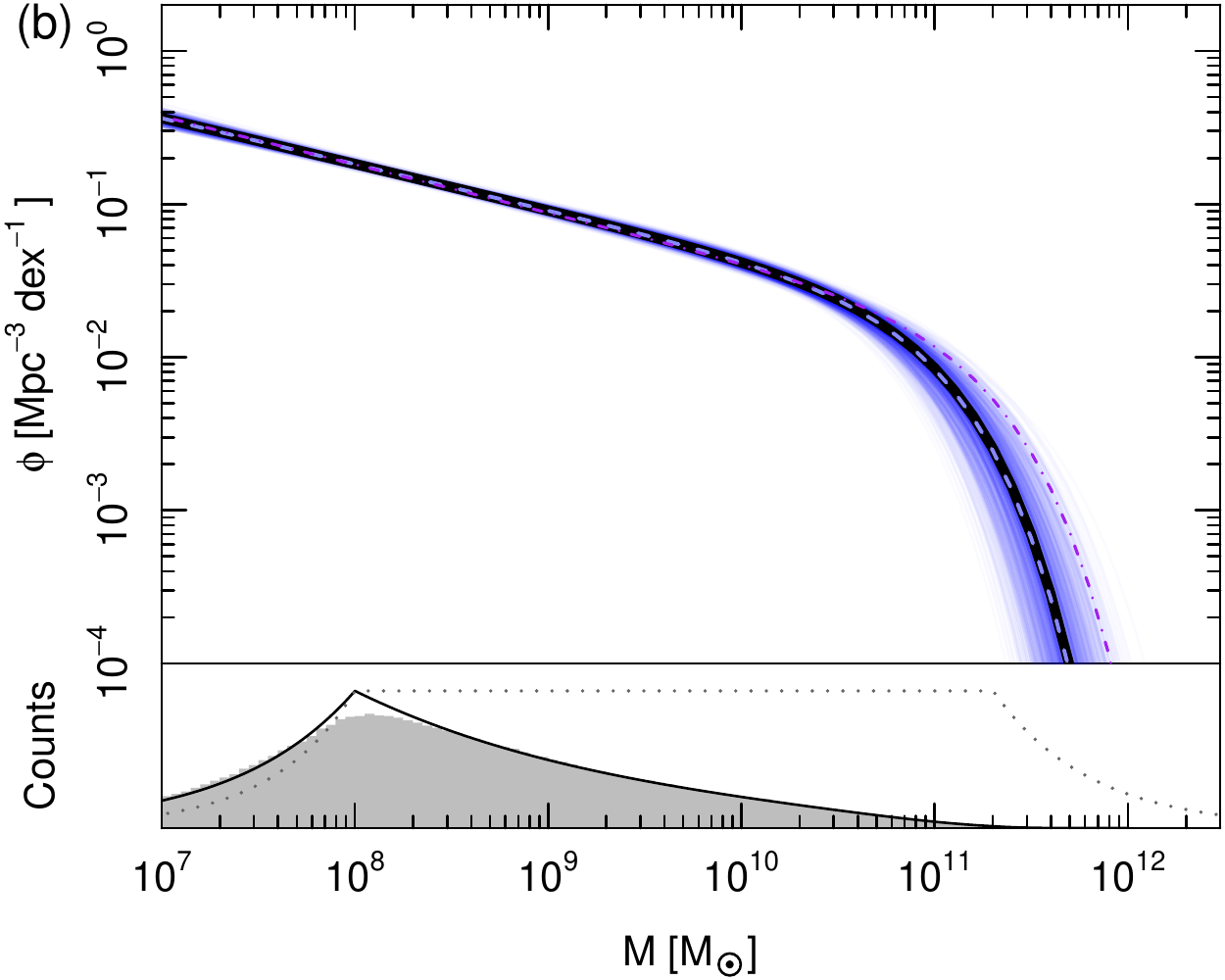}\\ \vspace{2mm}\hspace{0.8mm}
\includegraphics[width=1\columnwidth]{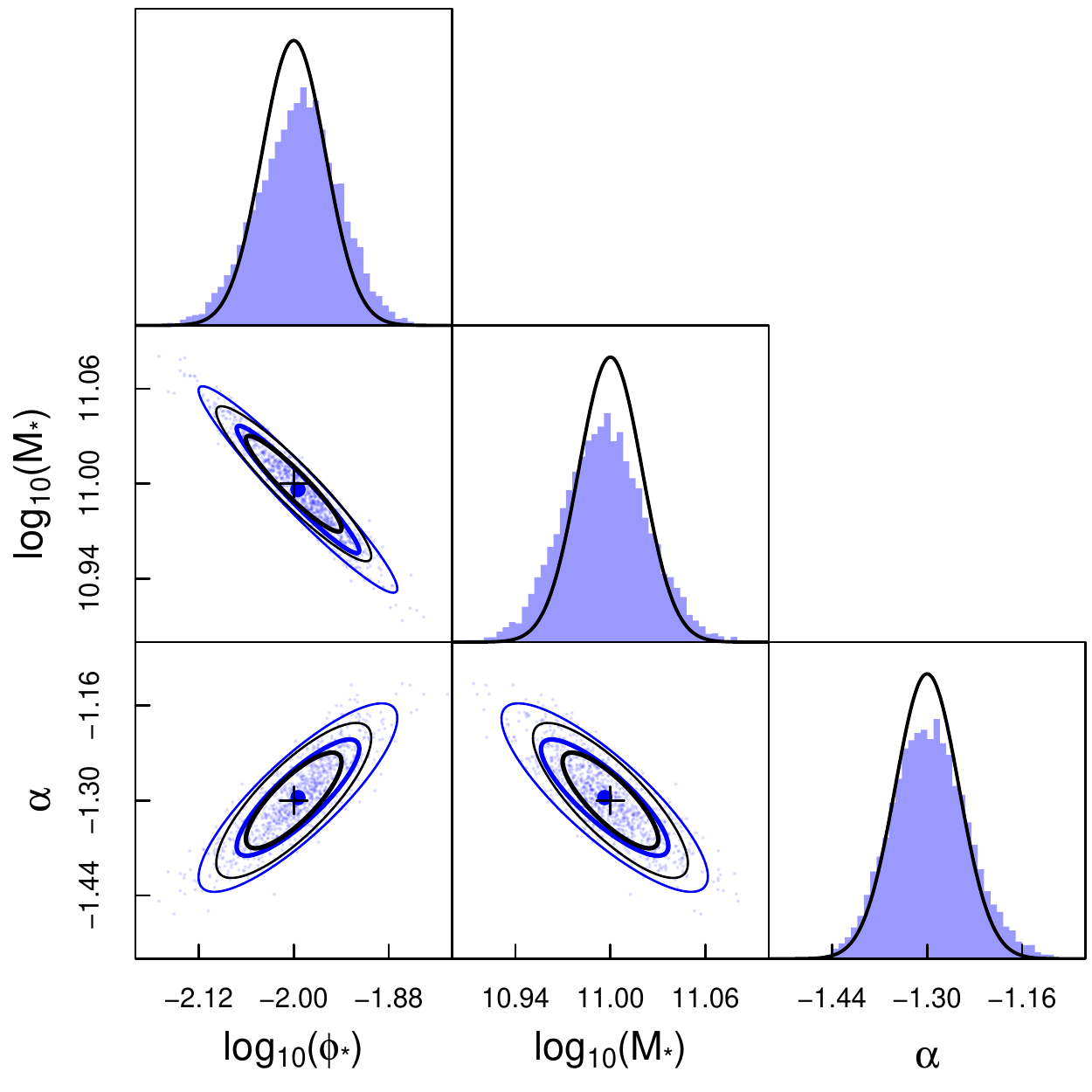}\hspace{5.1mm}
\includegraphics[width=1\columnwidth]{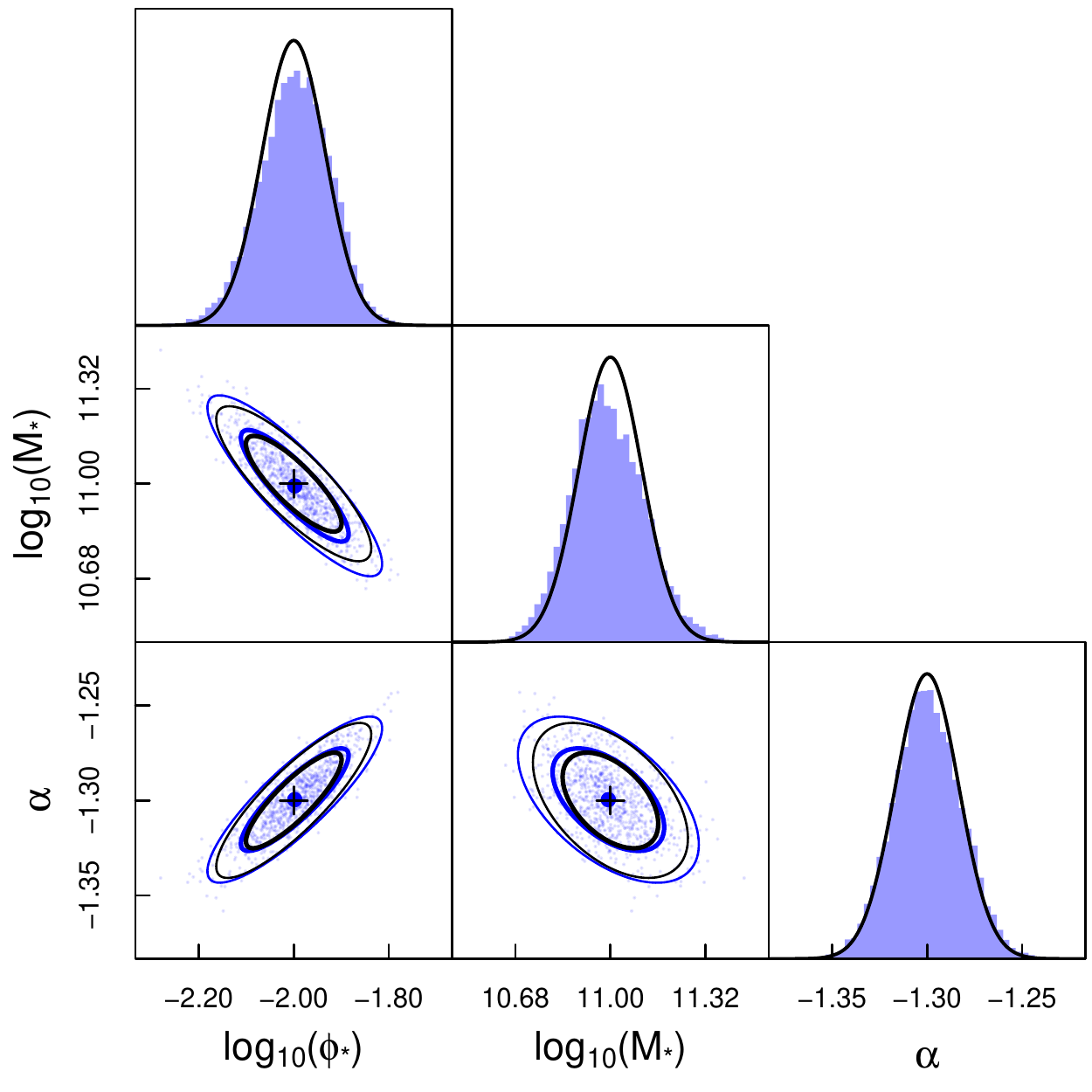}\vspace{-3mm}
\caption{Examples of recovering a Schechter function from mock surveys with $10^3$ galaxies, in the presence of widely different selection functions, biased towards high masses (a) and low masses (b). In each case, the MML fit is repeated with $10^4$ random mock samples, resulting in a zoo of MFs with parameter distributions shown in the bottom panels. The mean fits are in excellent agreement with the input model and the parameter distributions (blue ellipses) roughly match the Hessian predictions (black ellipses). Details are given in \ss{medium}.}\label{fig:medium}
\end{center}
\end{figure*}

\subsection{Medium galaxy samples}\label{ss:medium}

The previous example was so large that the fitting errors vanish on the scale of \fig{large}. To discuss these errors, their covariance and dependence on the selection function, we now transition to smaller samples of $N\approx10^3$ galaxies, where the expected model uncertainties are much larger. The samples are again drawn from the input Schechter function with parameters in \eq{true}, but using two selection functions: one that is purely sensitivity-limited ($\veff\propto M^{3/2}$) and one that is approximately volume-limited, providing a constant effective volume in the mass range $10^8\msun\leq M\leq2\cdot10^{11}\msun$ with a deliberate exponential cut-off for higher masses. The effective volumes of these two mock surveys are shown as the dotted lines in the `selection' panel of \fig{medium}. The respective expected source count densities $\phi(x|\para_{\rm true})\veff(x)$ are shown as grey shading. The galaxy masses are randomly drawn from this source count density and then perturbed with random observing errors from a log-normal distribution with standard deviation $\sigma$, where $\sigma$ itself is different for each galaxy and drawn from a uniform random distribution between 0 and $\sigma_{\rm max}=0.5$. Hence, the mean error scale is $\sigma_{\rm mean}=0.25$. We chose to vary the uncertainty scale for each source to verify the MML method in this case.

For both selection functions, $10^4$ random mock samples of $N\approx10^3$ galaxies are generated and fitted with a Schechter function using \dftools, while accounting for the different measurement uncertainties. The actual number of galaxies in each sample is itself drawn from a Poisson distribution with expectation $\langle N\rangle=10^3$ to mimic the shot noise inherent to any real survey. The fits are plotted as light blue lines in the upper panel of \fig{medium}. The distribution of these blue lines are very different in \fig{medium}a and b: in the first case, the mock sample is biased towards massive galaxies, leaving the low-mass end of the model poorly constrained; the second case shows the opposite situation. We deliberately chose these two extreme cases to illustrate the robustness of the MML method irrespective of the selection function.

The distributions of the $10^4$ fitted Schechter parameters are displayed as blue histograms in the bottom panels of \fig{medium}. They are approximately Gaussian, as expected to the extent that the Laplace approximation applies, \ie that the log-likehood is described by a second-order Taylor expansion around its maximum. The parameter covariances are shown as blue point-clouds with 68\% (thick blue lines) and 95\% (thin blue lines) elliptical contours. These contours are centred on the average fitted parameters (blue dots). For comparison, the black crosses show the input parameters $\para_{\rm true}$ and the black lines, centred on these parameters, show the average Gaussian uncertainties and covariances predicted from the averaged inverse Hessians of the log-likelihoods (see \ss{uncertainties}).

The covariance figures convey two messages: First, the agreement between mean fitted parameters and the input parameters is excellent in the sense that their difference is small ($<10\%$) relative to the mean parameter uncertainties. This can also be seen in the visual overlap of the Schechter function associated with the mean parameters (dashed blue line) and the input Schechter function (thick black solid line). Statistically, the differences between the average fits and input parameters are consistent with being equal to zero. In other words, the MMLE behaves nearly like an unbiased estimator (\ie its expectation matches the true population value) in these examples. Secondly, the agreement between the parameter covariances determined from the $10^4$ runs and those predicted from the Hessian is good in the sense that the difference is smaller than the actual variances. Hence, for practical purposes, the Hessian approximation of the parameter uncertainties normally suffices. This statement does not apply in general, but counter-examples are rare and rather unphysical (see \ss{resampling}), except when accounting for cosmic LSS (see \ss{lssexample}).

Finally, we emphasize that Eddington bias -- the focus of \ss{large} -- also affects the examples of \fig{medium}, but has been dealt with automatically by the MML method. To show this, \fig{medium} also displays the Schechter functions associated with the average parameters obtained when ignoring observational errors (dash-dotted purple lines). These MFs significantly deviate from the input function, especially in the poorly constrained parts, due to Eddington bias.

\begin{figure}
\begin{center}
\includegraphics[width=\columnwidth]{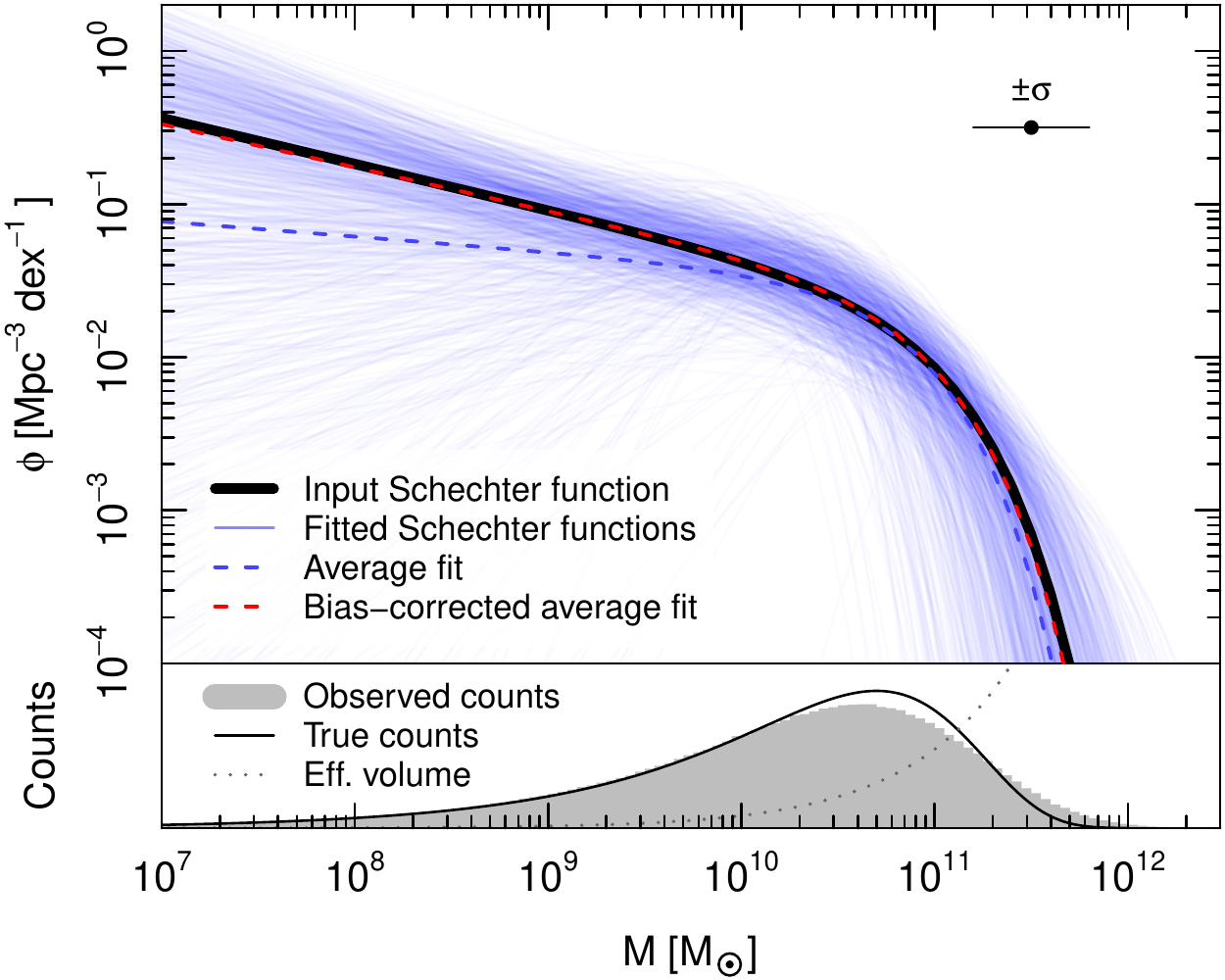}\vspace{0.1cm}
\includegraphics[width=\columnwidth]{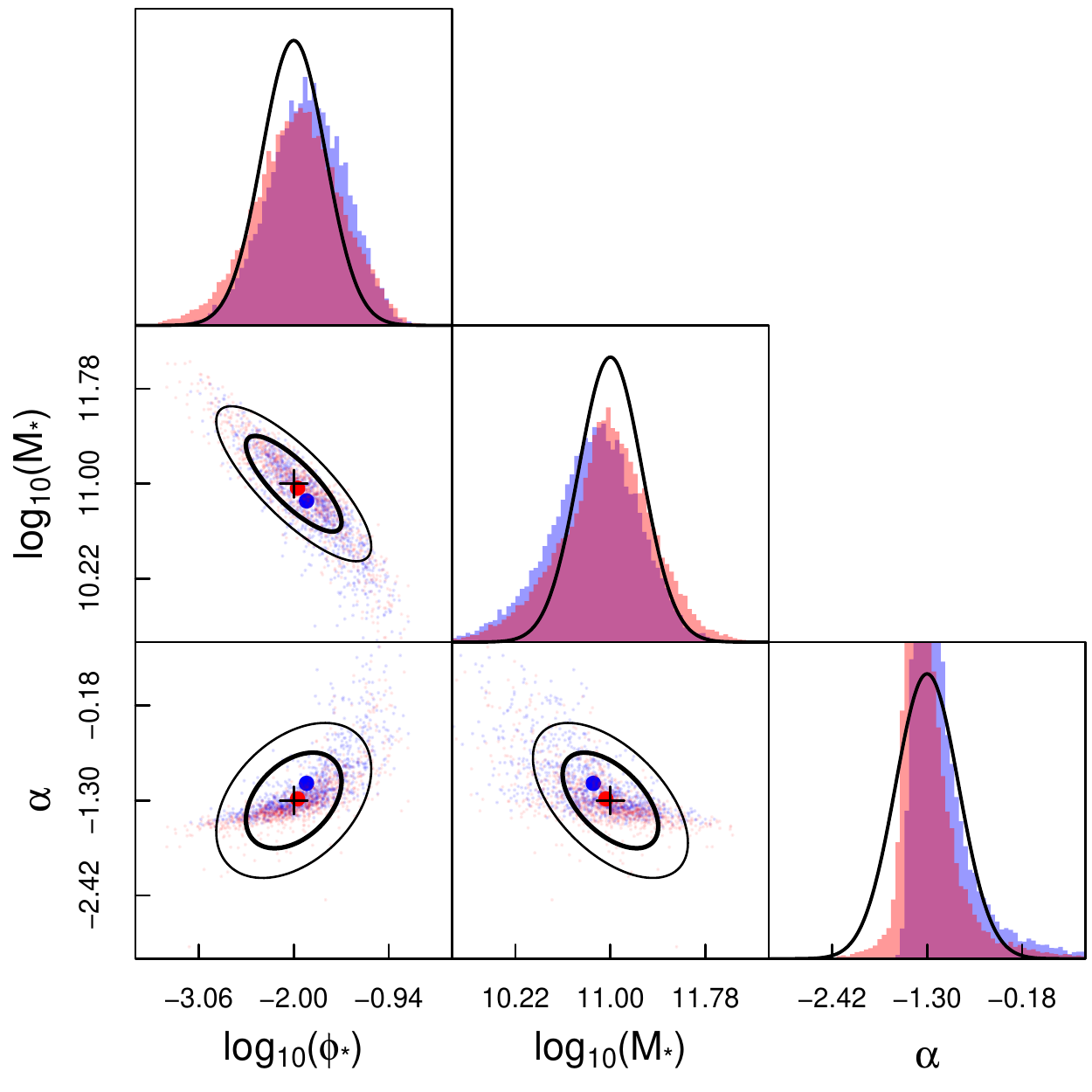}\vspace{0.0cm}
\caption{Example of recovering a Schechter function from mock surveys with only $10$ galaxies. The MML fit is repeated with $10^4$ random mock samples, resulting in a zoo of MFs with parameter distributions shown in the bottom panel. The average fit deviates significantly from the input model, revealing the classical estimator bias of the ML method for small samples. This bias is corrected using jackknifing. The parameter distributions of all $10^4$ runs are shown in the bottom panel for the normal (blue) and bias-corrected (red) MML estimator. Details are given in \ss{small}.}\label{fig:small}
\end{center}
\end{figure}

\subsection{Small galaxy samples}\label{ss:small}

As explained in \ss{bias}, the MMLE, like the MLE, is expected to be slightly biased, \ie its expectation differs from the true population model. This estimator bias vanishes as $N^{-1}$ as the samples size $N\rightarrow\infty$ \citep{Kendall1979} and is hence most pronounced when fitting small samples. In the examples of Sections~\ref{ss:large} and \ref{ss:medium} based on $N=10^5$ and $N=10^3$ galaxies, the bias was neglegible. So let us consider mock surveys of only $N\approx5$ to $N\approx100$ galaxies. These surveys are drawn from the input Schechter function (with parameters in \eq{true}) using an effective volume varying as $\veff(x)\propto M^{0.8}$ (dotted line in \fig{small}), which is similar to a sensitivity-limited survey (\eq{sensitivitylimit}), but less biased towards high masses to ensure that the low-mass end of the MF is at least marginally constrained. The expected source count density is shown as grey shading in \fig{small}. The masses drawn from this distribution are perturbed with observing errors from a log-normal distribution of standard deviation $\sigma=0.3$ (error bar in \fig{small}).

\begin{figure}
\begin{center}
\includegraphics[width=\columnwidth]{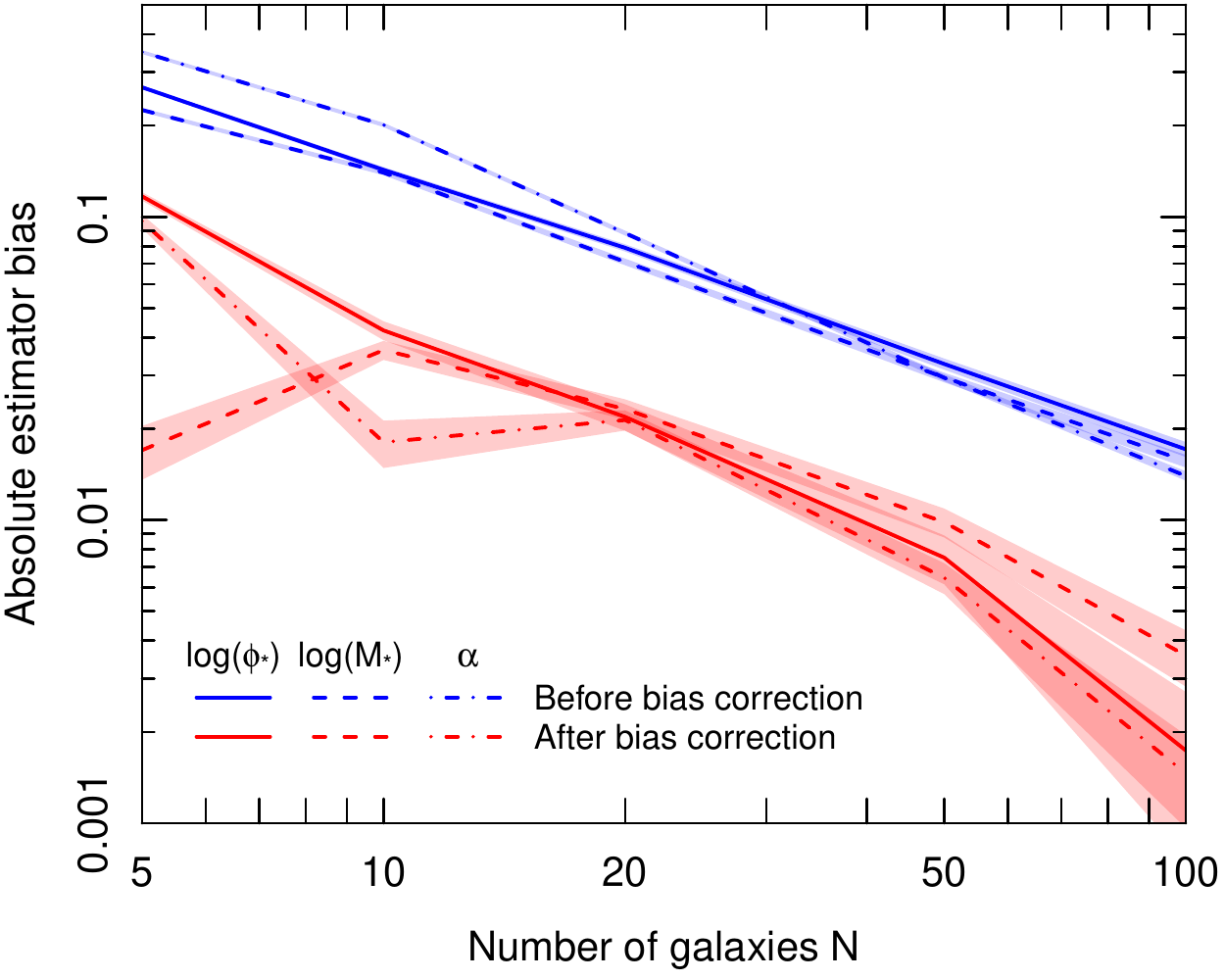}
\caption{The MML estimator is, like any ML estimator, biased. Plotting this bias as a function of $N$ reveals the typical bias scaling as approximately $N^{-1}$. The first-order jackkinfing method reduces the bias by an order of magnitude. Details are given in \ss{small}.}\label{fig:small_trend}
\end{center}
\end{figure}

We generate surveys of five different sizes with expected numbers of objects $\langle N\rangle=5,\ 10,\ 20,\ 50,\ 100$. For each size, $10^4$ mock surveys are generated, each with an actual number of $N$ galaxies, drawn from a Poisson distribution of mode $\langle N\rangle$. Surveys with only two or less galaxies are excluded (12\% for $\langle N\rangle=5$ and 0.3\% for $\langle N\rangle=10$), as they can admit unbound MLEs for three parameters. The MML method is then used to fit a Schechter function to each survey, while accounting for the observational uncertainties.

For  the $\langle N\rangle=10$ surveys, the Schechter fits are shown as light blue lines in \fig{small}. Naturally the scatter between these fits is much larger than in the previous example (\fig{medium}) due to the 100-times smaller sample size. The distributions and covariances of the fitted parameters are shown in blue in the bottom panel of \fig{small}. As in the previous example, black lines represent the covariances from the Hessian of the log-likelihood. The average fitted parameters (blue dots) lie significantly off the input values (black crosses). Analogously, in the upper panel, the Schechter function associated with the average parameters (dashed blue line) clearly differs from the input Schechter function (thick black line). This difference between expected MML fits and true parameters is the estimator bias we aimed to evidence. While clearly visible, this bias remains small compared to the parameter uncertainties. Hence correcting this bias might not be necessary.

If desired, the MMLE bias can be removed to first order in $N^{-1}$ by the jackknifing method presented in \ss{bias}. Applying this method results in the red parameter distributions in \fig{small}. The average best-fitting parameters produce the Schechter function shown as the red dashed line in the upper panel. This function is almost identical to the input Schechter function, demonstrating the effectiveness of this approach. The precise biases of the corrected (red) and uncorrected (blue) parameters is shown in \fig{small_trend} for all considered sample sizes. We find that jackknifing reduces the estimator bias by about an order of magnitude. The somewhat strange behaviour of the parameter errors at $N=5$ is due to the higher-order correction terms becoming important for such low numbers of galaxies.

%%%%%%%%%%%%%%%%%%%%%%%%%%%%%%%%%%%%%%%%%%%%%%%%%%%%%%%%%%%%%%%%%%%%%%%%%%%%%%%%%%%%%%

\section{Advanced examples}\label{s:advanced}

So far, all examples focused on a homogenous universe without LSS, where the galaxy population is fully described by a three-parameter Schechter function. This section expands the view towards additional complications encountered when working with real data. All of these complications can be dealt with using \dftools.

\subsection{Cosmic large-scale structure}\label{ss:lssexample}

\begin{figure}
\begin{center}
\includegraphics[width=\columnwidth]{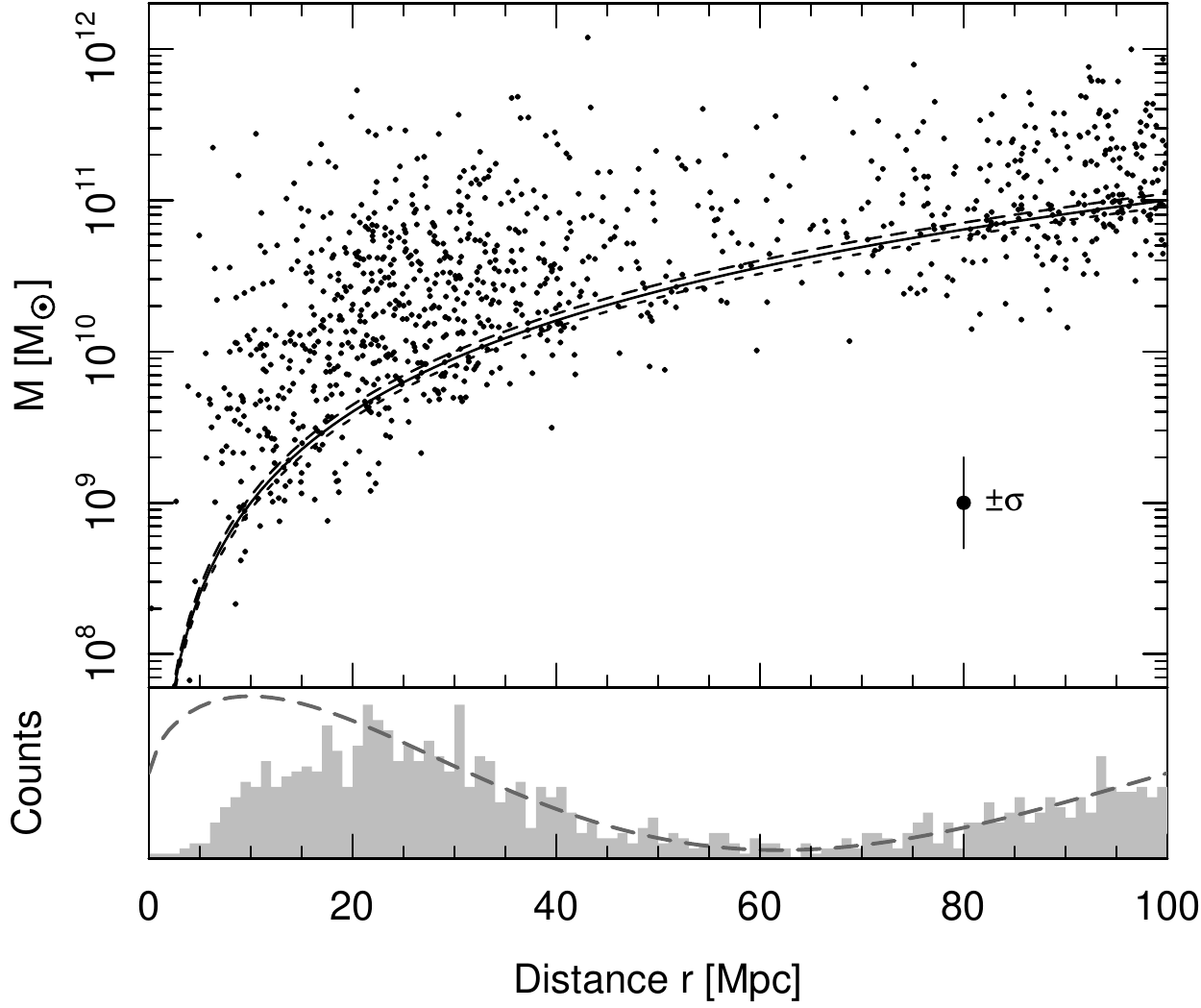}
\caption{Distance$-$mass distribution of a mock sample with a sensitivity-limited selection function (solid line) assuming the non-uniform cosmic LSS represented by the function $g(r)$ shown as long dashed line in the bottom panel. Details are given in \ss{lssexample}.}\label{fig:lss_distance}
\end{center}
\end{figure}

\begin{figure}
\begin{center}
\includegraphics[width=\columnwidth]{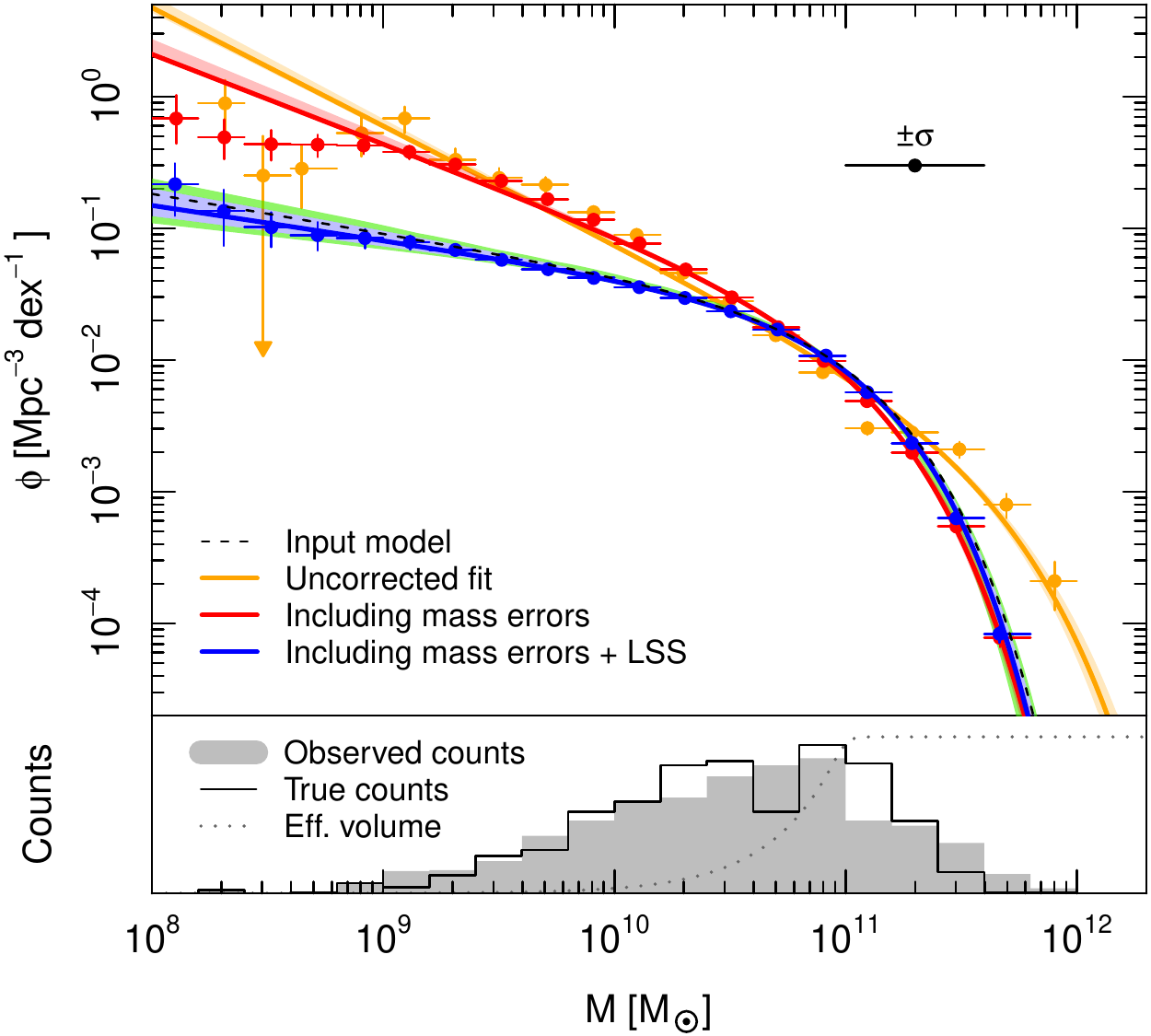}
\caption{Illustration of fitting a MF to a mock sample (shown in \fig{lss_distance}) with non-uniform LSS. If LSS-correction is activated, the input function is accurately recovered. Details are given in \ss{lssexample}.}\label{fig:lss_fit}
\end{center}
\end{figure}

Galaxy surveys are inevitably subject to cosmic LSS, which can bias the reconstruction of the galaxy MF as explained in \ss{lsstheory}. To test the removal of this bias (via \eq{veffgradualautolss}), we consider a typical, sensitivity-limited survey with a fuzzy detection limit. Isocontours of the selection function $f(x,r)$ are shown in the upper panel of \fig{lss_distance} for $f=0.1$ (short-dashed line), $f=0.5$ (solid line) and $f=0.9$ (long-dashed line). Next, we pick a non-uniform function $g(r)$ (dashed line in the bottom panel of \fig{lss_distance}), representing the number density contrast due to cosmic LSS. Using the resulting overall selection function $f(x,r)g(r)$ and our reference Schechter function (with parameters $\para_{\rm true}$ of \eq{true}), we draw a sample of $N=10^3$ galaxies and perturb their masses by random, log-normal observing errors of standard deviation $\sigma=0.3$ (in $x=\log_{10}M/\msun$). The resulting mock sample (black points in \fig{lss_distance}) then admits the oscillating source counts shown as grey histogram in \fig{lss_fit}.

This mock sample is subject to two major biases: Eddington and LSS bias. If neither is accounted for when fitting a Schechter function (using a $1/\vmax$ or simplistic ML approach), the best-fitting solution (yellow line in \fig{lss_fit}) differs widely from the input MF (dashed black line).

\begin{figure}
\begin{center}
\includegraphics[width=\columnwidth]{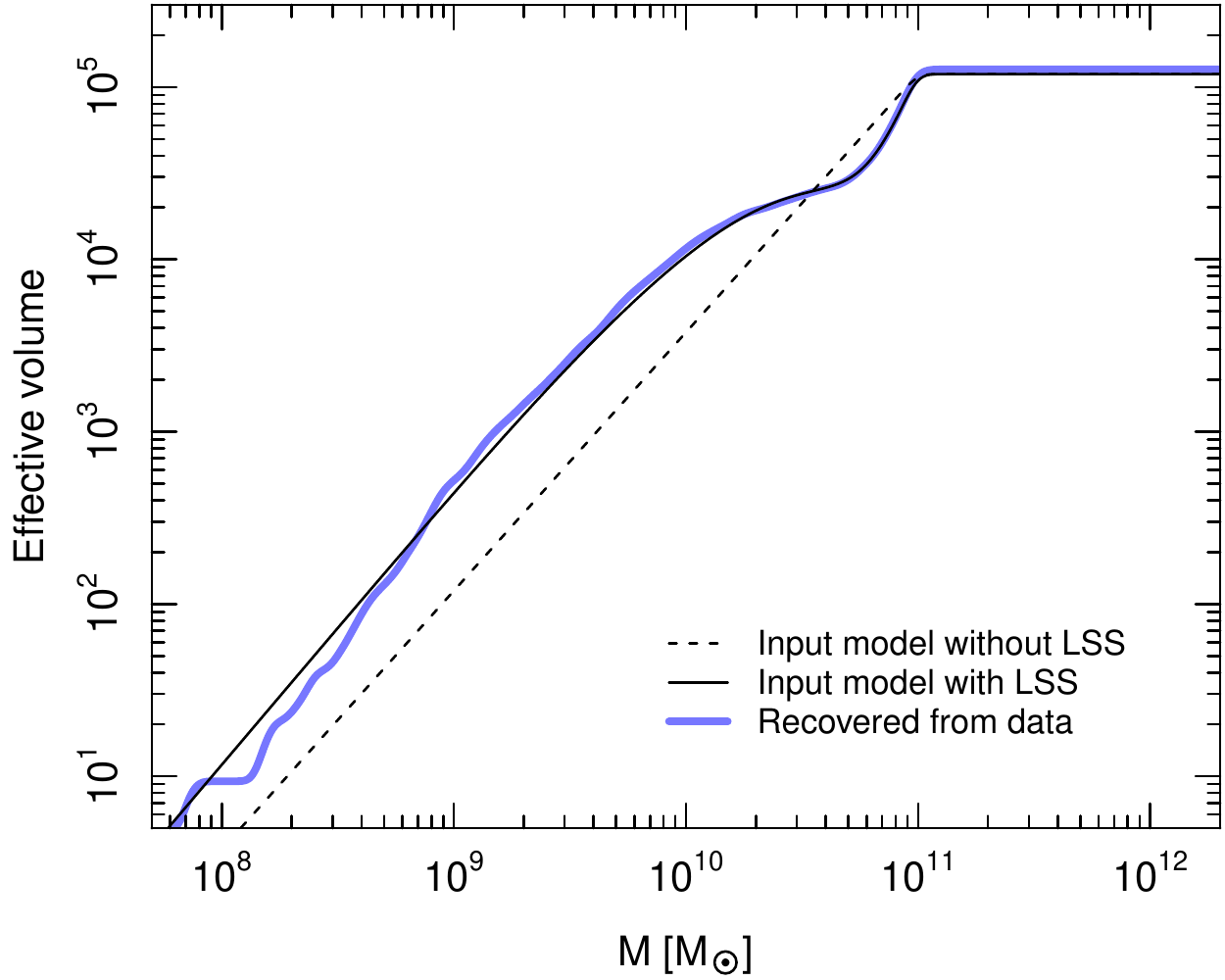}
\caption{Effective volume functions. The \emph{unknown} effective volume with LSS $\vefflss(x)$, needed to recover the MF in the presence of LSS, is approximately recovered using the known selection function $f(x,r)$ without LSS and the distance distribution of the survey (\fig{lss_distance}). Details are given in \ss{lssexample}.}\label{fig:lss_veff}
\end{center}
\end{figure}

Let us first look at the solution, if only Eddington bias is corrected, \ie mass observing errors are accounted for but not LSS. In this, case the effective survey volume is computed (automatically by \dftools) using \eq{veffgradual}. The effective volume is shown as the dashed line in \fig{lss_veff}. Solving the MML method using the `fit-and-debias' algorithm in \dftools, results in the Schechter function fit shown as the red line in \fig{lss_fit}. This fit works well in the high-mass end, but fails on the low-mass side, dominated by LSS bias.

The LSS bias can be approximately removed by using \eq{veffgradualautolss} instead of \eq{veffgradual} to compute the effective volume. To do so in \dftools, it suffices to set \texttt{correct.lss.bias = TRUE} when calling \texttt{dffit}. The resulting effective volume is shown as the blue line in \fig{lss_veff}. This effective volume accounts for LSS to the extent that this LSS is imprinted in the distance distribution of the galaxies in the sample. It strongly resembles the `true' effective volume with LSS (black solid line in \fig{lss_veff}), given in \eq{veffgraduallss}, which requires the input function $g(r)$ that is unknown to the observer. In fact, the best-fitting parameters $\hat\para=(\log_{10}\phi_\ast,\log_{10}M_\ast,\alpha)=(-1.97\pm0.07,10.96\pm0.04,-1.28\pm0.06)$ are statistically consistent with the input parameters $\para_{\rm true}$.

The parameter uncertainties quoted above are standard deviations, \ie square-roots of the diagonal covariance elements, computed from Hessian matrix (see \ss{uncertainties}). These covariant parameter uncertainties are represented by the light blue envelope around the solid blue line in \fig{lss_fit}. As detailed in \ss{uncertainties}, uncertainties can also be computed by resampling the data. In \texttt{dffit} this is achieved by specifying an integer value for the argument \texttt{n.bootstrap} (equal to $Q$ in \ss{uncertainties}). The user can choose whether to refit $\vefflss(x)$ at each resampling iteration via the logical argument \texttt{lss.errors}. If set to \texttt{FALSE} (no refitting of $\vefflss(x)$), the resulting parameter covariances are statistically consistent with the values computed from the Hessian matrix. If set to \texttt{TRUE}, $\vefflss(x)$ is refitted at each iteration. This approach results in $\sim2$-times larger standard errors, represented by the light green envelope in \fig{lss_fit}. The parameter uncertainties are bound to increase if $\vefflss(x)$ is refitted at each iteration, because the variance of $\vefflss(x)$ implies an additional uncertainty in the model parameters. The bootstrap method allows us to evaluate this additional uncertainty.

This error analysis reveals that the uncertainties of the most likely MF parameters increase significantly if the uncertainty of the LSS is accounted for. As emphasized in \ss{uncertainties}, it is therefore advisable to always quote parameter uncertainties with and without LSS uncertainties, when fitting real galaxy data.

\subsection{Quasi non-parametric modelling}\label{ss:nonpara}

So far all examples have used galaxy populations generated and fitted by a Schechter function. The MML formalism can nonetheless deal with any real MF model $\phi(x|\para)$, including so-called `non-parametric models', which parametrize the MF in bins rather than a single analytical function. In the literature, the ML formalism for fitting such binned MFs is often called the stepwise ML (SWML) formalism. It was first introduced by \cite{Efstathiou1988}. Unlike classical SWML methods, our MML formalism fully accounts for observing errors (Eddington bias).

\begin{figure}
\begin{center}
\includegraphics[width=\columnwidth]{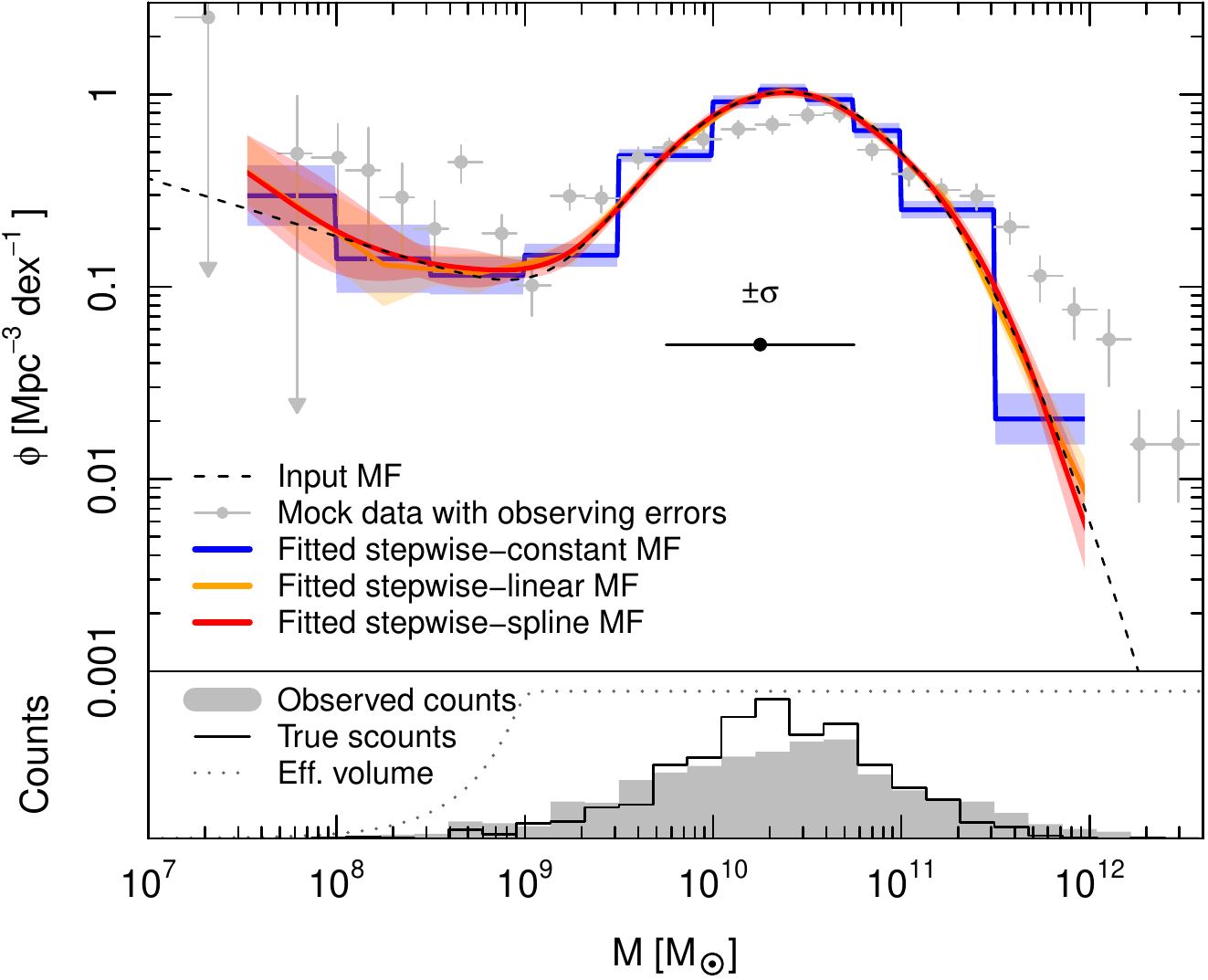}\vspace{0.1cm}
\caption{Illustration of fitting quasi non-parametric MF models to mock data generated from a survey drawn from the dashed input MF. Details are given in \ss{nonpara}.}\label{fig:sw}
\end{center}
\end{figure}

The \dftools package includes the function \texttt{dfswmodel} to generate stepwise MFs, which can then be fitted using \texttt{dffit} in exactly the same way as when fitting other MFs (\cf example in \ss{package}). To test this approach we generate a mock sample of $N=10^3$ galaxies in the same way as in previous examples. However, this time the input MF (short-dashed black line in \fig{sw}) differs from a Schechter function in that is has a second turning point with opposite curvature. Using this arbitrary input MF with the effective volume shown as the dotted line in the bottom panel and random observing errors of $\sigma=0.5$ (in $x=\log_{10}M/\msun$) results in the number counts of the grey histogram. The binned MF obtained using the $1/\vmax$-method is shown as grey data points. These points differ significantly from the input MF due to Eddington bias.

As in all previous examples, we then fit the mock data using the MML method. As fitting functions we use three stepwise MFs generated by \texttt{dfswmodel}:
\bi
\item A stepwise MF made of $N_{\rm bins}$ constant bins (blue line in \fig{sw}) is the simplest quasi non-parametric form.
\item Better fits can be obtained by choosing the model as a power law in each bin (yellow line). Upon requiring this stepwise power law to be continuous, this function is fully specified by its values at the $N_{\rm bins}$ bin centres.
\item The MF can also be modelled by a continuous cubic spline (red line) connecting vertices at the $N_{\rm bins}$ bin centres.
\ei
The function \texttt{dffit} automatically extrapolates these functions linearly outside their domain while finding the MML solution. This extrapolation avoids issues with measurements that lie on the edge of the MF domain. In the present example, the three stepwise MFs have been parametrized on $N_{\rm bins}=10$ bins of different sizes between $x=7.5$ and $x=12$. \fig{sw} highlights two clear advantages of the stepwise power law and spline functions relative to the stepwise constant model (blue). First, they are statistically consistent with the input model (black dashed-line) at any point, not just somewhere along each mass bin. Secondly, by construction, they satisfy the continuity (and smoothness for spline) condition, which one would expect for most physically meaningful MFs. Therefore, it seems advisable to use these continuous MF models when performing stepwise fits.

\subsection{Model evidence}\label{ss:evidence}

Which is the right MF model to fit? The MML method can fit (nearly) any MF model $\phi(x|\para)$, including quasi non-parametric ones (\cf \ss{nonpara}). So how can we decide, solely from the data, on the best model, at least amongst a finite set of proposals? Bayesian inference offers a powerful tool to answer this question: The conditional probability $Z$ of a model given the data is proportional to the integral of the likelihood function over the full parameter space. We here compute this integral in the Laplace approximation \citep{Daniels1954}, which treats the likelihood in the Gaussian approximation. In other words, the log-likelihood function is approximated at second order around its maximum. In the MML nomenclature, this approximation reads
\be\begin{split}
	Z & = \int \L(\hat\para,\hat\para)\exp\left[-\frac{1}{2}(\para-\hat\para)^\dag\cov(\hat\para)^{-1}(\para-\hat\para)\right]d^P\theta\\
	& = \L(\hat\para,\hat\para) \sqrt{(2\pi)^P|\cov(\hat\para)|},
\end{split}
\ee
where $\L(\hat\para,\hat\para)$ is the modified likelihood at the MML solution (\ss{likelihood}), $P$ is the number of scalar parameters (\ie the number of elements in the vector $\para$) and $\cov(\hat\para)$ is the covariance of the optimal parameters (\ss{uncertainties}). The value of $\ln Z$ is automatically computed when calling \texttt{dffit} and provided in the output list of this routine. If two competing models are \emph{a priori} (\ie before using the data) equally likely, then the odds of the first model over the second is given by the ratio $B=Z_1/Z_2$, known as Bayes factor.

As an illustration, we use an extension of the Schechter function to the four-parameter, $\para=(\log_{10}\phi_\ast,\log_{10}M_\ast,\alpha,\beta)$, MRP function \citep{Murray2017},
\be
	\phi(x|\para)_{\rm MRP} = \ln(10)\phi_\ast\mu^{\alpha+1}e^{-\mu^\beta},
\ee
where $\mu= M/M_\ast=10^x\msun/M_\ast$. The only difference to the Schechter function is the additional parameter $\beta$, which modulates the steepness of the exponential cut-off at the high-mass end. A Schechter function is recovered if $\beta=1$.

\begin{figure}
\begin{center}
\includegraphics[width=\columnwidth]{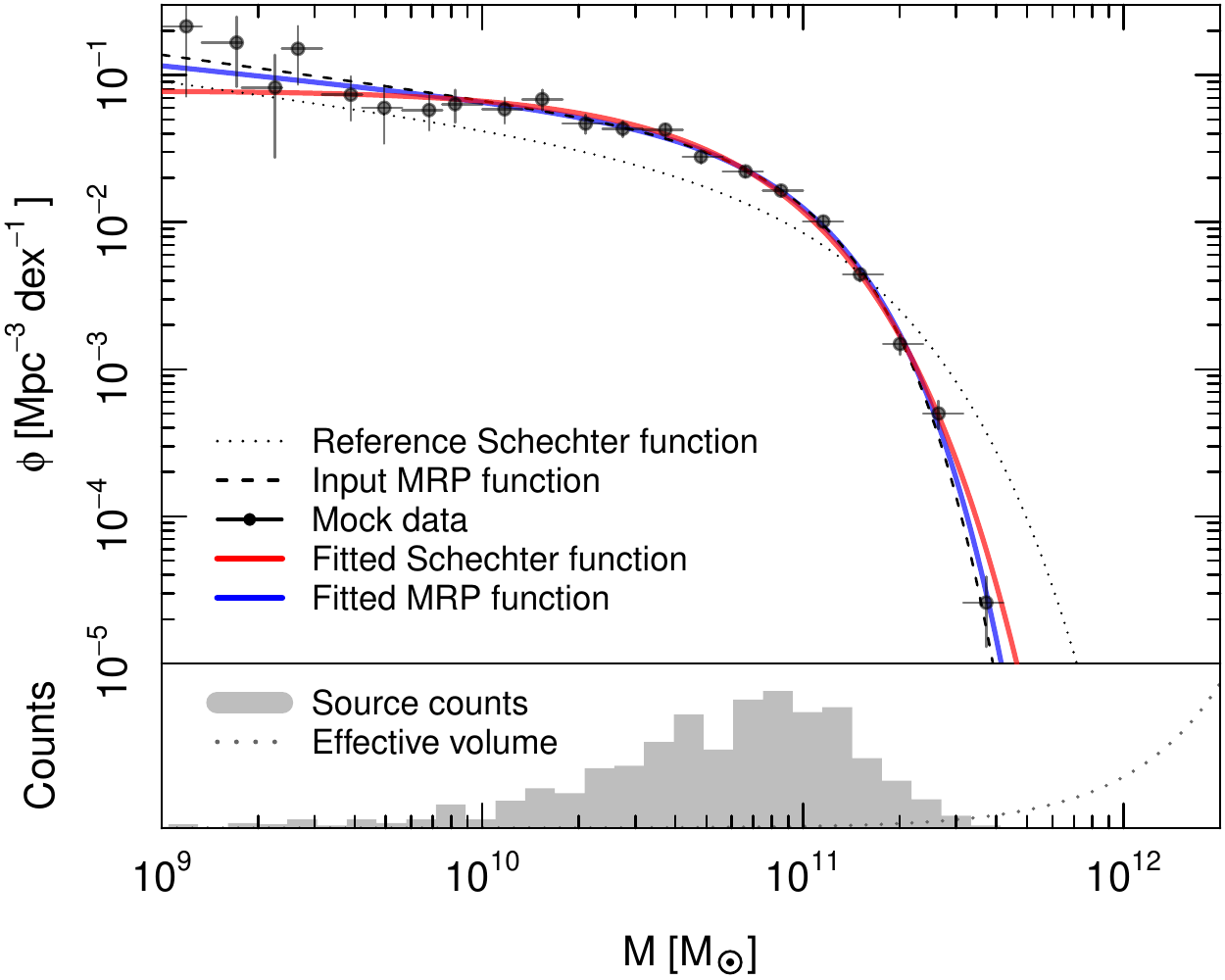}\vspace{0.1cm}
\includegraphics[width=\columnwidth]{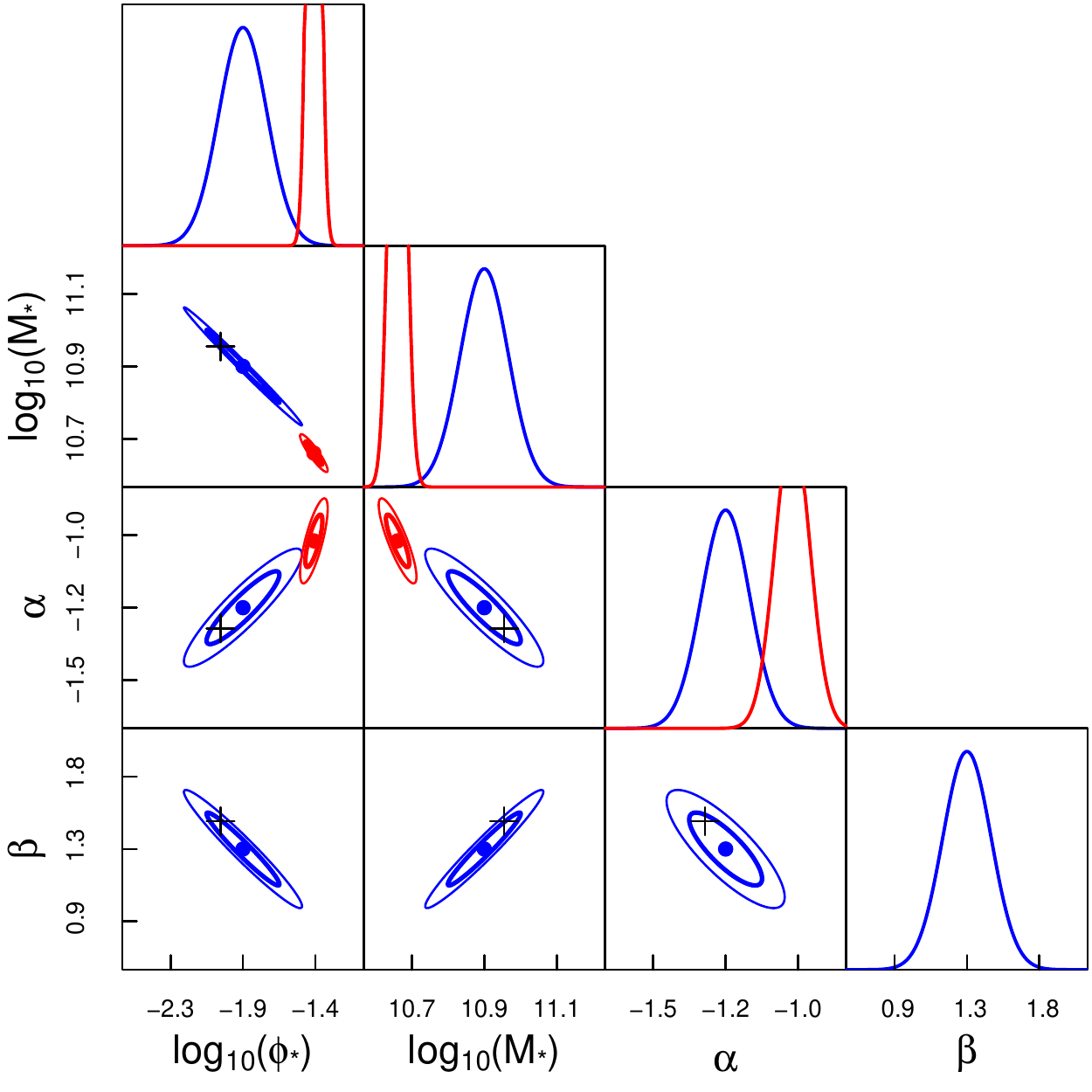}\vspace{0.6cm}
\caption{Example of fitting an MRP function and a Schechter function to a mock survey generated from an MRP function. The input MRP function can be approximately mimicked by a Schechter function with parameters $\phi_\ast$, $M_\ast$, $\alpha$ differing from those used in the MRP function, as can be seen from the parameter covariances in the bottom panel. Details are given in \ss{evidence}.}\label{fig:mrpfit}
\end{center}
\end{figure}

\fig{mrpfit} shows the reference Schechter function (dotted line, parameters in \eq{true}), compared to an MRP function (dashed line) with the same Schechter parameters and $\beta=1.3$. Comparing the dotted to the dashed line, it is clear that the value $\beta>1$ steepens the high-mass end and lowers the low-mass amplitude. From this MRP function we draw a mock sample of $N=10^3$ galaxies, adopting a sensitivity-limited effective volume function (\eq{sensitivitylimit}) without LSS. For illustration, these data are binned by mass and shown as black dots in \fig{mrpfit} -- without observing errors, hence matching the input model (dashed line).

\begin{figure}
\begin{center}
\includegraphics[width=\columnwidth]{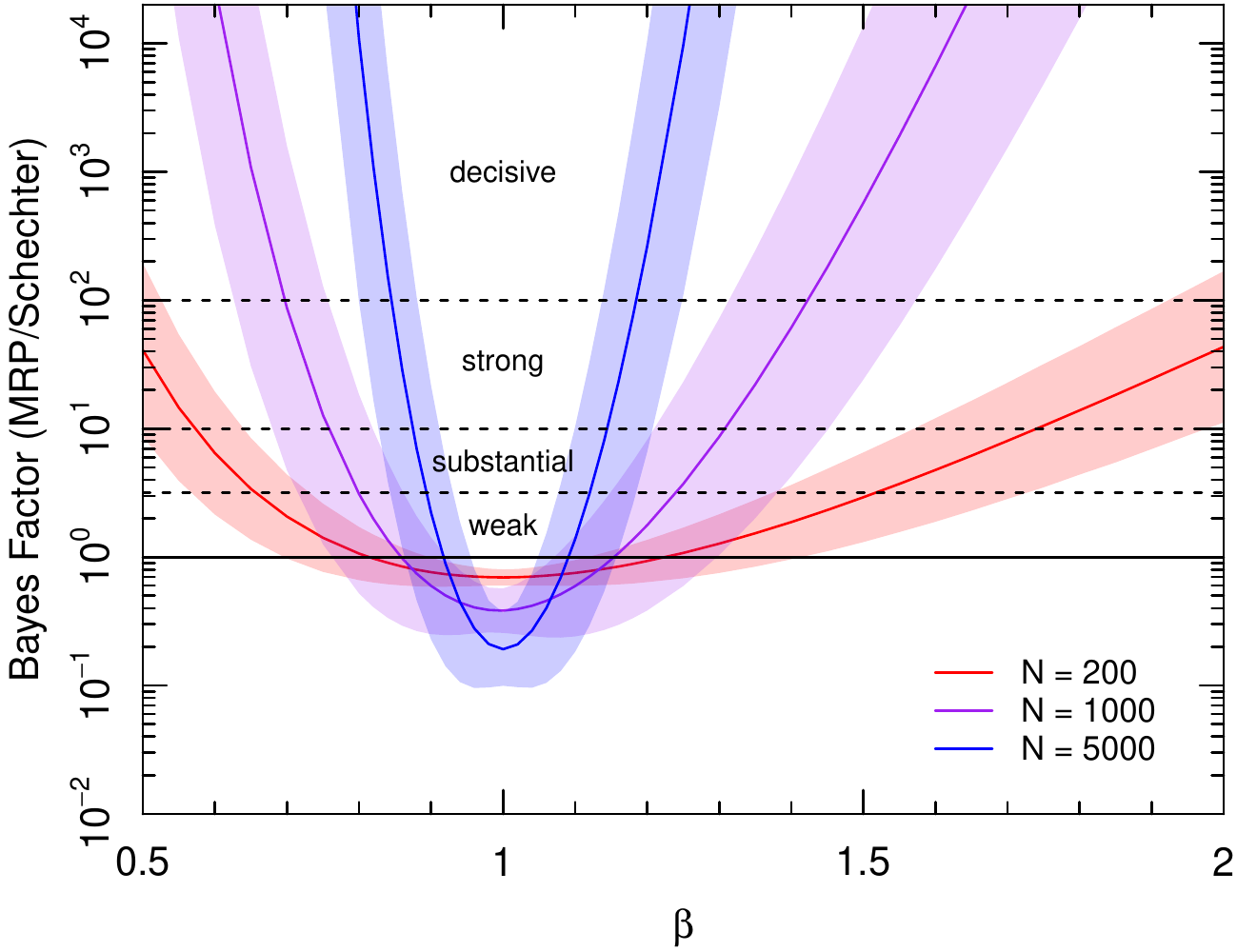}\vspace{0.1cm}
\caption{Evidence for the MRP model against the Schechter model as a function of the $\beta$-parameter in the input MRP function (where $\beta=1$ is a Schechter function). Lines and shaded regions show the mean Bayes factors and standard deviations determined from $10^3$ realizations for each $\beta$, for three different survey sizes. Details are given in \ss{evidence}.}\label{fig:mrpevidence}
\end{center}
\end{figure}

Using the MML method, we fit the mock data with both an MRP function (blue line in \fig{mrpfit}) and a Schechter function (red line). The respective model parameters and covariances are shown in the bottom panel. While the MRP function is closer to the input model (as expected), the Schechter function provides a surprisingly good fit. This is because values of $\beta\neq1$ in the MRP function, can be partially compensated by adjusting the three Schechter function parameters -- a statement that is also obvious from the strong covariances of the Schechter parameters with $\beta$. Naturally, the best fitting Schechter parameters differ significantly from the MRP parameters, because of this compensation of a $\beta\neq1$.

If we do not know whether the data in \fig{mrpfit} are drawn from an MRP or a Schechter function, finding the true population model is graphically quite tricky. The Bayes factor of the MRP model over the Schechter model is $B=10.3>1$ in this example, meaning that the MRP model is favoured. Note that this factor drops to $B\approx4$ if the LSS is considered to be unknown, because LSS could also be responsible for a deviation from the Schechter model (see \ss{lssexample}).

One would expect that it becomes easier to distinguish between the MRP and Schechter model, if $\beta$ deviates more strongly from $\beta=1$ and if  more data is available. This expectation is tested in \fig{mrpevidence}, which shows the Bayes factor (not accounting for LSS and without mass uncertainties) as a function of $\beta$ for three different survey sizes $N$. For each pair $\{\beta,N\}$, we generated and fitted $10^3$ random mock surveys. The distribution of their Bayes factors (one standard deviation) is shown as transparent shading. Interestingly, the Bayes factor implicitly penalizes models with more free parameters -- a property sometimes referred to as the Bayesian version of Ockham's razor. Therefore, if $\beta=1$, the Bayes factor favours the three-parameter Schechter model over the four-parameter MRP model, despite the fact that the MRP function is identical to the Schechter function for $\beta=1$. This penalization also implies that smaller samples require larger deviations from $\beta=1$ in order to favour the MRP model.

In the terminology of \cite{Kass1995}, a Bayes factor $B>10$ (or $B<0.1$) is called ``strong evidence'' (other denominations shown \fig{mrpevidence}). According to \fig{mrpevidence}, a sensitivity-limited galaxy survey must detect at least $N=200$ ($10^3$, $5\cdot10^3$) galaxies to provide such strong evidence for a value of $\beta\approx0.6$ (0.8, 0.9) or $\beta\approx1.7$ (1.3, 1.1). Through explicit calculations we found that roughly 10-times more galaxies are required if the (unknown) LSS is accounted for and mass uncertainties of $\sigma=0.3$ (in $\log_{\rm 10}M$) are assumed. % check!

\subsection{Mass-dependent measurement uncertainties}\label{ss:varsigma}

The mock data considered so far included statistical errors, whose magnitude was independent of the measured mass. In most real measurements, the statistical uncertainty of a datum nonetheless depends on its value. Such systematic variations of uncertainties are naturally dealt with by the MML framework, given a correct handling of the prior PDFs $\rho_i(x)$ of the observed data. To illustrate this point, let us assume that a datum of true value $x$ yields a measured value $x_{\rm obs}$ with probability $\varrho(x_{\rm obs}|x)$. Hence, an observation $i$ with measured value $x_i$ has a true value $x$ with probability
\be\label{eq:rhoderived}
	\rho_i(x) = \frac{\varrho(x_i|x)}{\int\varrho(x_i|\tilde x)d\tilde x}.
\ee

\begin{figure}
\begin{center}
\includegraphics[width=\columnwidth]{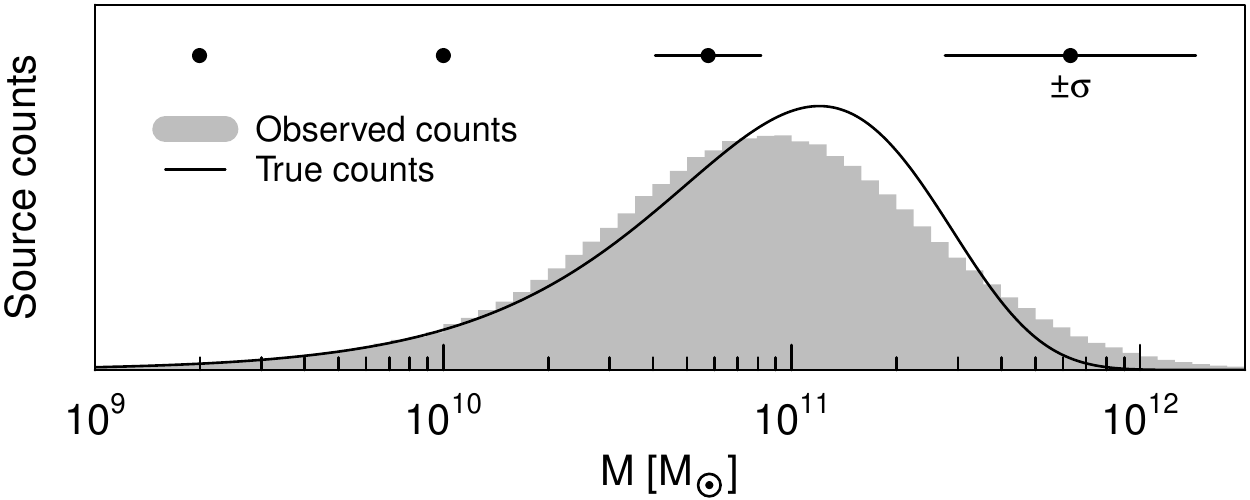}\vspace{-0.3cm}
\caption{Source counts of the sensitivity-limited mock survey of \fig{medium}a, when subjected to observational uncertainties $\sigma$ that depend on the mass. Details are given in \ss{varsigma}.}\label{fig:varsigma_scd}
\end{center}
\end{figure}

As an explicit example, we reconsider the sensitivity-limited survey with $N=10^3$ galaxies of \fig{medium}a, but assume the Gaussian uncertainty model
\be
	\varrho(x_{\rm obs}|x) = \frac{1}{\sqrt{2\pi}\sigma(x)}\exp\left(-\frac{(x-x_{\rm obs})^2}{2\sigma(x)^2}\right),
\ee
with a $\sigma$ that depends on the true value $x$ via,
\be
	\sigma(x) = \max(0,0.2x-2).
\ee
With this choice of $\sigma(x)$ the range and mean of $\sigma$ in the samples is similar to that in the example of \fig{medium}a. The true source counts and those perturbed by this mass-dependent error model are shown in \fig{varsigma_scd}.

\begin{figure}
\begin{center}
\includegraphics[width=\columnwidth]{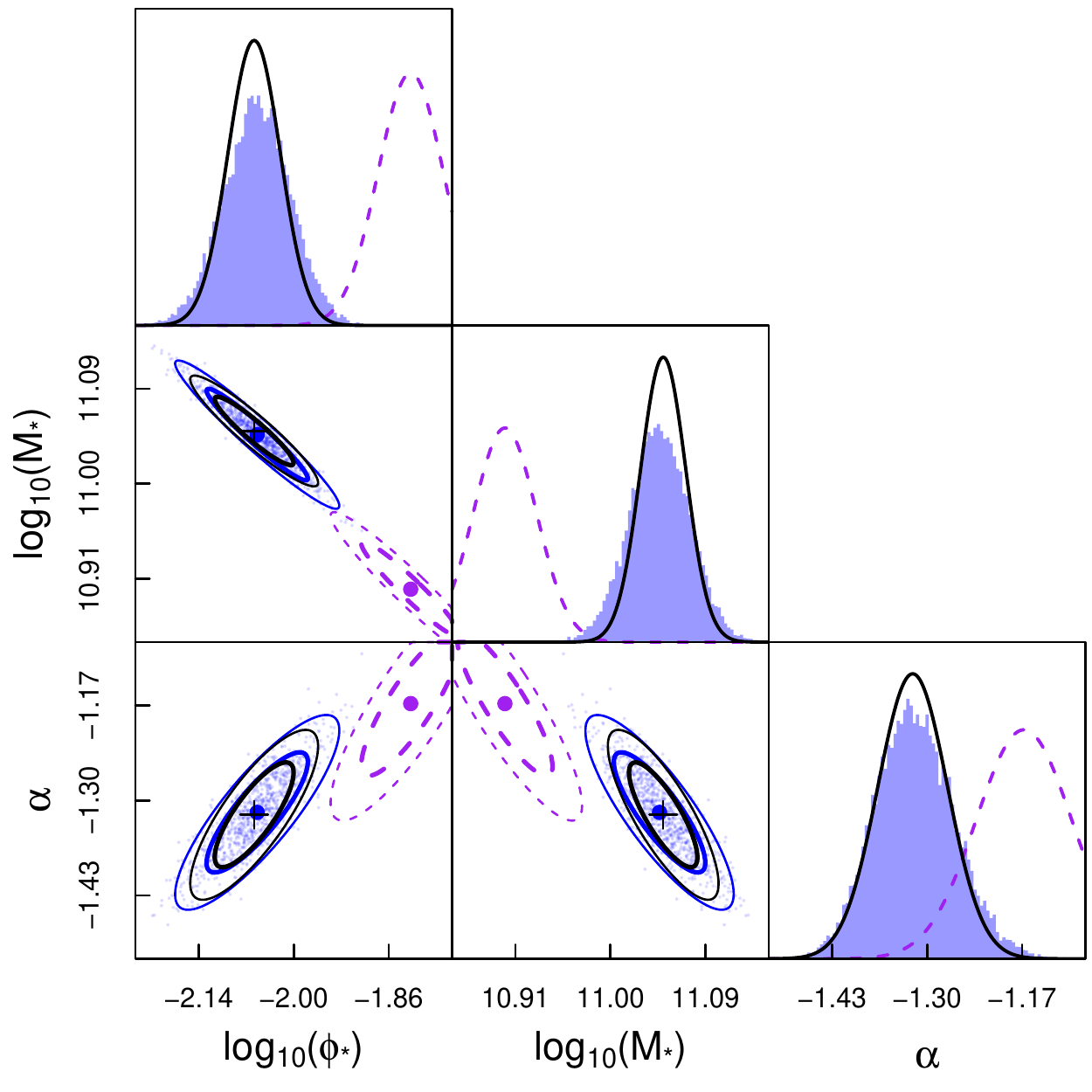}\vspace{-0.3cm}
\caption{Schechter function parameters of a sensitivity-limited mock survey with mass-dependent uncertainties. Black crosses show the input parameters. Their covariances, computed from the Hessian of the modified likelihood function, are shown as black thick (68\% confidence) and thin (95\%) ellipses. Blue point-clouds and histograms represent the fitted parameters for $10^4$ random mock samples. Their means and covariances are shown as big blue dots and ellipses. Purple dots and dashed ellipses indicate the parameter solutions, if the observational uncertainties are incorrectly assumed to be Gaussian. Details are given in \ss{varsigma}.}\label{fig:varsigma_cov}
\end{center}
\end{figure}

We then evaluate the observed PDFs $\rho_i(x)$ via \eq{rhoderived} and apply the MML algorithm to recover the most likely parameters of the Schechter function generating the data. This experiment is repeated with $10^4$ random mock samples drawn from the same population, resulting in the fitted parameter distributions shown in blue in \fig{varsigma_cov}. The excellent match between the input parameters (black crosses) and the numerical expectation of the MML solution (big blue dots) demonstrates the applicability of the MML method to such non-trivial error models.

Note that the observed probabilities $\rho_i(x)$, computed via \eq{rhoderived}, are \emph{not} Gaussian in this example, despite the Gaussian form of $\varrho(x_{\rm obs}|x)$. This is a subtle, but crucial point. If, instead of using \eq{rhoderived}, we incorrectly forced the observed probabilities $\rho_i(x)$ to be Gaussians with standard deviations $\sigma(x_i)$, the most likely fitted parameters (purple in \fig{varsigma_cov}) were no longer consistent with the input model. This comparison emphasizes the general point that correct error models of the data are important for an accurate recovery of the population model -- a statement that is not specific to the MML formalism.

\begin{figure}
\begin{center}
\includegraphics[width=\columnwidth]{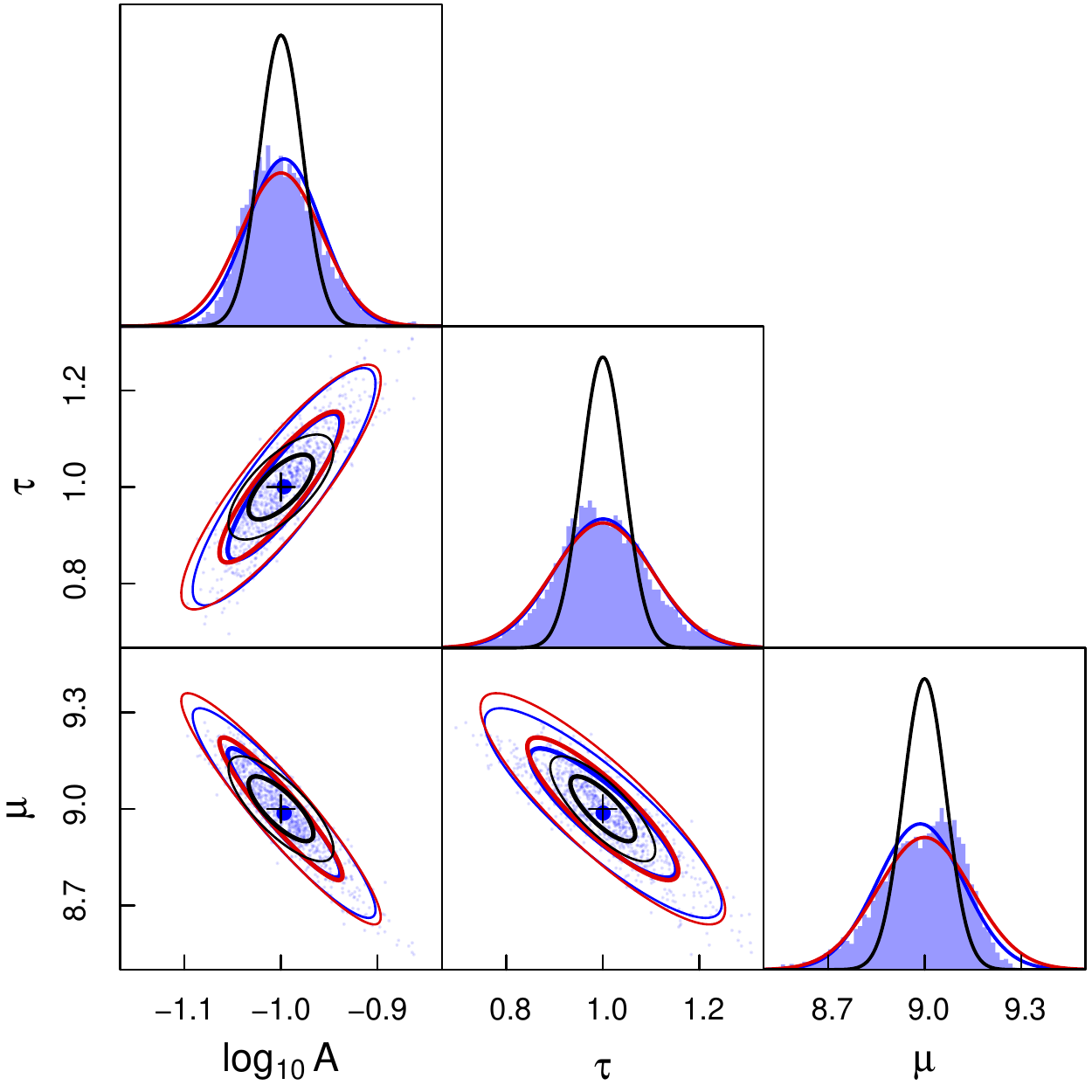}\vspace{0.1cm}
\caption{Illustration of parameter covariances in the case of very large observing errors that are comparable to the width of the MF given in \eq{gauss}. The adopted input parameters are shown as black crosses. The black ellipses centred on these crosses show the 68\% (thick lines) and 95\% (thin line) confidence intervals of fitted parameters predicted from the Hessian of the modified likelihood function. These predictions differ significantly from the best-fitting solutions of $10^3$ random mock samples (blue). However, the covariance of the fitted parameters is well predicted using a resampling method (red lines). Details are given in \ss{resampling}.}\label{fig:resampling}
\end{center}
\end{figure}

\subsection{Resampling uncertainties}\label{ss:resampling}

The computations of parameter (co)variances and Bayes factors presented so far relied on estimating the covariance matrix from the Hessian matrix of the modified likelihood function (Laplace approximation). As noted in \ss{uncertainties}, this approach is only valid if the likelihood function is approximately Gaussian and if the uncertainties of the data are smaller than their range (standard deviation). In most realistic examples this is indeed the case, and the straightforward Hessian covariance estimations work well (\cf bottom panel of \fig{medium}) -- except that they exclude LSS uncertainties, as illustrated in \ss{lssexample}.

The Hessian covariances become inaccurate if the observational uncertainties of the log-masses $x$ are close to or larger than the range (standard deviation) of the log-masses themselves. This is best illustrated by adopting a Gaussian MF with a controlled standard deviation $\tau$,
\be\label{eq:gauss}
	\phi(x|\para)_{\rm Gaussian} = \frac{A}{\sqrt{2\pi\tau^2}}\exp\left(-\frac{(x-\mu)^2}{2\tau^2}\right).
\ee
The parameters $A$ and $\mu$ set the amplitude and mode of the MF. As an example, we pick the parameters $\para=(\log_{\rm10}A,\tau,\mu)=(-1,1,9)$ and sample the MF adopting a constant effective volume without LSS. We generate $10^3$ mock surveys, each with an expected number of $N=10^3$ galaxies. The log-masses $x$ are perturbed by Gaussian random errors of standard deviation $\sigma=1$. Hence, the data uncertainty is identical to the width of the MF $\tau$.

The distribution of the fitted model parameters is shown as blue histograms and point-clouds in \fig{resampling}. The blue ellipses are the 1-sigma (68\%) and 2-sigma (95\%) contours fitted directly to these points. For comparison, the black Gaussians and ellipses show the average (co)variances computed from the Hessian matrices. The mean values of the fits (big blue dots) agree with the input parameters (black crosses), showing that the expectation of the MMLE is correct, \ie the estimator bias (\ss{bias}) is negligible for $N=10^3$ galaxies. However, the (co)variances estimated from the Hessian are clearly too small, showing the failure of the Hessian approximation in the presence of large observing errors.

It is possible to compute accurate parameter covariances for any data errors via the bootstrapping method described in \ss{uncertainties}. This technique is implemented in \dftools and activated by setting \texttt{n.bootstrap} when calling \texttt{dffit}. The argument \texttt{n.bootstrap} is the integer number of resampling iterations, called $Q$ in \ss{uncertainties}.

The average covariances obtained by bootstrapping are shown as the red lines in \fig{resampling}. They agree with the numerical expectations (blue), within the statistical uncertainties of these expectations. This example demonstrates the power of bootstrapping in computing the covariances. Moreover, bootstrapping allows an accurate sampling of the parameter posterior, even if the parameter correlations are highly non-linear, \ie if the covariance matrix provides a poor description of the parameter uncertainties. If, in a particular instance, the user does not know whether the Hessian parameter uncertainties are good enough, it suffices to activate the bootstrapping mode and compare the covariance matrices of the two approaches.

\begin{figure}
\begin{center}
\includegraphics[width=\columnwidth]{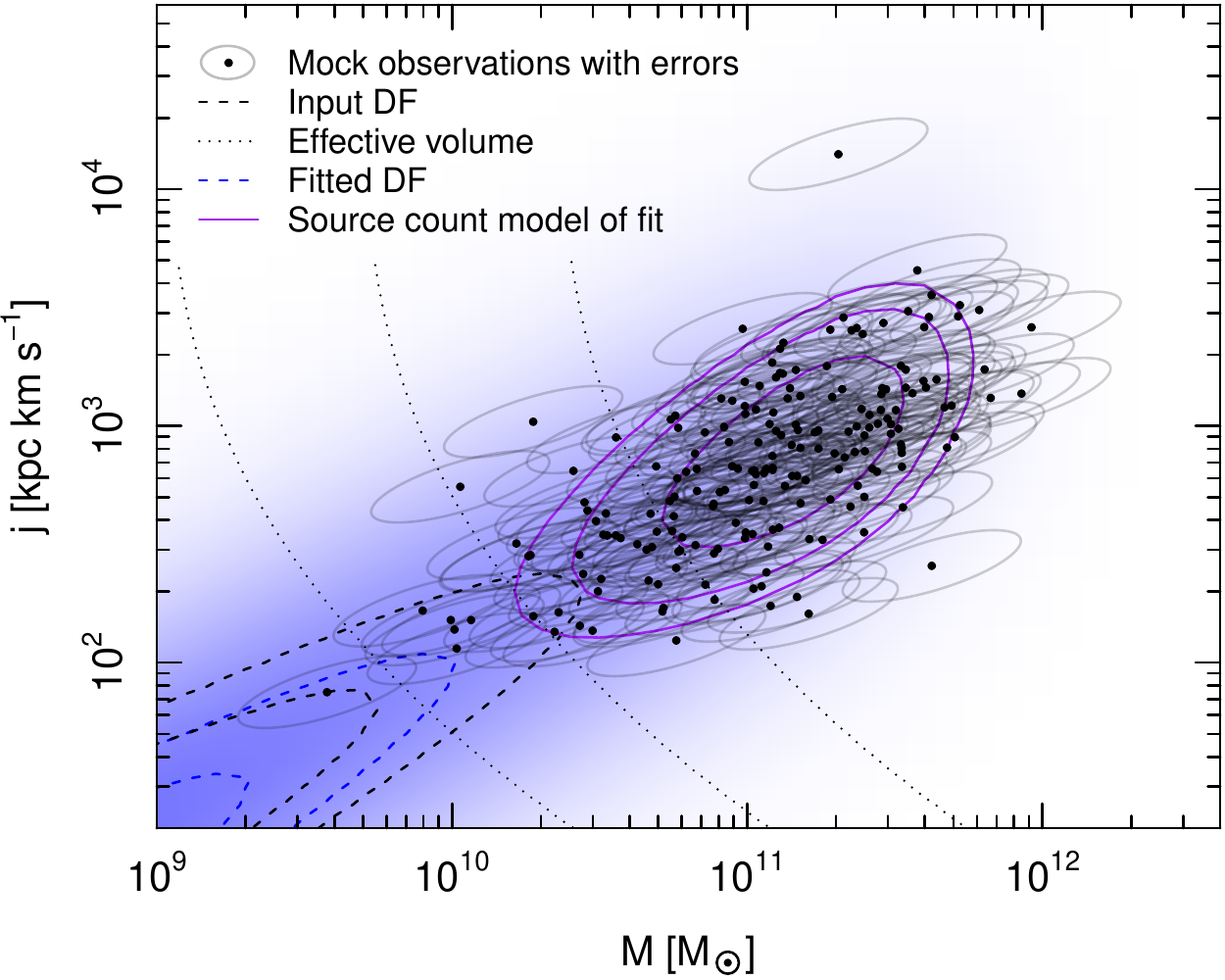}\vspace{-0.1cm}
\includegraphics[width=\columnwidth]{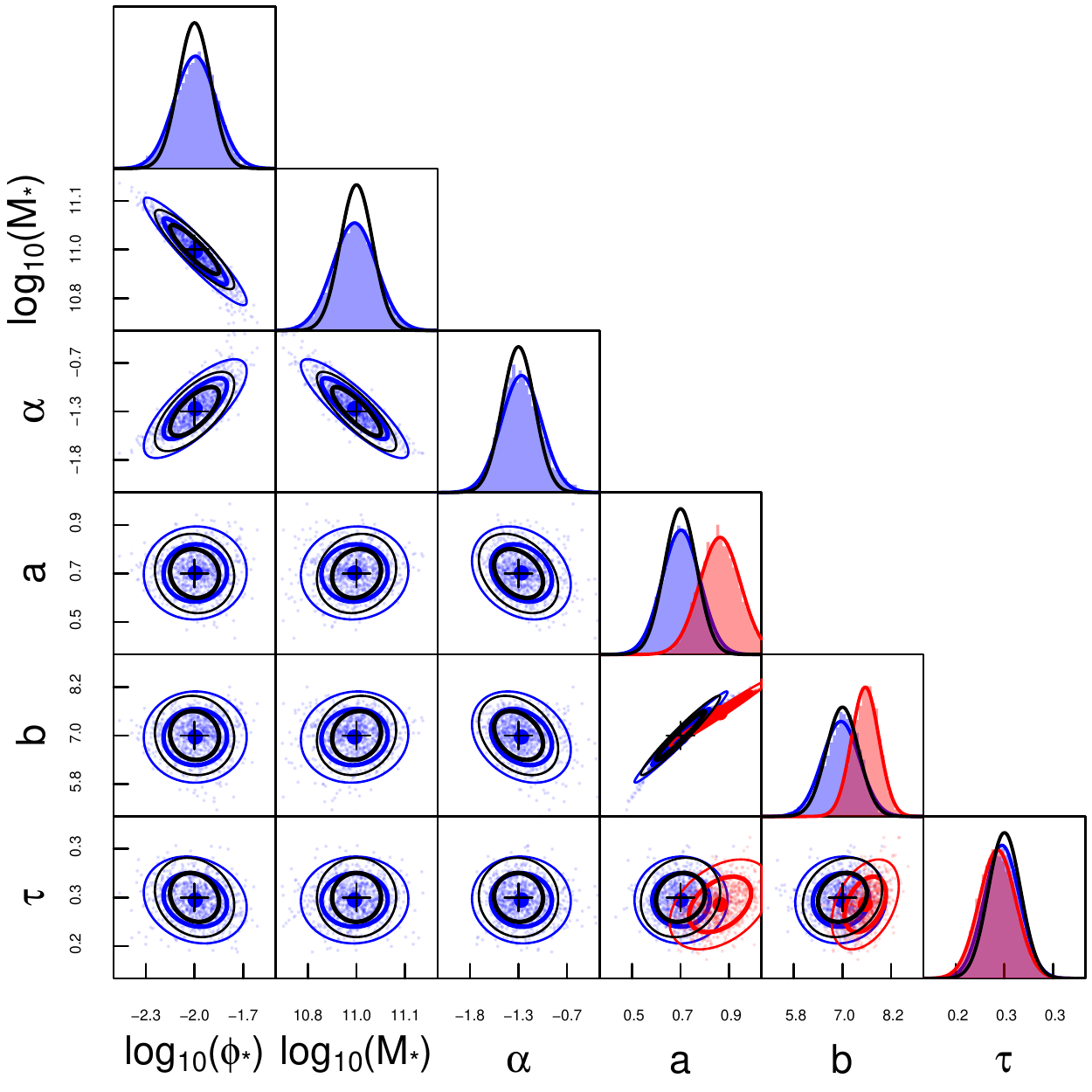}\vspace{-0.3cm}
\caption{Example of fitting a 2D mass-angular momentum distribution function to a mock survey. The top panel shows the data of a single random survey, drawn from the 2D DF in \eq{phi6} using the effective volume of \eq{veffmj}, both represented by isocontours spaced by factors of 2. The blue shading and dashed blue isocontours represent the best MML fit recovered from this particular mock survey. The MML fit is repeated with $10^4$ random mock samples, resulting in the distribution of best-fitting parameters (blue) shown in the bottom panels. The mean fits (big blue dot) are in excellent agreement with the input model (black crosses), and the parameter distributions (blue ellipses) roughly match the Hessian predictions (black ellipses). The \hyperfit solution is shown in red. Details are given in \ss{2dexample}.\vspace{-0.3cm}}\label{fig:mj}
\end{center}
\end{figure}

\subsection{Two-dimensional distribution}\label{ss:2dexample}

Finally, this section illustrates the MML fitting of a multi-dimensional DF, theoretically discussed in \ss{multi}. We limit the example to the two-dimensional (2D) mass-angular momentum distribution of galaxies: each galaxy has two observables, its mass $M$ and specific angular momentum $j$. As in all previous cases, we won't further specify the type of matter to which these quantities apply. The two observables are summarized in the vector $\x=(x_1,x_2)$ with components $x_1=\log_{10}{M/\msun}$ and $x_2=\log_{10}{j/[\unitj]}$. We assume that the population is described by the 2D generative DF, first introduced by \cite{Choloniewski1985},
\be\label{eq:phi6}
	\phi(\x|\para)_{\rm Mj} = \frac{\ln(10)\phi_\ast\mu^{\alpha+1}}{\sqrt{2\pi\tau^2}}\exp\left(-\frac{[x_2-a(x_1-b)]^2}{2\tau^2}-\mu\right),
\ee
which is a Schechter function for the mass distribution, combined with a power law $M$--$j$ relation of slope $a$, zero-point $b$ and Gaussian scatter $\tau$. The `true' parameters are assumed to be $\para_{\rm true}=(\log_{10}\phi_\ast,\log_{10}M_\ast,\alpha,a,b,\tau)=(-2,11,-1.3,\frac{2}{3},7,0.3)$, adopting the same three Schechter parameters as before (\eq{true}) and the rounded values of the observed $M$--$j$ distribution of baryons (stars and cold gas) in disk galaxies \citep{Obreschkow2014}. Isocontours of this DF are shown in \fig{mj}.

To draw galaxy samples from this $M$--$j$ distribution, we adopt an effective volume that depends both on $M$ and $j$: $\veff(\x)$ scales with $M$ as in a sensitivity-limited survey with constant mass-to-light ratio (\eq{sensitivitylimit}) and it depends on $j$ following an error-function that is roughly constant for $x_2>3$, but decreases rapidly for smaller angular momenta, mimicking a natural decrease in low-$j$ detections due to the difficulty of measuring the sizes of these small galaxies. Explicitly,
\be\label{eq:veffmj}
	\veff(\x)\propto 10^{1.5x_1}\left[{\rm erf}(x_2-3)+1\right].
\ee
Three isocontours of $\veff(\x)$ are shown in \fig{mj}.

A random mock sample of $N=200$ galaxies is drawn from $\phi(\x|\para_{\rm true})_{\rm Mj}\veff(\x)$ and perturbed by random errors drawn from a fixed covariant Gaussian distribution (shown as grey ellipses in \fig{mj}). The resulting randomized data are shown as black points in \fig{mj}. Fitting the six-parameter DF of \eq{phi6} to these mock data results in the model shown in blue. The `source count model of the fit', also shown in \fig{mj} represents the distribution of galaxies expected if no observing errors were made.

To check the statistical accuracy of the fitting solution, we generate and fit $10^4$ independent mock samples. The distributions of the best-fitting parameters are shown in blue in the bottom panel of \fig{mj}. For comparison, we also show the input parameters and expected Gaussian uncertainties from the average Hessian matrix of the modified likelihood functions (black). As in the case of fitting a MF (\eg \fig{medium}), the expectation of the fit (big blue dots) is statistically consistent with the input parameters (crosses). The expected covariances (blue ellipses) are approximately consistent with the Hessian prediction (black ellipses) -- the slight deviation being due to the slight inaccuracy of the Hessian approach, explained in \ss{uncertainties} and corrected in \ss{resampling}.

\cite{Robotham2015} derived a general method for fitting $D$-dimensional data ($D\geq2$) with a $(D-1)$-dimensional linear model, \ie a straight line if $D=2$. This method, implemented in the \hyperfit package for $R$, accounts for heteroscedastic errors that are correlated between the different dimensions. Thus, \hyperfit normally outperforms standard regression or bisector techniques, frequently used in astronomy. However, nor \hyperfit or these other techniques account for the fact that the data were sampled via a known or unknown selection function (for extensions see, \citealp{Pihajoki2017}). The MML approach overcomes this limitation: the example of \fig{mj} can be regarded has fitting a linear model, $x_2=a(x_1-b)$, with intrinsic scatter $\tau$ to the $M$--$j$ relation, while accounting for the \emph{known} effective volume and \emph{unknown} MF. Applying \hyperfit to the same data results in the red parameters in \fig{mj}. It appears that ignoring the non-uniform distribution of the data leads to overestimating the slope $a$, which also affects the offset $b$. Hence, accounting for the MF and effective volume is important when fitting the $M$--$j$ relation in this example. In conclusion, \dftools can be used as a generalization of \hyperfit to account for non-uniform data. 

%%%%%%%%%%%%%%%%%%%%%%%%%%%%%%%%%%%%%%%%%%%%%%%%%%%%%%%%%%%%%%%%%%%%%%%%%%%%%%%%%%%%%%

\section{Conclusion}\label{s:conclusion}

The purport of this work reaches beyond astrophysics. The central aim was to develop, implement and test a general method to determine the most likely $P$-dimensional ($P\geq1$) parameter $\para$ of a distribution function model $\phi(\x|\para)$, generating a sample of $N$ objects with $D$-dimensional ($D\geq1$) properties $\{\x_1,...,\x_N\}$. This inference problem is subject to two simultaneous complications: First, the data (\ie the values of $\x_i$) can have arbitrary, heteroscedastic measurement uncertainties, known only up to the prior probability distributions $\rho_i(\x)$, also called `belief functions' in statistics \citep{Denoeux2013}. Secondly, the data sample can be biased in that the probability $\veff(\x)$ of detecting an object of the underlying population depends on its true value $\x$. This problem is very generic, because data uncertainties and sample biases naturally appear in many applications \citep[\eg][]{Aggarwal2009}. In fact, they are almost inevitable whenever the data are gathered using subsamples of larger sets, imperfect sensors or mathematical approximations, such as extrapolation.

We found that the solution to this challenging inference problem is provided by the implicit \eq{masterequation}, which relies on our modified likelihood function $\L(\para,\hat\para)$. The solution of \eq{masterequation} was shown to be unique and identical to the maximum of the `true' likelihood function, \ie the likelihood of the full Bayesian hierarchical model that treats the uncertain measurements as $N\cdot D$ additional model parameters. However, \eq{masterequation} can be solved orders of magnitude faster than this hierarchical model using the iterative fit-and-debias algorithm of \ss{alg}. Its implementation in the $R$-package \dftools was tested thoroughly, demonstrating a quick convergence towards the correct solution. Gaussian uncertainties and covariances of the model parameters can be estimated using the Hessian matrix of the modified likelihood function (\ss{uncertainties}). However, this approach fails if the uncertainties of the data are larger than their range. In this case, parameter covariances either require evaluating the full Hessian (\eq{htrue}) or, more conveniently, bootstrapping the data (\ss{resampling}).

In astrophysics, the most prominent application of the MML method is the fitting of space DFs, quantifying the number of astrophysical objects per unit cosmic volume as a function of some intrinsic property $\x$. The detection probability $\veff(\x)$ can then be interpreted as the effective volume (hence the symbol) in which the objects can be detected.

By far the most common space DFs are MFs (or LFs), which therefore dominate the terminology of this article. MFs are one-dimensional generative DFs ($D=1$) with scalar observables $x=\log_{10}(M/\msun)$. The natural steepness of galaxy MFs makes their recovery prone to Eddington bias -- an effect that many modern galaxy surveys tend to neglect. However, various mock examples (Sections~\ref{ss:large} and\ \ref{ss:medium}) demonstrate that Eddington bias is very significant compared to the otherwise shot noise and LSS limited fitting uncertainties. The same examples prove that the MML method robustly removes Eddington bias, given a model of the observational uncertainties.

With the fast development of galaxy redshift surveys with imaging capabilities or even integral field spectroscopy modes, analyses of higher-dimensional DFs ($D\geq2$) are on the rise. Prominent examples include the 2D mass--size \citep{Lange2015}, mass--angular momentum \citep{Romanowsky2012} and spin--ellipticity \citep{Emsellem2011} distributions, as well as the 3D mass--size--velocity \citep{Koda2000} and mass--spin--morphology \citep{Obreschkow2014} distributions. The implementation of the MML method in \dftools accurately handles such higher-dimensional DFs as illustrated in one example (\ss{2dexample}). In particular, the method can also be used to fit linear models of any dimension, similarly to the \hyperfit method \citep{Robotham2015}, but accounting for arbitrary selection functions.

A specific problem with MFs and other space DFs of astrophysical objects is that the detectability of these objects not only depends on their intrinsic properties $\x$, but generally also on the distance $r$ to the observer, sometimes even on the 3D position $\r$. This addition to the problem is further complicated by the inevitable presence of unknown cosmic LSS. Often, the detectability also depends on hidden properties (\eg the galaxies' inclination or colour) that are not part of the fitted observables $\x$. All these effects can be accounted for in the definition of the effective volume (\ss{gradual}), which, in the case of LSS, depends on the best-fitting model parameters $\hat\para$ (\ss{lsstheory}).

Let us now turn to some limitations of the current presentation of the MML method, which might require further investigation. First, there are a number of secondary uncertainties, not yet included in the MML method or any other fitting algorithm to our knowledge:
\bi
\item \textit{Uncertain selection functions:} In principle, the formalism could be extended to include such uncertainties, probably at the cost of slowing down the algorithm considerably. At the moment, we recommend to adopt a bootstrapping technique, \ie a wrapper around \dftools that resamples the selection function and refits a MF at each iteration.
\item \textit{Uncertainties in the measurement uncertainties}: In practice, the functions $\rho_i(x)$ are themselves subject to both systematic and random errors. The former are hard to address, but the effect of random errors can again be estimated by refitting MFs to different choices of $\{\rho_i(\x)\}$.
\item \textit{Distance uncertainties for LSS:} We have not included distance uncertainties in estimating and removing LSS bias. (Of course distance uncertainties can be included in the uncertainties of $\x$, such as in mass uncertainties.) In the case of spectroscopic redshift measurements, these uncertainties are negligible relative to the typical scales of density fluctuations that dominate LSS. Only photometric redshift measurements might require accounting for distance uncertainties in the removal of LSS bias.
%To our knowledge, this rather marginal problem has not yet been addressed in the literature.
\ei

Another important aspect not considered in this work is cosmic evolution, which makes DFs, such as galaxy MFs depend on redshift $z$ or comoving distance $r$. Of course, it is possible to subdivide a galaxy sample into different redshift bins and fit a MF individually to each of them to evidence trends in the parameter evolution. However, sometimes it is (arguably) desirable to fit just one or two additional MF parameters of an analytic evolution model. ML methods dealing with this case have been presented \citep{Lin1999,Loveday2012}, but they do not simultaneously account for observational errors (Eddington bias). Fitting evolution models in the context of the MML method is possible, but it is not as straightforward as including redshift or distance as an additional observable in $\x$. Fitting evolution models requires extending the formalism to redshift-dependent DFs $\phi(\x|z,\para)$, which will lead to a redshift-integral.

The natural next step is to apply the MML method to real galaxy data from existing and future surveys. This brings all the benefits of the standard ML method, while fully accounting for major empirical unknowns, especially mass errors and LSS. Therefore, MML fits allow a robust comparison of different data sets and, within the Laplace approximation, a clean identification of the best DF model. More generally, we hope that the MML estimator and \dftools will spread to other fields within and outside astrophysics, where measurement uncertainties and sample biases play a significant role in statistical inference.

%%%%%%%%%%%%%%%%%%%%%%%%%%%%%%%%%%%%%%%%%%%%%%%%%%%%%%%%%%%%%%%%%%%%%%%%%%%%%%%%%%%%%%

\section*{Acknowledgements}
Parts of this research were conducted by the Australian Research Council Centre of Excellence for All-sky Astrophysics (CAASTRO), through project number CE110001020. We thank the anonymous referee for their in-depth feedback.

%%%%%%%%%%%%%%%%%%%%%%%%%%%%%%%%%%%%%%%%%%%%%%%%%%%%%%%%%%%%%%%%%%%%%%%%%%%%%%%%%%%%%%

%\bibliography{../Bibliography/astro}
%\bibliographystyle{mnras}

%%%%%%%%%%%%%%%%%%%%%%%%%%%%%%%%%%%%%%%%%%%%%%%%%%%%%%%%%%%%%%%%%%%%%%%%%%%%%%%%%%%%%%

\appendix

\section{Proof of MML solution}\label{a:exactsolution}

The MML method presented in this article is able to produce rapid parameter fits due to a subtle approximation, namely that the posterior PDF $\tilde{\rho}_i(x|x_i,\para)$ -- the probability of a given observation $x_i$ to have true properties $x$ -- is fully specified by its prior PDF, $\rho(x|x_i)$, and the generative DF model \textit{at the ML solution}. 
With this approximation, one is able to define the ``posterior" PDF of each observation as (\eq{debias})
\begin{equation}
  \tilde{\rho}_i(x|\para) = \frac{\rho_i(x|x_i)\phi(x|\para)V(x)}{\int \rho_i(x|x_i) \phi(x|\para)V(x)dx}.
\end{equation}
Effectively, this equation makes the point that if we just knew the true generative distribution (via its parameters $\para$), then we would know the 
true probability of each observation having underlying properties of $x$.
Of course, we don't know the true generative distribution a priori. This suggests splitting the problem into two layers of iteration. First, in a meta-iteration, 
	make a choice of parameters, $\hat{\para}_0$, calculate the expensive factor $\tilde{\rho}_i$, and then solve (by downhill iteration) for the best set of parameters,
	$\para'_0$ using the likelihood \eq{likelihood}.
This inner optimization is cheap, since the expensive calculation is only performed once. Following this, set the ``guess", $\hat{\para}_1$, to $\para'_0$, and so on
	until $\para'_i = \hat{\para}_i$, at which point we postulate that $\para\equiv\hat{\para}_i$.

It is clear that this is an approximation. The true posterior PDF for each datum must involve marginalizing over the posterior of $\para$, that is, the true posterior probability of any observation having true value $x$ is a weighted sum over the distribution-corrected prior for all possible values of the parameters, not just their most likely value.
It may be surprising then that this approximation `works'.

Here we present the true likelihood, based on full marginalization over the posterior, and show that it analytically produces the same MLE as the modified likelihood. 
For completeness, we also show that it does \textit{not} produce the same Hessian (and therefore covariance) as the modified likelihood.

Let the underlying data, $\hat{x}_i$, be drawn from $\lambda(x|\para) = \phi(x|\para)V(x)$, and let the observed data, $x_i$, include some uncertainty in its measurement, $x_i \sim \rho'_i(x|\hat{x}_i)$ (note that $\rho'_i \neq \rho_i$ in general, except for symmetric distributions, but that one can be computed from the other, cf. Appendix \ref{a:ordering}). 
A Bayesian hierarchical model will require an estimation of both the model parameters $\para$ and the underlying value $\hat x_i$. 
In particular, for a Poisson-distributed set of infinitesimal bins in $x$, the likelihood is the product of the expectation, $\lambda$, at each estimate $\hat{x}$, normalized by the expectation of the total number of observations, and then multiplied by the prior probability of measuring the set of $x_i$ given the estimates $\hat{x}_i$. Succinctly, the log-likelihood is
\begin{equation}
	\ln \L(\hat{x}, \para) = \sum_i \ln \lambda(\hat{x}_i|\para) + \ln \rho_i(x_i| \hat{x}_i) - \int \lambda(x|\para) dx,
\end{equation}
where $\hat{x}=\{\hat x_i\}$.

We are not typically interested in the values of $\hat{x}_i$, but rather in $\hat{\para}$, so we may marginalize over the former by integrating over the likelihood (not the log-likelihood!). 
Doing so involves a highly multi-dimensional integral (as many dimensions as there are data). 
However, so long as the uncertainties on each are independent, the integral is completely factorizable, and the final likelihood is
\begin{equation}
	\label{eq:full_likelihood}
 \ln \L(\para) = -\int \lambda(x|\para) dx + \sum_i \ln \left(\int \lambda(x|\para) \rho_i(x) dx \right).
\end{equation}
We note that while this equation is very similar to \eq{likelihood}, the logarithm is outside the integral in this case, and there is no $\tilde{\rho}$ factor involved.

The ML solutions are defined as $\hat{\para} = \para$ at which the Jacobian of $\ln \L$ is zero (note that there may in general be more than one solution). We find that the Jacobian of \eq{full_likelihood} is
\begin{equation}
	\label{eq:full_jac}
	\mathbf{J}_{\rm true}(\para) = \frac{\partial \ln \L}{\partial \para} = -\int\!\!\lambda_\para (x)dx + \sum_i \frac{\int \lambda_\para(x) \rho_i(x) dx}{\int \lambda(x|\para) \rho_i(x) dx},
\end{equation}
where a subscript denotes partial differentiation. 
Suppose that an MLE solution exists and further that it is unique (multiple peaks will be a challenge for either likelihood), and let such a solution be denoted $\hat{\para}$, i.e. $\mathbf{J}_{\rm true}(\hat{\para}) = 0$ and $\mathbf{H}_{\rm true}(\hat{\para})$ negative definite (with $\mathbf{H}_{\rm true}$ the Hessian). 
To show that the MML method will yield the same solution, our task is then threefold: (i) show that the MML likelihood admits an identical MLE, (ii) show that this solution is unique, and (iii) show that the iterative procedure always converges towards this solution.

In an iterative solution with $n$ meta-iterations, let $0\leq j <n$ be the current meta-iteration, and $\hat{\para}_j$ be the current ``meta-estimate" of $\para$. The Jacobian, with respect to the inner-iteration parameters, $\para$,  of \eq{likelihood} is
\be\begin{split}
\label{eq:mml_jac}
	& \mathbf{J}_{\rm MML}(\para, \hat{\para}_j) = -\int\!\!\lambda_\para (x)dx + \sum_i \int dx \tilde{\rho}_i(x|\hat{\para}_j) \frac{\lambda_\para(x)}{\lambda(x|\para)} \\
    &= -\int\!\!\lambda_\para (x)dx + \sum_i  \frac{\int dx \rho_i(x)\lambda_\para(x) \lambda(x|\hat{\para}_j)/\lambda(x|\para)}{\int dx \rho_i(x) \lambda(x|\hat{\para}_j)}.
\end{split}
\ee

The existence of the solution is shown easily by substituting $\para = \hat{\theta}_j = \hat{\para}$. In this case, the $\lambda$ in the numerator and denominator of \eq{mml_jac} cancel one another, to leave precisely $\mathbf{J}_{\rm true}(\hat{\para})$ which by construction is zero.

The uniqueness of the solution is easily shown by contradiction. 
Suppose an alternative solution, $\tilde{\para}$, was found. 
To be the final solution to the fit-and-debias algorithm, this requires that the solution within the iteration, $\para = \para'$, is the same as the ``meta-estimate'' $\hat{\para}_j$, i.e. we suppose that $\para' = \hat{\para}_j = \tilde{\para}$, and that $\mathbf{J}_{\rm MML}(\tilde{\para},\tilde{\para}) = 0$. 
However, whenever $\para = \hat{\para}_j$, we have $\mathbf{J}_{\rm MML}(\para',\para') \equiv \mathbf{J}_{\rm true}(\para')$.
Thus we imply that $\mathbf{J}_{\rm true}(\tilde{\para}) = 0$, which is in contradiction to the original statement that $\hat{\para}$ is a unique solution to the true likelihood. 
Conversely, if $\hat{\para}$ were \textit{not} a unique solution to the true likelihood, but rather multiple peaks existed, then it follows that $\tilde{\para}$ must be \textit{one of them}, which shows that the \textit{set} of solutions to the MML likelihood are the same as the true solution (but does not guarantee that the same solution will be reached given the same initial estimate).

Our final task is to show that the iterative scheme always converges towards the solution. 
One could imagine that even though a true solution, $\hat{\para}$, exists, the fit-and-debias algorithm nevertheless diverges or eternally oscillates around this solution while the true likelihood converges. Stated mathematically, our task is as follows. 
%Let $\para'$ be the trial solution, as used in the MML -- this may be the initial guess or the solution from a previous iteration. 
Let $\hat{\para}_j$ and $\para'_j$ retain their meanings from previous arguments, so that $\mathbf{J}_{\rm MML}(\hat{\para}_j,\para'_j) = 0$ (but note that we may not have converged yet, so that we may have $j < n$ and $\hat{\para}_j \neq \para'_j$). 
Then we require that  $|\hat{\para}_j - \para'_j| \leq |\hat{\para} - \hat{\para}_j| \ \ \forall j$. That is, we must show that every meta-iteration returns an estimate which is closer to the true solution. 
We can simplify this picture by noting that the next meta-iteration ($j+1$) will \textit{begin} at $\para = \para'_j$, and use the meta-estimate $\hat{\para}_{j+1} = \para'_j$. 
For the solution of this iteration to be closer to $\hat{\para}$ only requires that the Jacobian at this point be positive if $\tilde{\para} < \hat{\para}$ and vice versa (i.e. gradient must roll towards the true solution from this point). 
Furthermore, it suffices to show that the Jacobian at such a starting point has these properties for just one starting point to the left and right of $\hat{\para}$, as the uniqueness proved above ensures that the Jacobian cannot cross through zero except at $\hat{\para}$. Taking these offset points to the infinitessimal limit, we require only the gradient of the Jacobian at the solution itself, \ie we need only show that
\begin{equation}
	\frac{\partial}{\partial \para} \left. \mathbf{J}_{\rm MML}(\para,\para) \right|_{\hat{\para}} < 0.
\end{equation}
To this end we simply note that $\mathbf{J}_{\rm MML}(\para,\para) = \mathbf{J}_{\rm true}(\para)$ so that the derivative above is merely the true Hessian. This Hessian, by construction, is negative definite at $\hat{\para}$, so we have proved the MML method.

Despite the above proof, the Hessians for both the true likelihood and the MML are not in general the same. For completeness, we here give the Hessian at the MLE in each case:
\begin{align}
	\mathbf{H}^{lm}_{\rm true} &= Q(\hat{\para})\!-\!\sum_i\!\frac{\int\!\!\lambda_{\para_l}(x|\hat{\para}) \rho_i(x) dx \int\!\!\lambda_{\para_m}\!(x|\hat{\para}) \rho_i(x) dx}{\left[\int \lambda(x|\hat{\para}) \rho_i(x) dx\right]^2},\label{eq:htrue} \\
   \mathbf{H}^{lm}_{\rm MML} &=  Q(\hat{\para}) -\sum_i \frac{\int dx \rho_i(x)\lambda_{\para_l}(x|\hat{\para})\lambda_{\para_m} (x|\hat{\para}) /\lambda(x|\hat{\para})}{\int dx \rho_i(x) \lambda(x|\hat{\para})},\label{eq:hmml}
\end{align}
where
\begin{equation}
Q(\hat{\para}) = -\int \lambda_{\para_l \para_m}(x|\hat{\para}) + \sum_i \frac{\int \lambda_{\para_l \para_m}(x|\hat{\para}) \rho_i(x) dx}{\int \lambda(x|\hat{\para}) \rho_i(x) dx}.
\end{equation}

\section{Effective volume for LSS}\label{a:lss}

This appendix derives the equation to estimate $g(r)$ directly from the redshift-distribution of the galaxy sample. In most MF work, the galaxy distances $r_i$ are inferred from spectroscopic redshifts. For the purpose of correcting for LSS bias, we assume that these distances are exact, thus neglecting redshift-measurement errors and redshift-space distortions due to peculiar velocities. In principle both uncertainties can be incorporated in the MML formalism, but in the majority of cases this will likely complicate the formalism without measurable improvement. Note that this assumption of exact distances only applies to the LSS bias correction. Of course, distance uncertainties can be accounted for in the mass uncertainties $\phi_i(x)$, which are a central part of the MML formalism.

To the benefit of intuition, let us temporarily introduce distance bins $k$ of mean redshifts $r_k$ and width $\Delta r$. Let $n_k$ be the observed number of galaxies in each bin $k$. The relative density $g_k$ can then be calculated as the ratio of $n_k$ and the expected number of galaxies in bin $k$ in the absence of LSS,
\be\label{eq:LSSdeltak}
	g_k=\frac{n_k}{\int\int_{r_k-\Delta r/2}^{r_k+\Delta r/2}\phi(x|\hat\para)V^{\prime}(r)f(x,r)dr dx}.
\ee
As in the removal of Eddington bias (\eq{debias}), we here used the most likely model parameters $\hat\para$ to estimate the expected counts. We now let $\Delta r$ become infinitesimal while introducing the densities $g(r)=g_k/\Delta r$ and $n(r)=n_k/\Delta r$. In the isotropic case ($d^3r=V'(r)dr$), \eq{veffgraduallss} then becomes
\be\label{eq:veffgraduallss2}
	\vefflss(x) = \int \frac{f(x,r)n(r)}{\int\phi(x|\hat\para)f(x,r)dx}dr.
\ee
Note that the terms $V^\prime(r)$ have cancelled out. Since we are assuming precisely known distances $r_i$, the source density simplifies to $n(r)=\sum_i\delta_D(r_i-r)$, where $\delta_D$ is the Dirac delta function. Hence, the integral over $r$ reduces to a sum over the galaxies $i$, \ie to \eq{veffgradualautolss}.

\section{Mean effective volumes}\label{a:harmonic}

This section gives a numerical example for the statement that several types of galaxies of identical mass, but different effective volumes can be combined in the MF by constructing the harmonic mean of their effective volumes.

\begin{figure}
\begin{center}
\includegraphics[width=0.85\columnwidth]{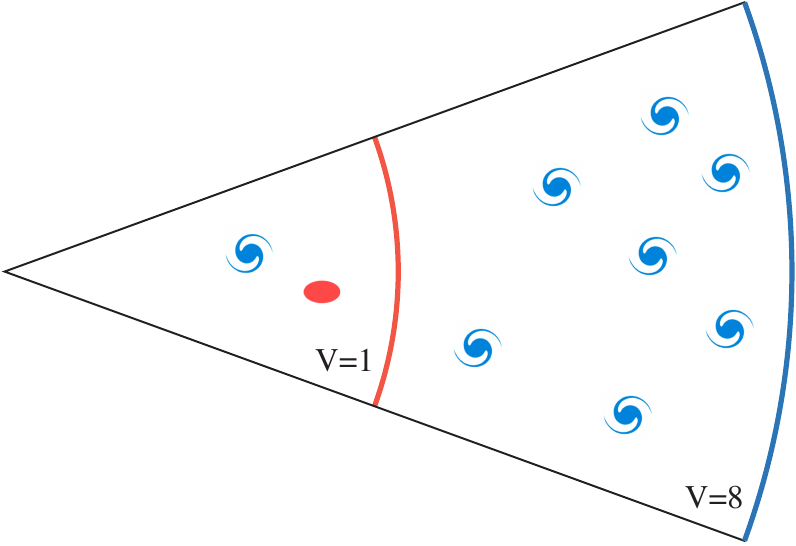}\vspace{0mm}
\caption{Sketch of surveying a simplified universe with red and blue galaxies. Both types are equally frequent, but red galaxies are only detectable in a smaller effective volume. In this case, the MF is recovered by taking the harmonic mean of the effective volumes of all detections, as explained in \a{harmonic}.}\label{fig:harmonic}
\end{center}
\end{figure}

Let us consider a simplified case (\fig{harmonic}) of two types of galaxies, blue and red ones, all of identical log-mass $x$. Our hypothetical universe contains exactly one blue and one red galaxy per each unit volume. Hence, the MF at $x$ has a value of $\phi(x)=2$. We survey a volume of 8 units, which contains 8 blue and 8 red galaxies. The blue galaxies can be detected across the whole survey volume, \ie their effective volume is $V_{\rm blue}=8$ and we count $n_{\rm blue}=8$ such galaxies. The red galaxies are much harder to detect, such they can only be detected in the most nearby volume unit, where there is just one such galaxy, \ie $V_{\rm red}=1$ and  $n_{\rm red}=1$.

In total, there are $n=n_{\rm blue}+n_{\rm red}=9$ galaxies. Since these galaxies must collectively have a MF value of $\phi(x)\equiv n/V=2$, their combined effective volume $V$ must be equal to 4.5. This value is indeed identical to the \emph{harmonic} mean of the effective volumes of all detected galaxies,
\be
	V = \left[\frac{1}{n}\Big(n_{\rm blue}V_{\rm blue}^{-1}+n_{\rm red}V_{\rm red}^{-1}\Big)\right]^{-1} = 4.5.
\ee
In other words, we can evaluate the MF directly by dividing the number of detections by the harmonic mean of their volumes. In doing so the different detectability of different galaxy types is accurately accounted for. This result readily generalizes to any number of galaxies and types of galaxies.

\section{Scatter, then select}\label{a:ordering}

\begin{figure}
\begin{center}
\includegraphics[width=\columnwidth]{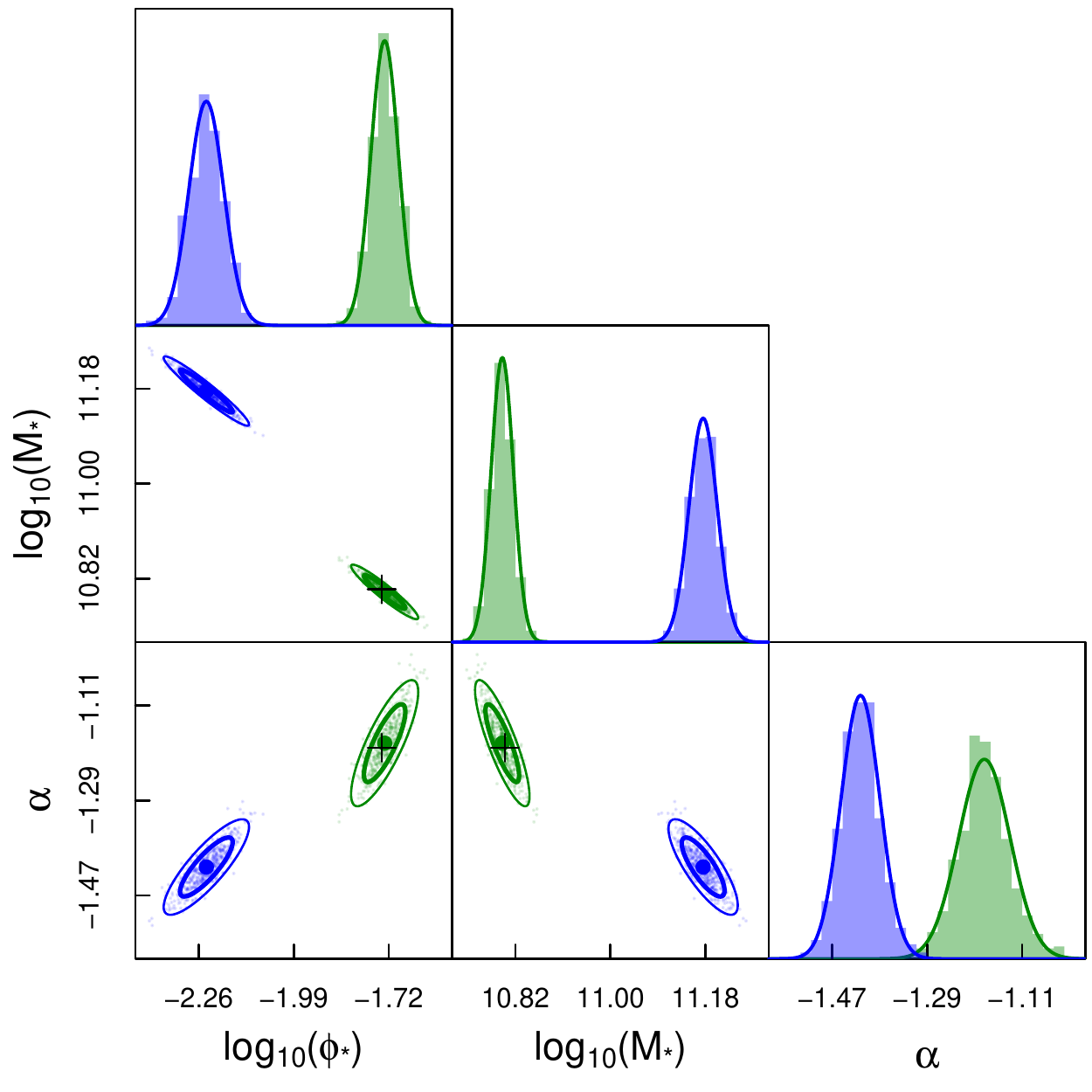}\vspace{-0.3cm}
\caption{Illustration of recovering the Schechter function parameters from mock data, drawn from a Schechter function, then scattered and then selected with a weight $V(x_{\rm obs})$. The true model parameters are shown as black crosses. The big blue \mbox{dots}/ellipses show the average and Gaussian 68\% (thick) and 95\% (thin) contours of $10^3$ random mock samples, fitted by incorrectly assuming that the data was first selected and then scattered. The green \mbox{dots}/ellipses show the fits that account for the correct ordering via the substitutions of equations~(\ref{eq:ordering1}) and (\ref{eq:ordering2}).}\label{fig:ordering_cov}
\end{center}
\end{figure}

This work focused on the case, where data are drawn from a population with weights $\veff(x)$ and then scattered by measurement uncertainties. There are other cases where this order is (partially) reversed. For instance, astronomical observations at the sensitivity limit of the telescope might only detect a few photons per object, subject to Poisson statistics. If we impose a fixed number of photons as the detection limit, the Poisson scattering is followed by the selection. In this particular situation it is possible to approximately include the Poisson statistics in the completeness $C$ (\ss{gradual}), when calculating the effective volume of each galaxy.

There are nonetheless more complex cases. For instance, the photon counting subject to Poisson noise might be followed by an apparent magnitude cut, conversion to absolute magnitudes using uncertain distances, additional cut by absolute magnitude, conversion to masses using uncertain mass-to-light ratios, etc. Such layered uncertainties are typically handled by a hierarchical scheme, where the model MF is convolved with an uncertainty model and passed through a selection function in a repeated sequence; e.g. first convolution with Poisson noise, then apparent magnitude cut, convolution with distance errors, absolute magnitude cut and convolution with mass-to-light error. Currently, only the last uncertainty can then be accelerated by the MML scheme, using the backward correction of \eq{debias} instead of a forward convolution.

Let us provide one example of fitting for a MF, first scattered and then subjected to a single selection. Explicitly the data $x$ are drawn from $\phi(x|\para)$, then scattered ($x\rightarrow x_{\rm obs}$) and then selected with weight $V(x_{\rm obs})$. It suffices to note that scattering first and then selecting is equivalent to drawing data from a modified MF, which has been smoothed (convoluted) by the observing error, without adding further noise. Formally, this corresponds to the following substitution in the formalism of \ss{likelihood},
\bea
	\phi(x|\para) &\rightarrow& \phi'(x|\para) = \int \phi(s|\para)\varrho(x|s)ds,\label{eq:ordering1}\\
	\rho_i(x) &\rightarrow& \rho'_i(x) = \delta_{\rm D}(x-x_i),\label{eq:ordering2}
\eea
where $\varrho(x|s)$ is the probability of observing a value $x$ given a true value $s$, solely due to the uncertainty of the measurement process.

As an illustration, we adopt the advanced example of \ss{varsigma}, which exhibits measurement uncertainties that depend on the true value of the log-mass $x$. The mock data is generated exactly as described in \ss{varsigma}, except that the scattering ($x\rightarrow x_{\rm obs}$) is performed before the selection with weight $V(x_{\rm obs})$. We use $10^3$ random mock samples. As shown in \fig{ordering_cov}, their fitted Schechter function parameters are consistent with the true input values, if and only if the substitutions of equations~(\ref{eq:ordering1}) and (\ref{eq:ordering2}) are included.

%%%%%%%%%%%%%%%%%%%%%%%%%%%%%%%%%%%%%%%%%%%%%%%%%%%%%%%%%%%%%%%%%%%%%%%%%%%%%%%%%%%%%%

\section{Using real observations}

This paper intentionally used mock data instead of real observations to allow for a comparison of the fits against the \emph{known} true solutions. This section illustrates how, explicitly, \dftools is applied to real data. We use the data recently published in \cite{Westmeier2017}, where an early version of \dftools has already been applied to estimate the MF of neutral atomic hydrogen (\ha) in the nearby Sculptor filament, a loosely bound elongated group of galaxies.

In \textit{R}, load the \dftools library and the data of \cite{Westmeier2017}:
{\color{blue}\begin{lstlisting}
library(dftools)
data = read.table("http://quantumholism.co
m/dftools/westmeier2017.txt",header=TRUE)
\end{lstlisting}}
\noindent There are 31 galaxies in this sample, hence the dataframe \texttt{data} has 31 rows and three columns specifying the \ha masses (in $\msun$), the mass uncertainties (standard deviations) and the effective volume s(in $\rm Mpc^3$). We recast these data into separate vectors containing the log-masses $x_i$ ($i=1,...,31$), log-normal uncertainties (using linear error propagation) and effective volumes of each galaxy,
{\color{blue}\begin{lstlisting}
x = log10(data$MHI)
x.err = data$errMHI/data$MHI/log(10)
veff.values = data$Vmax
\end{lstlisting}}
\noindent To fit a Schechter function (the default MF of \dftools) and display the best-fitting parameters, it suffices to call
{\color{blue}\begin{lstlisting}
survey = dffit(x,veff.values,x.err)
dfwrite(survey)
\end{lstlisting}}
\noindent which outputs the text
{\color{blue}\begin{lstlisting}
dN/(dVdx) = log(10)*10^p[1]*mu^(p[3]+1)
*exp(-mu), where mu=10^(x-p[2])
p[1] =   -1.315 (+-0.276)
p[2] =    9.541 (+-0.308)
p[3] =   -1.103 (+-0.147)
\end{lstlisting}}

\begin{figure}
\begin{center}
\includegraphics[width=\columnwidth]{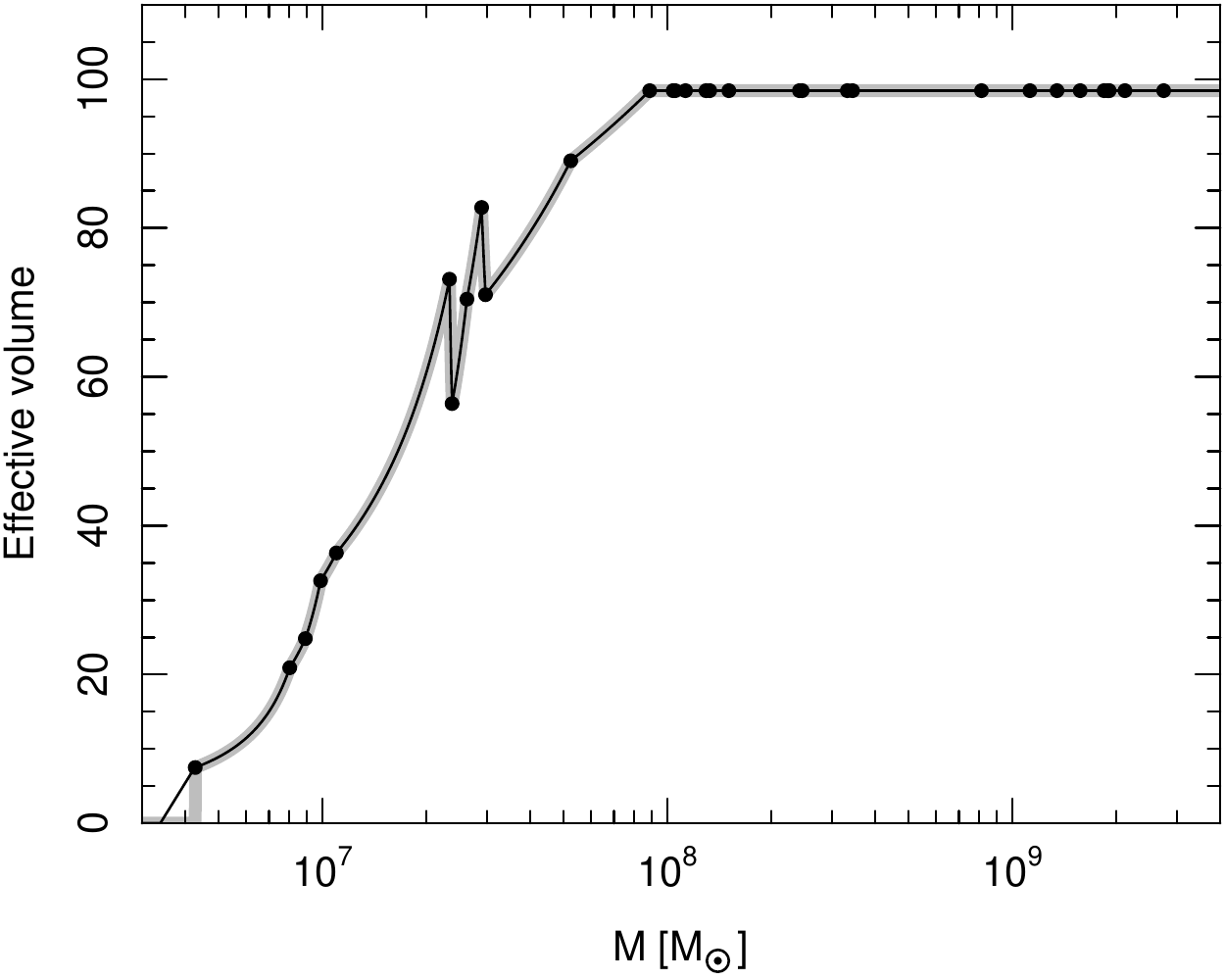}\vspace{-4mm}
\caption{Effective volume function of the Sculptor \ha survey. The individual values provided with the sample are shown as black dots. The default fit derived by \dftools (grey thick line) is slightly corrected manually (black thin line) based on additional knowledge of the survey, not contained in the data itself.}\label{fig:himfveff}
\end{center}
\end{figure}

\begin{figure}
\begin{center}
\includegraphics[width=\columnwidth]{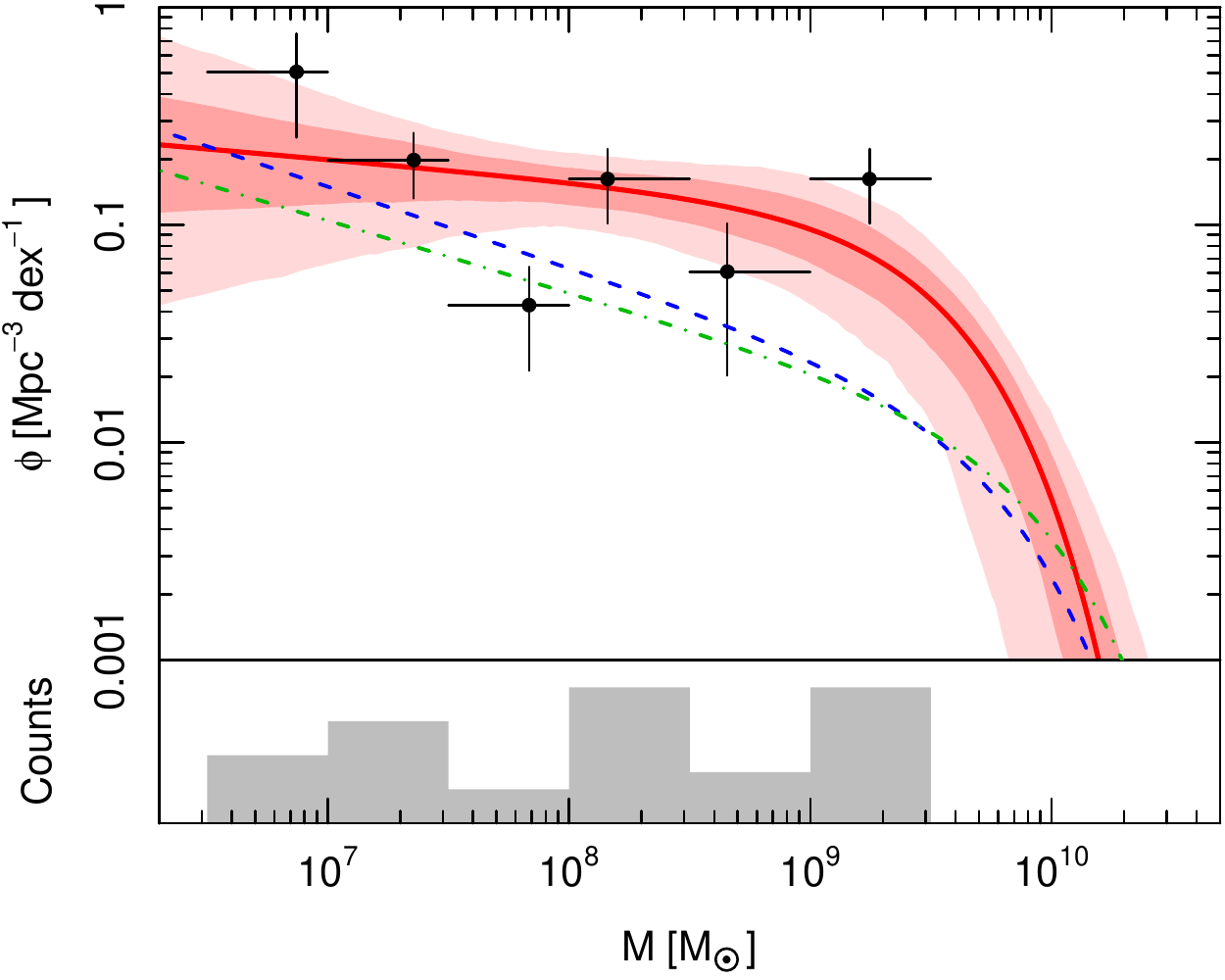}\vspace{-4mm}
\caption{\ha MF of the Sculptor group. The data (shown as binned black points and grey historgram) is fitted with a Schechter function, shown as red solid line with 68\% and 95\% confidence intervals as red shading. The mean \ha MF of the local universe as inferred from the HIPASS and ALFALFA surveys are shown as blue dashed and green dash-dotted lines, respectively.}\label{fig:himf}
\end{center}
\end{figure}

The thick line in \fig{himfveff} shows the effective volume as a function of mass, recovered from linear interpolation of the values $1/\veff(x_i)$ (see Section~\ref{sss:onlyvi}). Outside the observed mass range, the effective volume defaults to $\veff(x)=0$ for $x<\min\{x_i\}$ and $\veff(x)=\max\{\veff(x_i)\}$ for $x>\max\{x_i\}$. If a better model is available from survey-specific considerations, then this information can be exploited to improve the fit. Here, we like to replace the effective volume for $x<\min\{x_i\}$ by $\veff(x)=\max(0,75\cdot(x-6.53))$, while keeping the rest unchanged (thin line in \fig{himfveff}). To apply this modification to the fit, it suffices to define a list \texttt{selection}, composed of the vector \texttt{veff.values} and a new function \texttt{veff.fn}, specifying the effective volume outside the observed mass range:
{\color{blue}\begin{lstlisting}
veff.fn = function(x) {
  veff.max = max(veff.values)
  return(pmax(0, pmin(veff.max,
         (x - 6.53) * 75)))
}
selection = list(veff.values,veff.fn)
\end{lstlisting}}
\noindent In addition to using the modified effective volume (dashed line in \fig{himfveff}), we wish to determine the parameter uncertainties via bootstrapping, hence allowing for asymmetric uncertainties. To use $10^3$ bootstrap iterations with a fixed seed for the random number generator, call
{\color{blue}\begin{lstlisting}
set.seed(1)
survey = dffit(x,selection,x.err,
               n.bootstrap = 1e3)
\end{lstlisting}}
\noindent Finally, we produce the MF plot (\fig{himf}) with 68\% and 95\% confidence regions around the best fit. The chosen graphical arguments display the fitting function in red, display the observed data in black, remove posterior data, suppress the effective volume line and adjust the binning of input data.
{\color{blue}\begin{lstlisting}
mfplot(survey,xlim=c(2e6,5e10),
ylim=c(1e-3,1),uncertainty.type=3,
col.fit="red",col.data.input="black",
show.posterior.data=FALSE,col.veff=NULL,
nbins=6,bin.xmin=6.5,bin.xmax=9.5)
\end{lstlisting}}
\noindent To add the reference \ha MFs of the HIPASS \citep{Zwaan2005} and ALFALFA \citep{Martin2010} surveys, use
{\color{blue}\begin{lstlisting}
x = c(survey$grid$x)
y = dfmodel(x,c(log10(6.0e-3),9.80,-1.37))
lines(10^x,y,lty=2,lwd=1.5,col="blue")
y = dfmodel(x,c(log10(4.8e-3),9.96,-1.33)
lines(10^x,y),lty=4,lwd=1.5,col="#00bb00")
\end{lstlisting}}
\noindent In \fig{himf}, the Sculptor \ha MF clearly differs from the mean \ha MF in the local universe, as discussed by \cite{Westmeier2017}. The best-fitting parameters are displayed via
{\color{blue}\begin{lstlisting}
dfwrite(survey)
\end{lstlisting}}
\noindent resulting in the output
{\color{blue}\begin{lstlisting}
dN/(dVdx) = log(10)*10^p[1]*mu^(p[3]+1)
*exp(-mu), where mu=10^(x-p[2])
p[1] =   -1.308 (+0.252 -0.260)
p[2] =    9.535 (+0.143 -0.197)
p[3] =   -1.097 (+0.174 -0.141)
\end{lstlisting}}
\noindent Note that there are marginal differences to the uncertainty ranges quoted by \cite{Westmeier2017} due to a difference in the bootstrapping technique used in the previous (parametric bootstrapping) and the current (non-parametric bootstrapping) version of \dftools.

% Don't change these lines
\bsp	% typesetting comment
\label{lastpage}
\end{document}